\documentclass{jfm}
\usepackage{graphicx}
\usepackage{epstopdf, epsfig}
\usepackage{color}
\usepackage[normalem]{ulem}
\usepackage{bm}
\usepackage{comment}

\newcommand\Sch{\mbox{\textit{Sc}}}  

\shorttitle{Effects of surface tension reduction on wind-wave growth}
\shortauthor{K. Matsuda, S. Komori, N. Takagaki and R. Onishi}

\title{Effects of surface tension reduction on wind-wave growth and air--water scalar transfer}

\author{Keigo  Matsuda\aff{1}
  \corresp{\email{k.matsuda@jamstec.go.jp}},
  Satoru  Komori\aff{2,1},
  Naohisa  Takagaki\aff{3}
 \and Ryo  Onishi\aff{4,1}}

\affiliation{
\aff{1}Research Institute for Value-Added-Information Generation (VAiG), Japan Agency for Marine-Earth Science and Technology (JAMSTEC), Yokohama 236-0001, Japan
\aff{2}High-Performance Fine-Particle Research Center, Doshisha University, Kyotanabe 610-0394, Japan
\aff{3}Department of Mechanical Engineering, University of Hyogo, Himeji 671-2280, Japan
\aff{4}Global Scientific Information and Computing Center, Tokyo Institute of Technology, Tokyo 152-8550, Japan
}

\newcommand*\Matsuda[1]{\textcolor{black}{#1}}
\newcommand*\MatsudaA[1]{\textcolor{black}{#1}}
\newcommand*\MatsudaB[1]{\textcolor{black}{#1}}
\newcommand*\MatsudaC[1]{\textcolor{black}{#1}}

\begin{document}

\maketitle

\begin{abstract}
Effect of surface tension reduction on wind-wave growth is investigated using  
direct numerical simulation (DNS) of air--water two-phase turbulent flow. 
The incompressible Navier--Stokes equations for air and water sides are solved using an arbitrary Lagrangian--Eulerian method with boundary-fitted moving grids. 
The wave growth \MatsudaC{of finite-amplitude and non-breaking gravity--capillary waves, whose wavelength is 
less than 0.07 m,} is simulated for two cases of different surface tensions \MatsudaC{under a low--wind speed condition of several meters per second}. 
The results show that significant wave height for the smaller surface tension case increases faster than that for the larger surface tension 
case. 
Energy fluxes for gravity and capillary wave scales reveal that, 
when the surface tension is reduced, the energy transfer from the significant 
gravity 
waves to 
capillary waves decreases and the significant waves accumulate more energy supplied by wind.
This results in faster wave growth for the smaller surface tension case.
Effect on the scalar transfer across the air--water interface is also investigated. The results show that
the scalar transfer 
coefficient 
on the water side 
decreases due to the surface tension reduction. 
The decrease is caused by suppression of turbulence in the water side. 
In order to 
support the conjecture, 
the surface tension effect 
is compared   
with laboratory experiments in a small wind-wave tank.

%
\end{abstract}

\begin{keywords}
\end{keywords}

\section{Introduction}

Wind waves are generated by winds blowing over the air--water interface. 
A great number of studies have been published investigating the effects of wind waves on momentum, heat and mass transfers between the ocean and atmosphere, which strongly affect the local and global weather and climate. 
The mass transfer across the air--water interface is in particular an important issue for climate change studies because it is related to the air--sea CO$_2$ transfer \citep{Wanninkhof(2009)}.
\Matsuda{
The study on the growth of wind waves 
also has a long history. 
The pioneering works on the theoretical aspect of the wind wave growth were done by \citet{Phillips(1957)} and \citet{Miles(1957)}, and the understandings have been updated by a 
\MatsudaA{large}
number of analytical, experimental, and numerical studies, including recent studies, e.g., \citet{Gare(2013),Paquier(2015),Paquier(2016),Zavadsky&Shemer(2017),Perrard(2019),Buckley(2020),Wu&Deike(2021)}. 
\citet{Sullivan&McWilliams(2010)} and \citet{Wu&Deike(2021)} well reviewed and introduced the literature.
}
While most published studies have considered wind waves on uncontaminated sea water or fresh water with constant surface tension, 
some studies have discussed the effects of surfactants on the wind-wave structure and have shown that wind waves are suppressed by the surfactants in water
\citep{Scott(1972),Mitsuyasu&Honda(1986)}.
However, the detailed mechanism of the wind-wave suppression has not been confirmed.
The mechanism of wind-wave suppression due to surfactants has been hypothesised based on the Marangoni effect. 
More specifically, the interface deformation is suppressed by the surface tension gradient on the interface, which is referred to as the Marangoni stress  \citep[e.g.,][]{Scott(1972)}. Non-uniform surfactant concentration on the free surface causes non-uniform surface tension distribution because the surface tension decreases as the surfactant concentration increases.
However, this hypothesis is not proven since the Marangoni stress cannot be directly measured.

Furthermore, even in the case where surfactants are uniformly distributed on the free surface, the effect of uniform reduction of surface tension on the growth of wind waves remains unclear.
\citet{Kawai(1979b)} discussed the cause of increase of the critical wind speed for soap water, considering the effect of uniform surface tension reduction. Kawai analytically concluded that the critical wind speed does not change significantly when the surface tension is reduced to half the value of water.
\citet{Tsai&Lin(2004)} suggested that uniform surface tension reduction increases the wind-wave growth rate based on linear stability analysis. However, this result can be applied only to the initial wave growth at very early stages.
\citet{Zonta(2015)} investigated the growth of gravity-capillary waves driven by the shear at the air--water interface using \Matsuda{three-dimensional} numerical simulation and showed that the initial wave growth is faster for a larger Weber number (smaller surface tension). However, they did not clarify the pure surface tension effect because they changed both Froude and Weber numbers. 
\Matsuda{
Recently, \cite{Wu&Deike(2021)} 
numerically
investigated the fundamental mechanism of the growth of gravity-capillary waves with a laminar wind profile, changing the Bond and Reynolds numbers.  
However, \MatsudaA{they considered wind waves whose amplitude was smaller than the viscous sublayer thickness. Therefore,} their two-dimensional simulation results cannot directly be applied to 
three-dimensional air--water flows involving turbulent eddies.
}

In addition, the dependence of uniform surface tension reduction on the mass transfer across the air--water interface is also a subject of interest. 
Previous studies showed that the mass transfer is dominated by turbulence beneath the air--water interface \citep{Komori(1993a),Komori(2010)}. 
\Matsuda{
\citet{Veron&Melville(2001)} 
showed that the effect of scalar transfer is affected by
Langmuir circulations,  
which are streamwise vortices 
developing below a wind-sheared air--water wavy interface.
\MatsudaA{In contrast,} 
\cite{Takagaki(2015)} performed \MatsudaA{direct} numerical simulations \MatsudaA{(DNSs)} and concluded that the scalar transfer across a wind-driven air--water interface is mainly controlled by turbulent eddies, and the effect of the Langmuir circulations is relatively small.
\MatsudaA{Recently, \cite{Tejada-Martnez(2020)} performed DNS and large-eddy simulation and reported that turbulence on the water side is triggered by the Langmuir forcing, whereas the intensity of the turbulence and the scalar transfer induced by the turbulence 
are driven by the wind shear.}
If the uniform surface tension changes the wave growth and the turbulence beneath the interface, the mass transfer across the interface could be significantly affected by the surface tension. 
}

This study aims to clarify the effects of uniform surface tension reduction on wind-wave growth and scalar transfer across the interface under a comparatively low wind speed condition of several meters per second, where intensive wave breaking does not occur.
\MatsudaA{DNS} of air--water two-phase flow is performed to investigate the effects of the surface tension reduction on the mechanism of the wave growth 
and air--water scalar transfer. 
In addition, the surface tension effects 
obtained
by the DNS 
are discussed in Appendix \ref{appA} by comparing with the wave-height measurements for aqueous solutions 
with different surface tensions in a small wind-wave tank.

\section{Direct numerical simulation}
\subsection{Computational method}
The governing equations for 
air and water flows are the continuity and Navier--Stokes equations for incompressible flows:
\begin{eqnarray}
  \frac{\partial U_i}{\partial x_i} & = & 0 
\label{eq:continuity} \, , \\[3pt]
  \frac{\partial U_i}{\partial t} + U_j \frac{\partial U_i}{\partial x_j} & = &
    - \frac{1}{\rho}\frac{\partial p}{\partial x_i} 
    + \nu \frac{\partial^2 U_i}{\partial x_j \partial x_j} 
    - g \delta_{i3}
\label{eq:NS} \, , 
\end{eqnarray}
where $U_i$ is the velocity component in $i$th direction, 
$p$ is the pressure, 
$\rho$ is the density, 
$\nu$ is the kinematic viscosity, 
and $g$ is the 
gravitational acceleration. 
Different values of $\rho$ and $\nu$ were used for air and water flows:
$\rho=\rho_a$ and $\nu=\nu_a$ for the air flow, while $\rho=\rho_w$ and $\nu=\nu_w$ for the water flow.
The air--water interface was tracked by the height function $\eta(x_1,x_2)$, which represents the horizontal distribution of the vertical interface position. The transport equation of $\eta$ is given by 
\begin{equation}
  \frac{\partial \eta}{\partial t} 
  + U_1 \frac{\partial \eta}{\partial x_1} 
  + U_2 \frac{\partial \eta}{\partial x_2}
    = U_3
\label{eq:interface} \, .
\end{equation}
The dynamical balances at the interface in the normal and tangential directions are given by
\begin{eqnarray}
  p_{w} - \tau_{nw} + p_{s} & = & p_{a} - \tau_{na}
\label{eq:nbalance} \, , \\[3pt]
  \tau_{tw} & = & \tau_{ta}
\label{eq:tbalance} \, ,
\end{eqnarray}
where the subscripts $a$ and $w$ represent the quantities in the air and water, respectively. $\tau_n$ and $\tau_t$ are the normal and tangential components 
of the viscous stress tensor on the interface, 
respectively. 
$p_s$ is the pressure gap due to the surface tension $\sigma$ and is given by $p_s=\sigma \kappa$, where $\kappa$ is the interface curvature. 
\Matsuda{The definition of the curvature $\kappa$ is summarised in Appendix \ref{appB}.}

The transport equation of the 
non-dimensional 
passive scalar $C$ is given by
\begin{equation}
\frac{\partial C}{\partial t} + U_j\frac{\partial C}{\partial x_j} = 
D \frac{\partial^2 C}{\partial x_j \partial x_j}
\label{eq:scalar} \, ,
\end{equation}
where  
$D$ is the diffusion coefficient  
of the scalar.

The governing equations were solved using an arbitrary Lagrangian--Eulerian (ALE) method with boundary-fitted moving grids 
\citep{Komori(1993b),Komori(2010)}.
\Matsuda{
The grid points were allowed to move in the vertical direction following interface deformation, while they were fixed in horizontal direction.
The grid points on the interface moved along with the vertical interface motion, and 
the vertical positions of the other grid points on the air and water sides were updated at each time step 
adapting to the interface height change.
The governing equations described on the Cartesian coordinate were transformed to the equations on the generalized coordinate
and solved in separated domains for air and water flows; 
i.e., the density and the viscosity for each flow were constant at each grid point.
The detailed formulation of the ALE method used here is described in \citet{Komori(1993b)}.
}
The fifth-order upstream and fourth-order central finite difference schemes were adopted for the calculation of the advection and viscous terms, respectively. The velocity and pressure fields were coupled using the marker and cell (MAC) method. 
The time integration of each equation was calculated by the Euler implicit method. 
The reliability of the DNS code was confirmed with a comparison with laboratory experiment data \citep{Komori(1993b),Komori(2010),Kurose(2016)}.

\Matsuda{
As the interface shape is represented by the height function $\eta(x_1,x_2)$, the present DNS cannot consider the wave breaking with air entrainment or discontinuity of the surface slope.
Occurrence of wave breaking can cause a numerical instability for (\ref{eq:interface}). 
Deike and his colleagues \citep{Deike(2015),Deike(2016),Mostert&Deike(2020),Wu&Deike(2021)} performed DNS of two-phase flows with water waves using the volume of fluid (VOF) method with adaptive mesh refinement. Wave breaking can be captured by the VOF method explicitly. 
Here, we use the ALE method with boundary-fitted moving grids because breaking waves are not targeted.}
\Matsuda{
It should be also noted that the DNS of boundary-fitted approach in previous studies \citep{Yang&Shen(2011),Zonta(2015)} used spectral methods for the spatial discretization in the horizontal directions.
The differentiations by finite difference schemes are less accurate than those by spectral methods for large wavenumbers, and the difference between the spectral and the finite difference methods can be evaluated by the effective wavenumber $k_{\rm eff}$ \cite[e.g.,][]{Ferziger}. 
For the case of the finite difference schemes used in the present DNS,
we confirmed that the difference between the effective wavenumber $k_{\rm eff}$ and the exact wavenumber $k$ is less that 5\% for $k \le 0.47 k_{\rm max}$, where $k_{\rm max}$ is the maximum wavenumber and  $k_{\rm max} \approx 6.28\times10^3$ m$^{-1}$ for the present DNS. 
Therefore, the error in the present finite difference schemes is negligibly small for $k\lesssim 3.0\times10^3$ m$^{-1}$.
}

\subsection{Computational condition}\label{sec:DNScond}
\begin{figure}
  \centerline{
    \includegraphics[width=10cm]{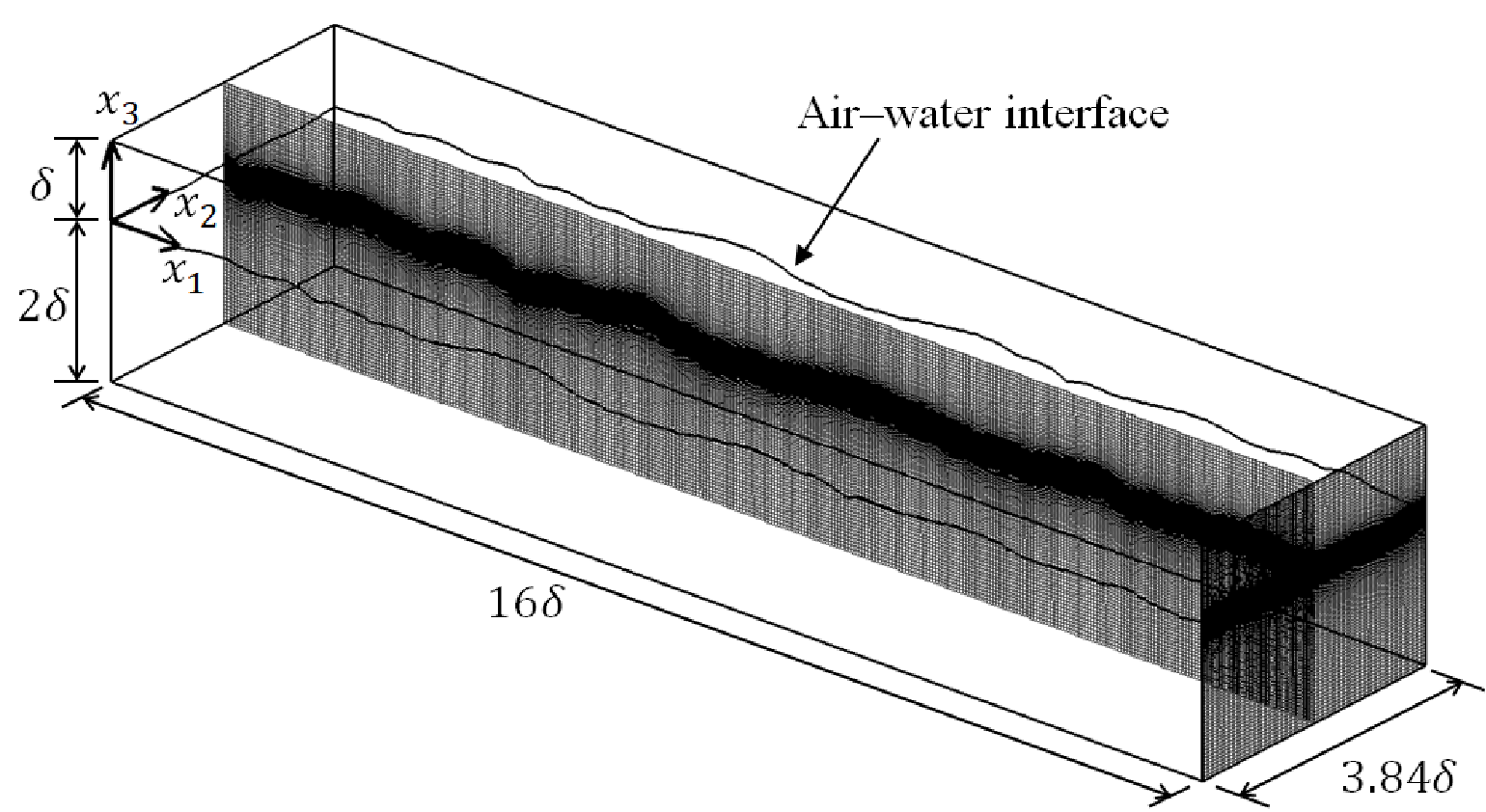}
  }
  \caption{Schematic diagram of the computational domain and grids. }
\label{fig:domain}
\end{figure}
Figure \ref{fig:domain} shows the schematic diagram of the computational domain.
The domain size was set to $L_1=16\delta$, $L_2=3.84\delta$ and $L_3=3\delta$ in the streamwise ($x_1$), spanwise ($x_2$) and vertical ($x_3$) directions, respectively, where $\delta$ was set to $1.25 \times 10^{-2}$ m. The initial interface was located $2\delta$ above the bottom, and this height was set to $x_3=0$.
The number of grid points was set to 400 in the streamwise direction, 96 in the spanwise direction, and 180 in the vertical direction (60 points on the air side and 120 points on the water side).
The uniform grid spacing was used in streamwise and spanwise directions; i,e, the streamwise position of $m_1$th grid point is $x_g(m_1)=m_1 \Delta x_1$, and the spanwise position of $m_2$th grid point is $y_g(m_2)=m_2 \Delta x_2$. The horizontal grid spacing is $\Delta x_1 = \Delta x_2 = 5.0 \times 10^{-4}$ m. 
In the vertical direction, the non-uniform grid spacing was adopted to set fine grids near the interface. 
\Matsuda{As explained in the previous subsection,}
the vertical grid positions change depending on the interface height $\eta$; i.e., 
the vertical position of $m_3$th grid point is given by 
$z_g(m_1,m_2,m_3)=-2\delta+\{\eta(x_g(m_1),y_g(m_2))+2\delta\}\zeta(M_w-m_3,M_w)$ for $m_3=0, \cdots, M_w$ (the water side and the interface), and $z_g(m_1,m_2,m_3)=\delta-\{\delta-\eta(x_g(m_1),y_g(m_2))\}\zeta(m_3-M_w,M_a)$ for $m_3=M_w+1, \cdots, M_w+M_a$ (the air side), 
where $\zeta(m,M)=\tanh \{ \alpha (1 - m/M) \} /\tanh \alpha$, and $\alpha$ is the scaling constant ($\alpha=2.8$). $M_w$ and $M_a$ are the number of vertical grid points in the water and air sides, respectively. 
When the interface is flat, i.e., $\eta(x_1,x_2)=0$, the vertical position of $m_3$th grid point is given by 
$z_{g0}(m_3)=-2\delta\{1-\zeta(M_w-m_3,M_w)\}$ for $m_3=0, \cdots, M_w$, and $z_{g0}(m_3)=\delta\{1-\zeta(m_3-M_w,M_a)\}$ for $m_3=M_w+1, \cdots, M_w+M_a$.
The minimum vertical grid spacing was $9.0\times10^{-6}$ and $8.8\times10^{-6}$ m in the air and water sides, respectively. These were sufficiently small to resolve the viscous sublayer thickness $\delta' \approx 5\nu/u_*$, which was $3.2\times10^{-4}$ and $6.0\times10^{-4}$ m for the air and water sides respectively.
The periodic boundary condition was applied to the velocity and pressure in the streamwise and spanwise directions, while the Neumann condition was applied in the vertical direction. 
A wall-bounded turbulent air flow had been developed for a flat and rigid surface until the velocity profile becomes a statistically steady state. The developed flow field was then used for the initial condition in the air side. 
The friction velocity on the air side was $u_{*a}\approx0.24$ m/s. The corresponding free-stream wind speed was $U_\infty \approx 5.2$ m/s based on the empirical relationship of \citet{Iwano(2013)}.
The water side was initially in static condition, and the initial displacement of the interface was set to zero.
A constant pressure gradient was applied to the air side in the streamwise direction for the driving force. 
The time integration was calculated for 7 seconds. 
The air and water densities were set to $\rho_a=1.2$ and $\rho_w=1.0\times10^{3}$ kg/m$^3$, respectively and the kinematic viscosities in the air and water sides were set to $\nu_a=1.5\times10^{-5}$ and $\nu_w=1.0\times10^{-6}$ m$^2$/s, respectively. The gravitational acceleration was set to $g=9.8$ m/s$^2$. 
This study considers three cases of surface tension $\sigma$: (1) the case where $\sigma$ is equal to that of 
water ($\sigma=1.0\sigma_w$), (2) the case where $\sigma$ is half to that of 
water ($\sigma=0.5\sigma_w$), and (3) the case where $\sigma$ is initially $\sigma=1.0\sigma_w$ and suddenly reduced to $\sigma=0.5\sigma_w$ at $t=4.0$ s.
The surface tension of 
water was set to $\sigma_w=7.2\times10^{-2}$ N/m.
\Matsuda{We have confirmed the grid resolution dependence of the DNS results (Appendix \ref{appNC}).}

\Matsuda{
The friction Reynolds number is defined for each side of air and water using the friction velocity and the boundary-layer thickness ($\delta_a$ and $\delta_w$ for the air and water sides, respectively). 
The friction Reynolds number on the air side, $Re_{\tau,a} = u_{*a} \delta_a/\nu_a$, is approximately 200, and it is sufficiently large to observe the logarithmic region.
The friction Reynolds number on the water side, $Re_{\tau,w} = u_{*w} \delta_w/\nu_w$, is dependent on the time as the boundary layer develops along with the time.
The DNS results show that, for $t=4.0$--7.0 s, 
$Re_{\tau,w}$ ranges from 165 to 193.
}

The scalar $C$ represents the non-dimensional concentration of the imaginary absorbed gas.
The scalar transport only in the water side was calculated because the concentration is uniform ($C=1$) in the air side. 
The initial scalar concentration in the water side was set to 0, and the scalar concentration on the interface was fixed to $C=1$ uniformly. The scalar transport calculation was started at $t=4.0$ s, when wind waves were well developed.
The Schmidt number, defined as $\Sch \equiv \nu_w/D$, was set to $\Sch=1$ in the water side.
Note that real gases usually have much larger $\Sch$; e.g., $\Sch \approx 600$ for CO$_2$. The scalar transfer coefficient 
$k_L$ (which is defined in section~\ref{sec:scalar} and also referred to as the transfer velocity) for such large $\Sch$ can be estimated from that for $\Sch=1$ using the relationship $k_L \propto \Sch^{-1/2}$  \citep{Jahne(1979)}.

\section{Results and discussion}
\subsection{Effect on wave growth} \label{sec:growth}

\begin{figure}
  \centerline{
    (\textit{a}) \includegraphics[width=6cm]{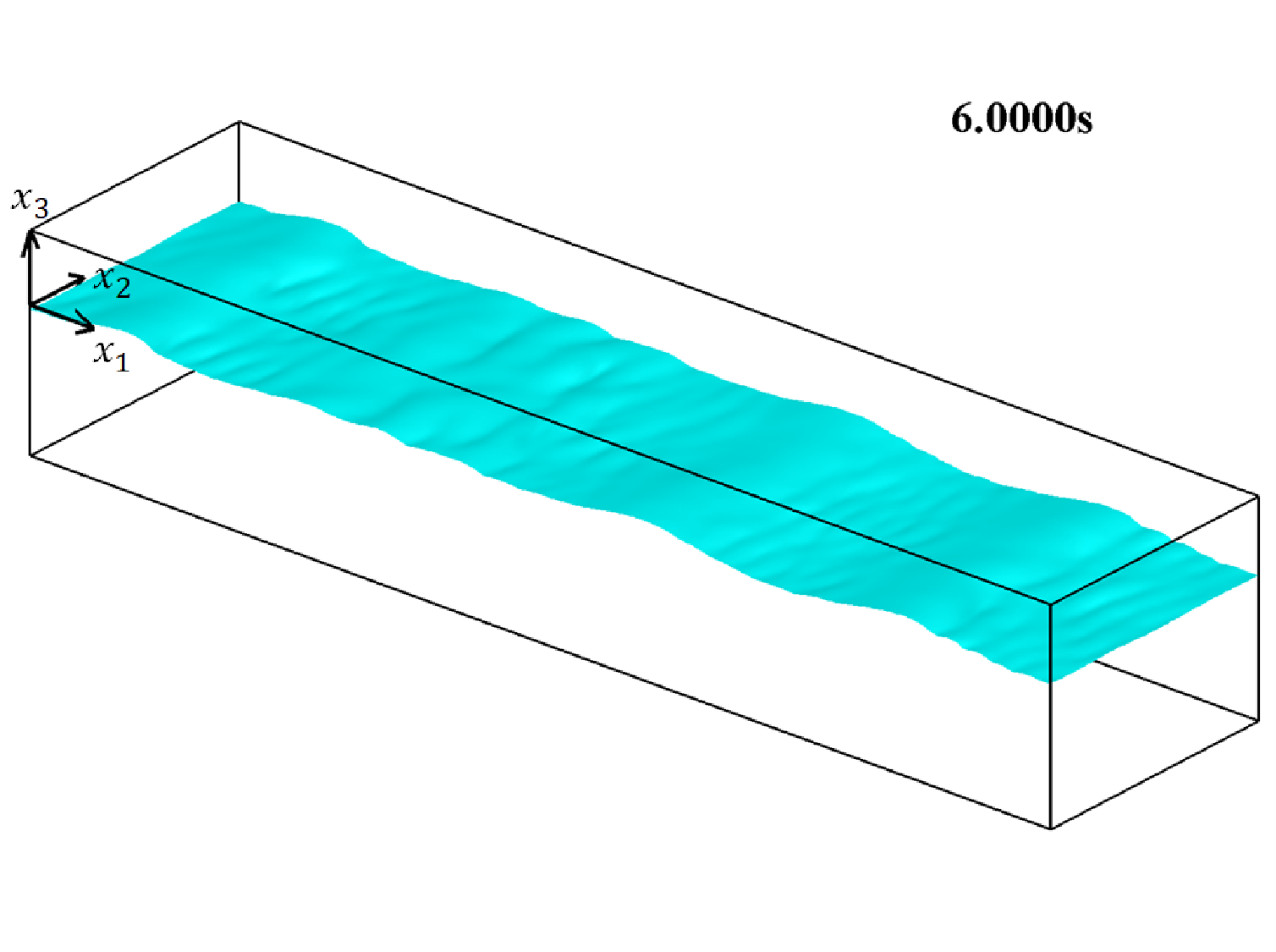}
    (\textit{b}) \includegraphics[width=6cm]{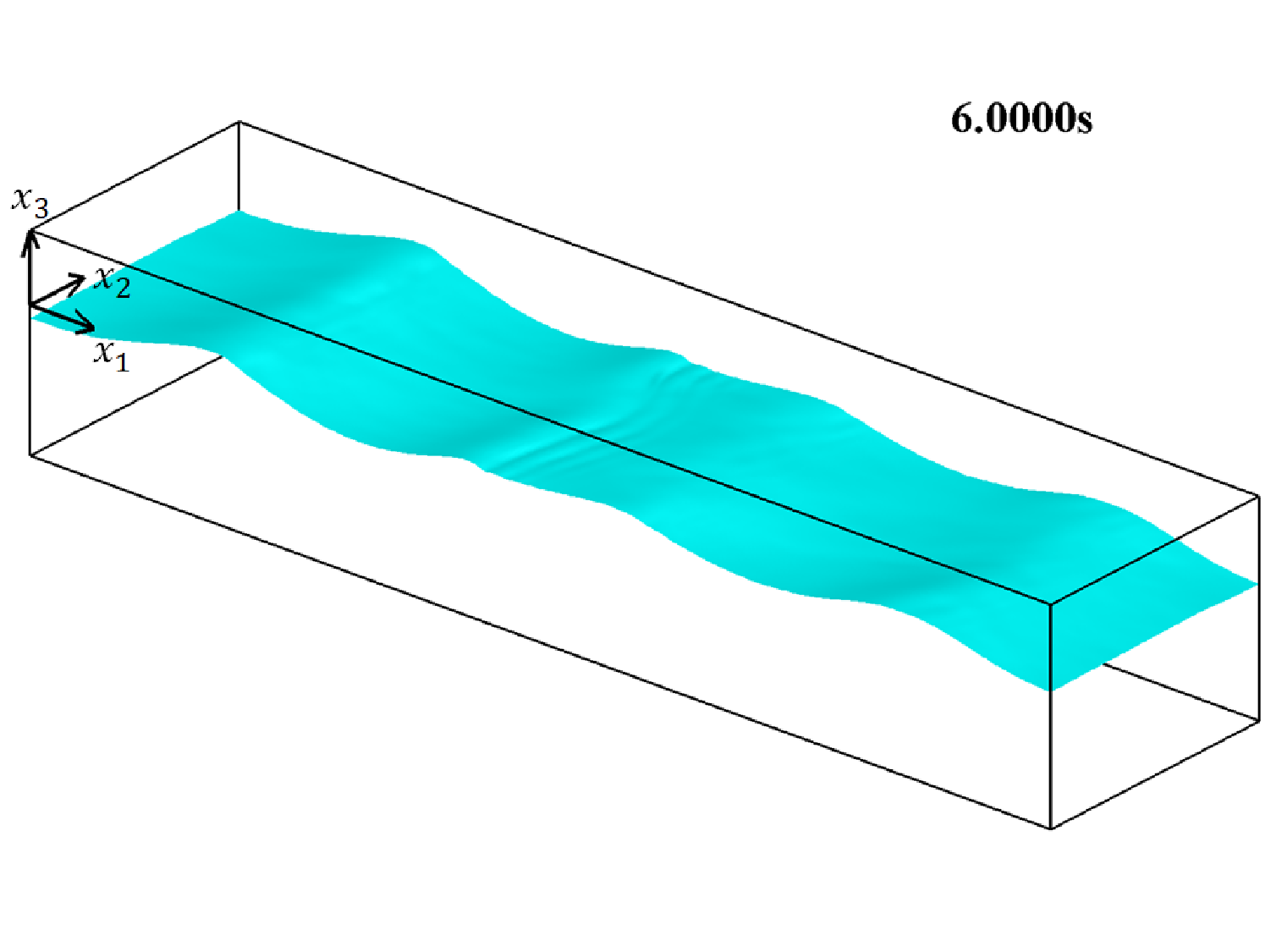}
  }
  \centerline{
    \Matsuda{
    (\textit{c}) \includegraphics[width=7cm]{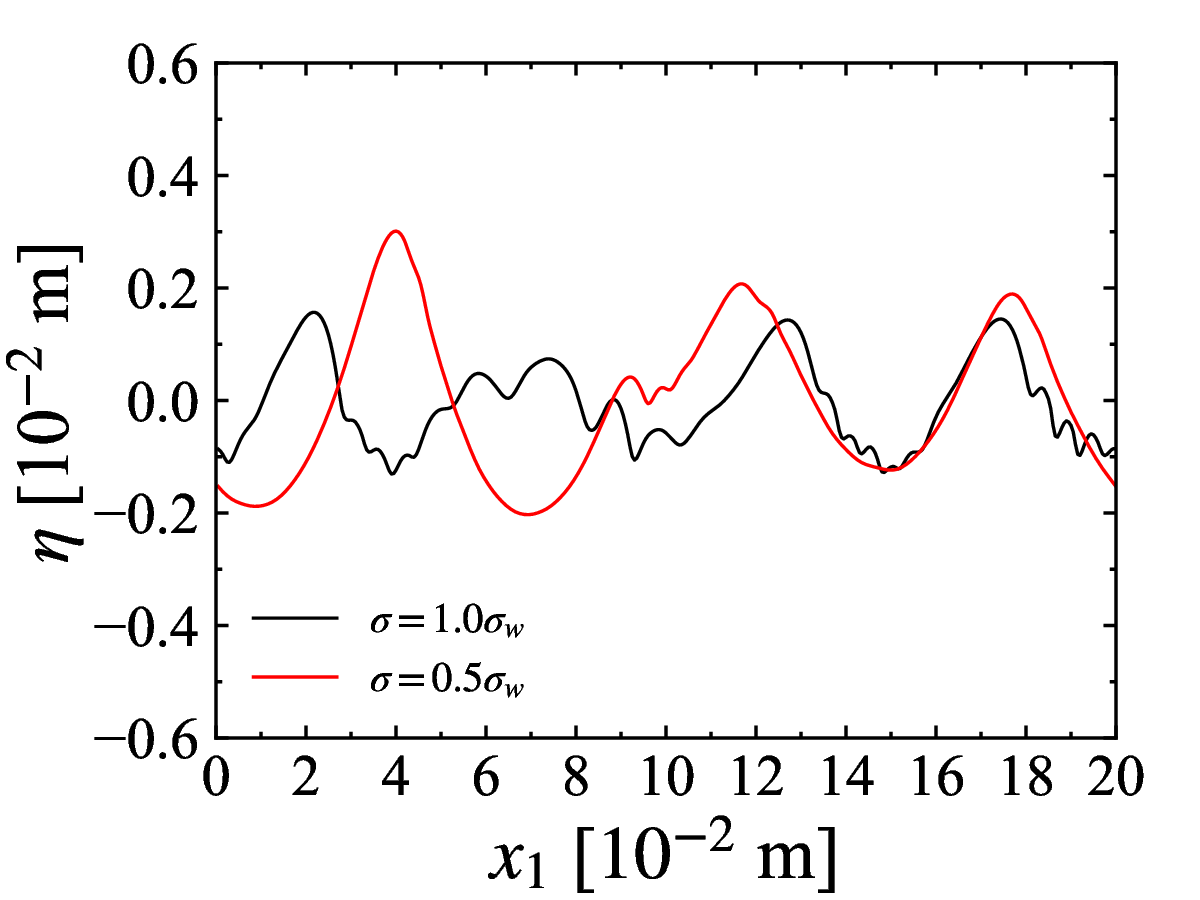}
    }
  }
  \caption{\Matsuda{Three-dimensional view of} air--water interface obtained by the DNS for the cases (\textit{a}) $\sigma=1.0\sigma_w$ and (\textit{b}) $\sigma=0.5\sigma_w$ at \MatsudaB{$t=6.0$ s}\Matsuda{, and (\textit{c}) streamwise distribution of interface displacement $\eta$ at $x_2=2.4\times10^{-2}$ m and \MatsudaB{$t=6.0$ s}}. 
  }
\label{fig:DNS_interface}
\end{figure}

Figure \ref{fig:DNS_interface} shows the shape of the air--water interface at 
\MatsudaB{$t=6.0$ s} for the water case (i.e., $\sigma=1.0\sigma_w$) and the reduced surface tension case (i.e., $\sigma=0.5\sigma_w$). 
The wind and the propagating waves move in the 
$x_1$
direction. 
For the case of $\sigma=1.0\sigma_w$, the wind waves form ripple waves on the 
downwind 
side of the wave crest. 
For the case of $\sigma=0.5\sigma_w$, the ripple waves are less significant than those for $\sigma=1.0\sigma_w$, and the wave crests are slightly sharper.
\Matsuda{
The above observation can be also confirmed by the streamwise distribution of instantaneous $\eta$ at the middle position of the spanwise width ($x_2=2.4\times10^{-2}$ m) in figure~\ref{fig:DNS_interface}(\textit{c}). For $\sigma=0.5\sigma_w$, the distribution of $\eta$ in $0.1\lesssim x_1 \lesssim 0.2$ m does not show clear ripple waves, while that for $\sigma=1.0\sigma_w$ shows ripple-like interface fluctuations in the downwind side of the wave crest.
}

\begin{figure}
  \centerline{
    (\textit{a}) \includegraphics[width=6cm]{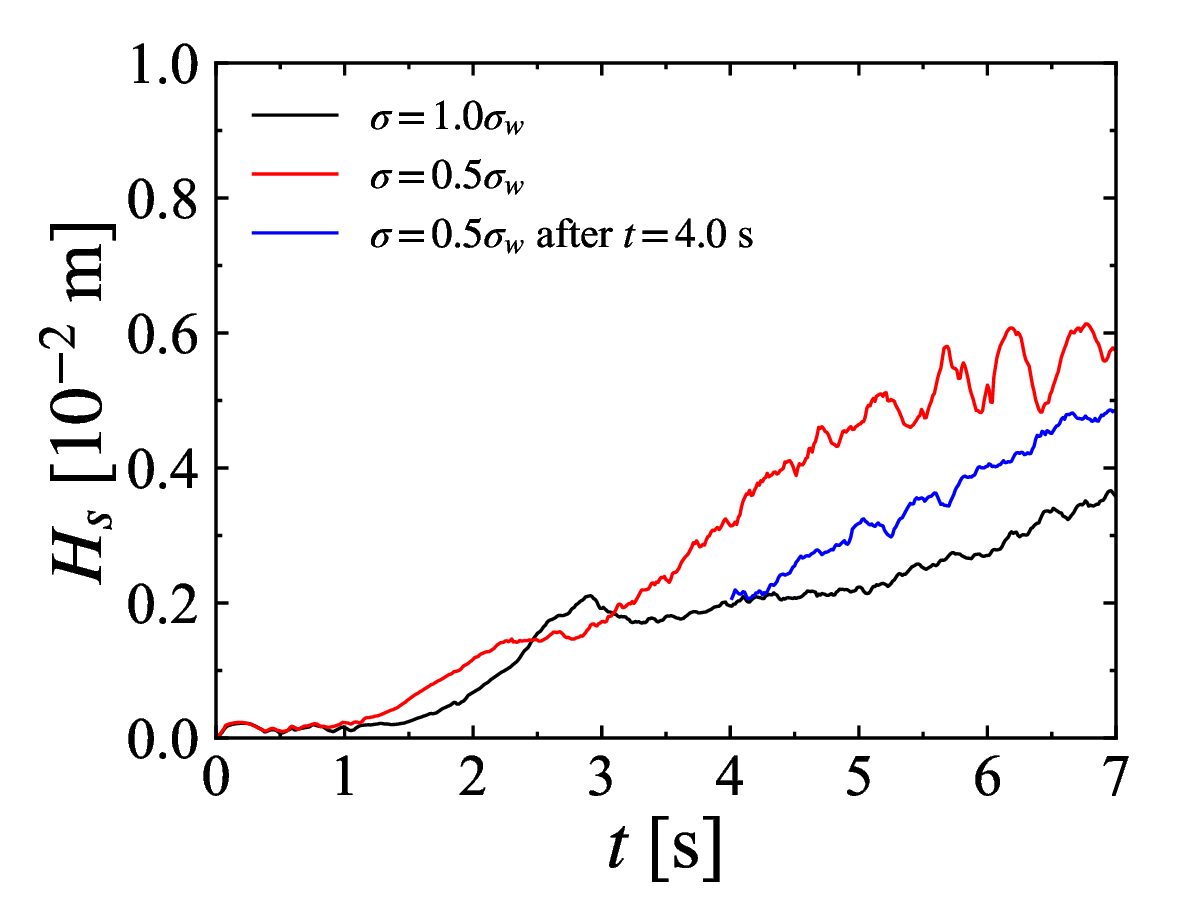}
    (\textit{b}) \includegraphics[width=6cm]{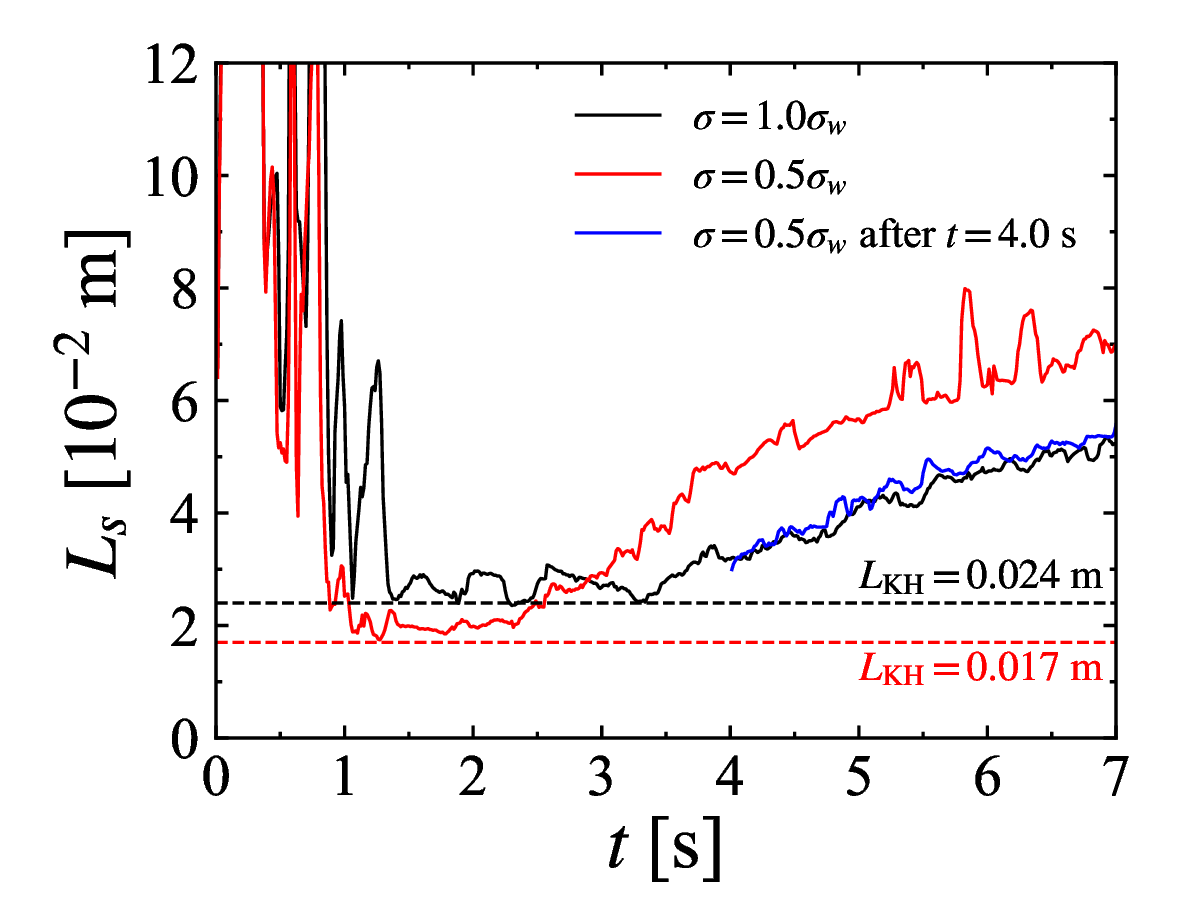}
  }
  \caption{Temporal variations of (\textit{a}) significant wave height, $H_s$, and (\textit{b}) significant wavelength, $L_s$, for the cases of $\sigma=1.0\sigma_w$, 
  $\sigma=0.5\sigma_w$, and  
sudden change from $\sigma=1.0\sigma_w$ to $\sigma=0.5\sigma_w$ at $t = 4.0$ s. 
Dashed lines are the wavelength amplified by the weakly non-linear Kelvin--Helmhltz instability, $L_{\rm KH}$.}
\label{fig:signif_wave}
\end{figure}

To quantify the 
effect of surface tension on the wave height, we have calculated the significant wave height, $H_s$, and the significant wavelength, $L_s$ \Matsuda{\citep[e.g.,][]{Toba(1972),Holthuijisen}}. 
The significant waves are defined as the waves whose height is the largest one-thirds among all individual waves, and the individual waves are identified using the zero-up crossing method \citep{Pierson(1954),Takagaki(2020)}. 
$H_s$ and $L_s$ are the average wave height and wavelength of the significant waves, respectively. 
Figure~\ref{fig:signif_wave} shows the temporal variations of $H_s$ and $L_s$. 
Note that, as shown 
in figure~\ref{fig:signif_wave}(\textit{b}), the significant waves do not capture wind waves at the initial period ($t \lesssim 1.0$ s) because the interface initially oscillates due to 
the initial impact of imposing the wall-bounded turbulent air flow driven by the pressure gradient. 
\MatsudaA{However, in the period of $t=$ 1.0--2.0 s, the wavelength is close to the minimum for both cases, and the wavelength for $\sigma=0.5\sigma_w$ is slightly shorter than that for $\sigma=1.0\sigma_w$.}
During and after this period, the significant wind waves are well captured, and
their wavelengths for both $\sigma$ cases remains less than \MatsudaB{or comparable to} 0.07 m until $t=7.0$ s.
This means that the wind waves simulated in this study are classified as gravity-capillary waves ($0.004 \lesssim L_s \lesssim 0.07$ m) according to the classification in \citet{Lin(2008)}.
Figure~\ref{fig:signif_wave}(\textit{a}) shows that, after the significant waves capture wind waves, waves for $\sigma=0.5\sigma_w$ start growing earlier than $\sigma=1.0\sigma_w$, and the wave height for $\sigma=0.5\sigma_w$ is remarkably higher than that for $\sigma=1.0\sigma_w$ after $t=3.0$ s. 
This result means that the waves are not suppressed by the uniform surface tension reduction in contrast to the effect of surfactants. 
We have also confirmed the effect of uniform surface tension reduction on the wave height by conducting laboratory experiments using a small wind-wave tank (see Appendix~\ref{appA}).

One could argue that the difference in wave height between the 
cases of $\sigma=1.0\sigma_w$ and $\sigma=0.5\sigma_w$ 
in figure~\ref{fig:signif_wave}(\textit{a}) is caused by hysteresis due to the difference in the initial wave development when the wind shear is loaded on the flat free surface. 
In order to clarify the dominance of hysteresis, we have performed the additional simulation with the surface tension suddenly changed to $\sigma=0.5\sigma_w$ from the wind waves at $t=4.0$ s developed using $\sigma=1.0\sigma_w$.
The results are shown by blue lines in figure~\ref{fig:signif_wave}.
When the surface tension was changed to $\sigma=0.5\sigma_w$, the significant wave height $H_s$ becomes higher than the case of $\sigma=1.0\sigma_w$ and becomes close to the case with the surface tension $\sigma=0.5\sigma_w$ from the initial time ($t=0$).
This indicates that the hysteresis is not critical, whereas the wave growth speed is increased due to the surface tension reduction. 

\MatsudaA{
The surface tension dependence of the significant wavelength in the period of $t=$ 1.0--2.0 s is similar to the analytical result of the Kelvin--Helmholtz (K--H) instability, based on which the initial wavelength is given by $L_m \equiv 2\pi\sqrt{\gamma/g}$ (where $\gamma = \sigma/\rho_w$), which is the wavelength for minimum phase velocity.
However, based on the theoretical analyses \citep{Jeffreys(1925),Miles(1957),Miles(1959)}, the classical K--H instability is not considered as the major mechanism that causes the initial wind-wave formation on an air--water interface.
\cite{Jeffreys(1925)} and \cite{Miles(1957),Miles(1993)} proposed theoretical models for the formation of initial waves under turbulent winds. Their models estimate that the wave energy grows exponentially, and the initial wavelength formed at 
the minimum wind speed is about 0.06-0.08 m, where the effect of the surface tension is considered to be small.
Miles' model is well accepted because the estimate agrees with observation data. 
However, this model cannot be applied to the present DNS results for the period of $t=$ 1.0--2.0 s.
\cite{Miles(1957),Miles(1993)} considered the critical layer, which has the same mean wind speed as the wave speed, in the logarithmic layer of a wall-bounded turbulent flow, whereas, in the the present DNS results, the height for 
the wind speed equivalent to the wave speed is in the viscous sublayer.
\cite{Miles(1959)} generalized the K--H model to take account the logarithmic layer of a wall-bounded turbulent flow, and concluded that the K--H instability is unlikely for air--water interfaces at commonly observed wind speeds.  
Apart from the initial wave formation,
\cite{Bontozoglou&Hanratty(1990)} examined the weakly nonlinear K--H instability \citep{Miles(1986)} to characterize the instability of finite amplitude waves close to resonance, and showed that the K--H instability becomes strongly subcritical for the waves with wavelength shorter than the resonant wavelength given by $L_{\rm KH} \equiv 2\pi\sqrt{2\gamma/g}$. 
Their postulate  
is that, for air--water flows, gravity-capillary waves can be generated by winds and grow by the subcritical K--H instability so that those become sufficiently high to trigger a finite amplitude bifurcation of doubling the wavelength \citep{Chen&Saffman(1979),Chen&Saffman(1980)}.
The resonant wavelength $L_{\rm KH}$ for $\sigma=1.0\sigma_w$ and $\sigma=0.5\sigma_w$ are 0.024 and 0.017 m, respectively.
In the present DNS results for both surface tension cases, the significant wavelength $L_s$ in the period of $t=$ 1.0--2.0 s almost corresponds to the wavelength $L_{\rm KH}$. 
Therefore, we can conjecture that the waves in this period are amplified due to the subcritical K--H instability for finite amplitude waves close to resonance.
}

\begin{figure}
  \centerline{ 
    (\textit{a}) \includegraphics[width=7cm]{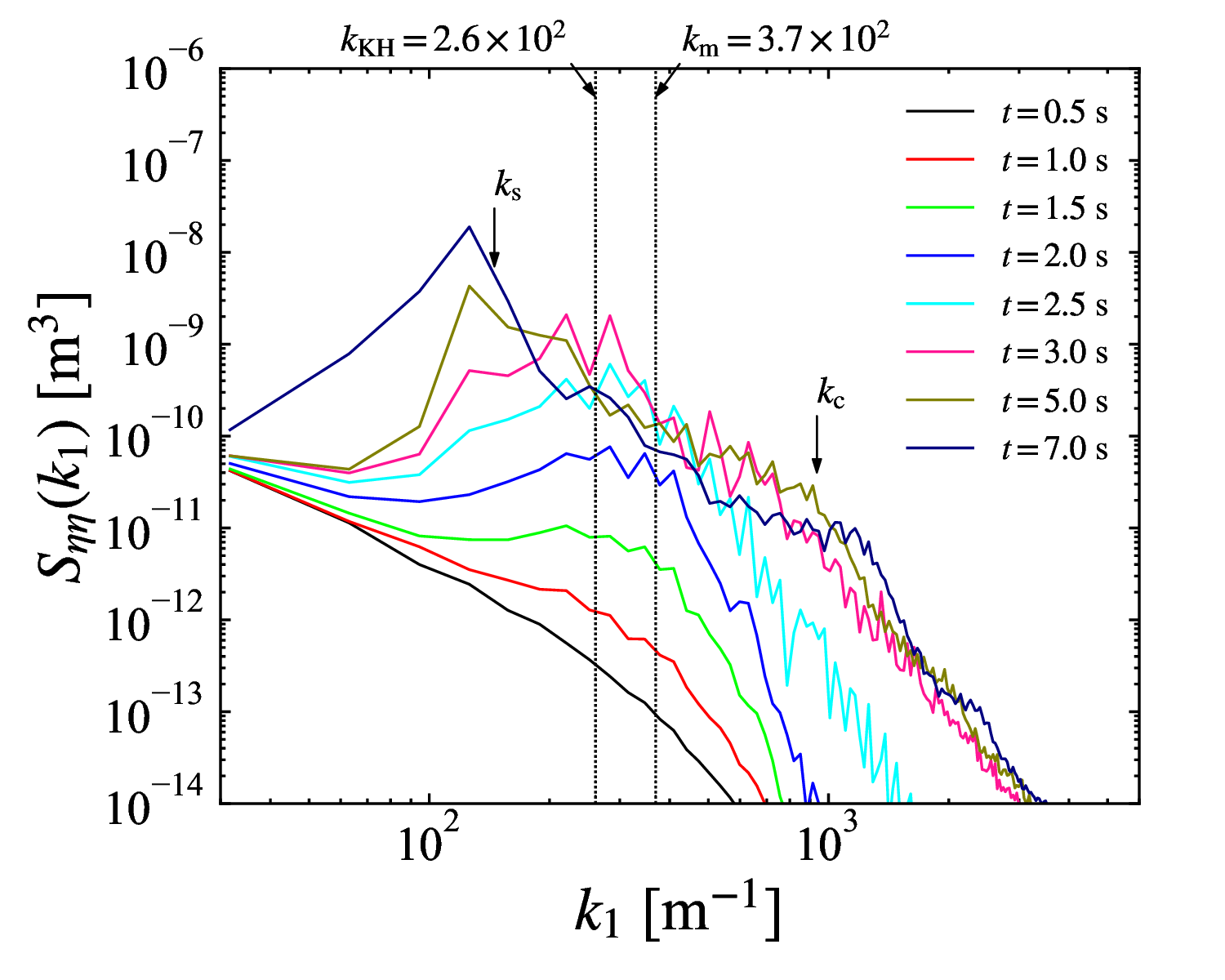} \hspace{-10mm}
    (\textit{b}) \includegraphics[width=7cm]{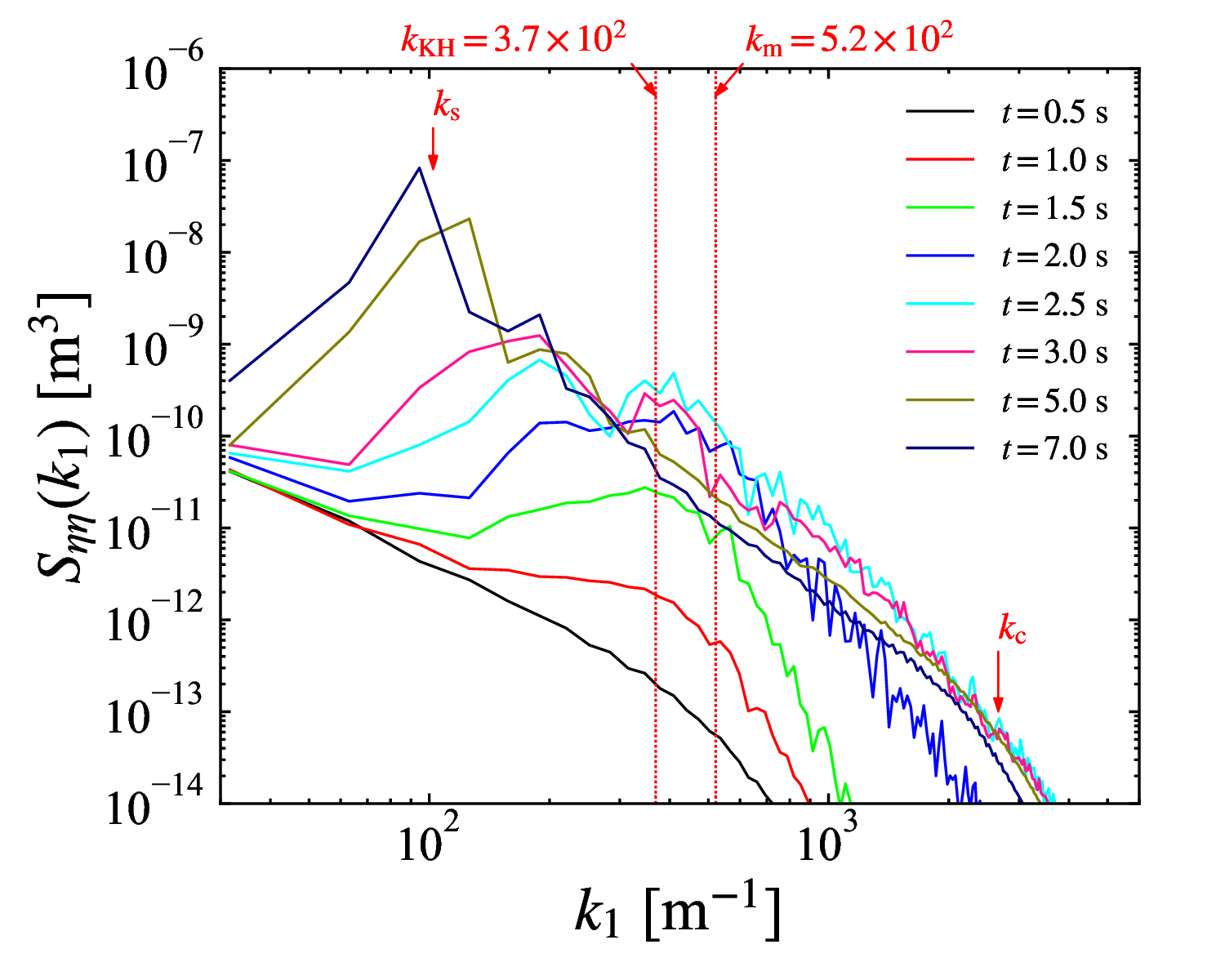} \hspace{-10mm}
  }
  \centerline{ (\textit{c}) \includegraphics[width=7cm]{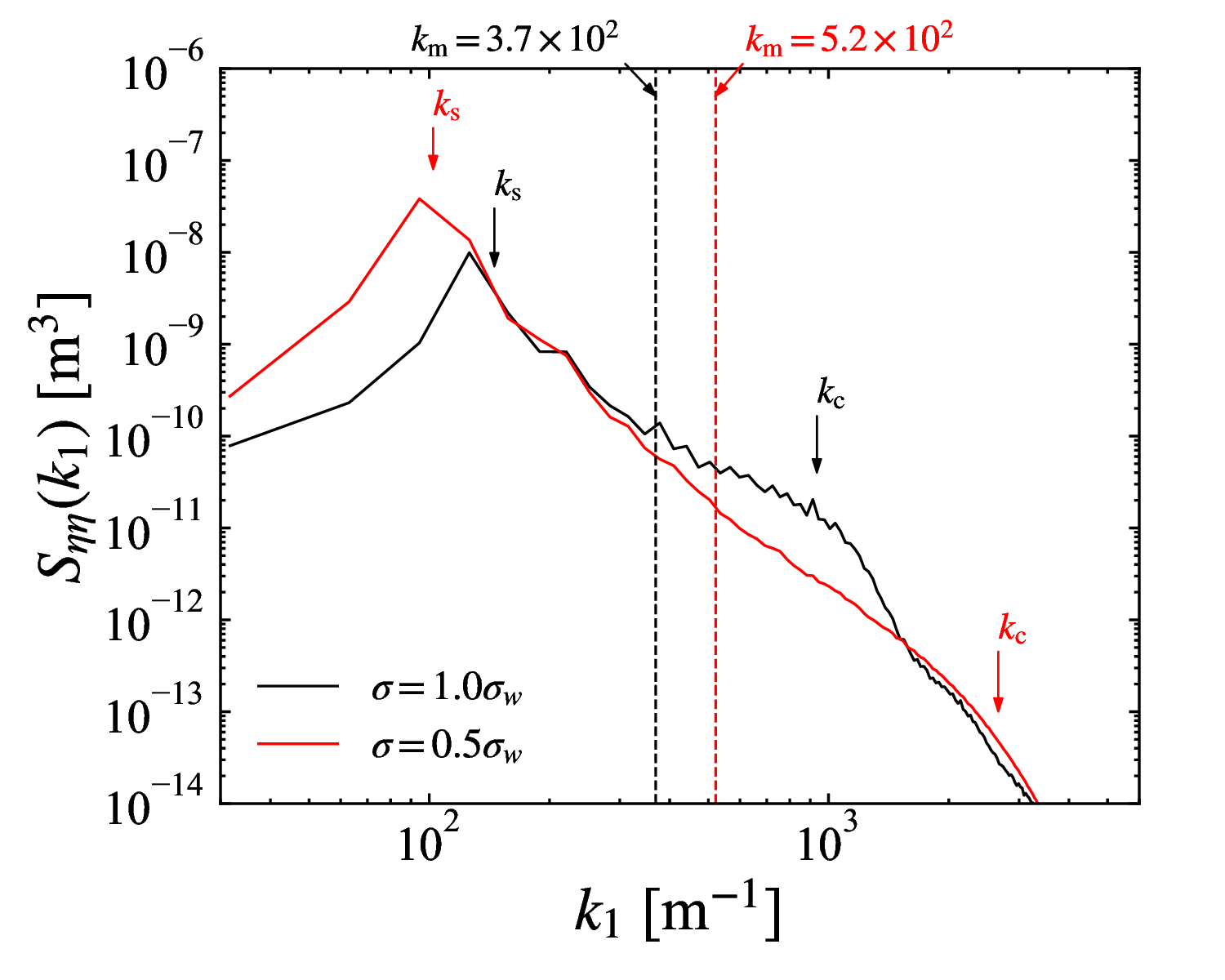} \hspace{-5mm} }
  \caption{Wave height spectra: 
Temporal evolutions of the spectra for (\textit{a}) $\sigma=1.0\sigma_w$ and (\textit{b}) $\sigma=0.5\sigma_w$ averaged for 0.5 s before the indicated time instant; 
(\textit{c}) comparison of spectra for $\sigma=1.0\sigma_w$ and $\sigma=0.5\sigma_w$ averaged for the period of $4.0<t\le7.0$ s. 
Dashed lines are the wavenumbers $k_{\rm KH}=2\pi/L_{\rm KH}$ and $k_m=2\pi/L_m$. 
The arrows indicate the wavenumber of mean significant wave $k_s=2\pi/\overline{L_s}$ and the resonant capillary wavenumber $k_c=k_m^2/k_s$, where $\overline{L_s}$ is the time-averaged $L_s$ for the period of 4.0-7.0 s. 
The wavenumbers for $\sigma=1.0\sigma_w$ and $\sigma=0.5\sigma_w$ are indicated in black and red, respectively.
}
\label{fig:wave_spectrum}
\end{figure}

The one-dimensional wave height spectrum $S_{\eta\eta}(k_1)$ in the streamwise direction has been calculated to understand the temporal change of $H_s$ and $L_s$.
The one-dimensional wave height spectrum is defined as $S_{\eta\eta}(k_1) \equiv \frac{1}{L_2}\int_0^{L_2} \widehat{\eta}(k_1,x_2)\widehat{\eta}^*(k_1,x_2) {\rm d}x_2$, where $\widehat{\eta}(k_1,x_2)$ is obtained by the Fourier transform in the streamwise direction, and $k_1$ is the streamwise wavenumber. The asterisk denotes the complex conjugate.
Figure \ref{fig:wave_spectrum}(\textit{a}) shows the temporal change of the wave height spectrum for $\sigma=1.0\sigma_w$. 
Note that the spectrum is temporally averaged for the past 0.5 s with respect to the indicated time instant. 
In the early growth period (up to $t=2.5$ s), as predicted by \citet{Bontozoglou&Hanratty(1990)}, 
waves with wavenumbers close to and larger than $k_{\rm KH} \equiv \sqrt{g/2\gamma} = 2\pi/L_{\rm KH}$ grow predominantly. 
This is because the spectrum for wavenumbers around $k_{\rm KH}$ and $2k_{\rm KH}$ are amplified due to 
\MatsudaA{the subcritical K--H instability for the waves close to the second harmonic resonance}
at an early stage of wave growth \citep{Bontozoglou&Hanratty(1990)}.
After $t=2.5$ s, the peak moves to a lower wavenumber as the wind waves grow, and the spectrum tail broadens on the high wavenumber side. 
The change of the peak location to lower wavenumbers corresponds to the increase in the significant wavelength in figure \ref{fig:signif_wave}(\textit{b}).
\Matsuda{
The decrease of the peak wavenumber is commonly observed for wind waves \citep[e.g.,][]{Imasato(1976)}, and these would be due to the nonlinear wave--wave interactions. 
}
The broadening of the spectrum tail on the high wavenumber side is due to 
a resonant nonlinear wave-wave interaction between the fundamental gravity wave and higher harmonics of order $N$, which forms ripple-like capillary waves \citep{Chen&Saffman(1979),Fedorov(1998),Caulliez(2013)}.
The resonant condition can be approximately satisfied when the phase velocity of the capillary waves matches that of the fundamental wave.
The wavenumber $k_c$ of the capillary waves that resonate with the significant waves with wavenumber $k_s$ 
is given by $k_c = k_m^2/k_s$, where  
\MatsudaA{$k_m=\sqrt{g/\gamma}=2\pi/L_m$},
when the non-linearity is negligible. 
The arrows in figure~\ref{fig:wave_spectrum}(\textit{a}) indicate the wavenumber $k_s=2\pi/\overline{L_s}$ for the time-averaged significant wavelength $\overline{L_s}$ for 4.0 s $<t \le$ 7.0 s and the wavenumber $k_c$ of the resonant capillary waves. 
The spectrum at $t=5.0$ and 7.0 s shows a gentle slope for $k_m < k < k_c$ and decreases steeply for $k > k_c$. 
This confirms that the harmonic resonance causes the broadened spectrum.
Figure \ref{fig:wave_spectrum}(\textit{b}) shows similar temporal change of the spectrum for $\sigma=0.5\sigma_w$:
Even though $k_{\rm KH}$ for $\sigma=0.5\sigma_w$ is larger than 
that for 
$\sigma=1.0\sigma_w$ by a factor of $\sqrt{2}$,
waves with wavenumbers around $k_{\rm KH}$ grow predominantly in the early growth period (up to $t = 2.0$ s), and then the spectrum broadens on low and high wavenumber sides.

Figure \ref{fig:wave_spectrum}(\textit{c}) compares the wave height spectra for the two cases of surface tension. The spectra are temporally averaged over the period of 4.0 s $<t \le$ 7.0 s. 
On the low wavenumber side, the peak for $\sigma=0.5\sigma_w$ is larger than that for $\sigma=1.0\sigma_w$, corresponding to the larger significant wave height. 
On the high wavenumber side, the wavenumber $k_c$ of the resonant capillary waves increases due to the surface tension reduction because $k_m$ is inversely proportional to the square root of the surface tension.
The bump on the spectrum also moves to the higher wavenumber following the increase of $k_c$. 
This indicats that the wavelength of the resonant capillary waves becomes shorter.
It is also observed that the amplitude at $k_c$ for $\sigma=0.5\sigma_w$ 
is about two order of magnitude smaller than that for $\sigma=1.0\sigma_w$. 
This means that ripple-like capillary waves are suppressed due to the surface tension reduction.

\begin{figure}
  \centerline{ \includegraphics[width=8cm]{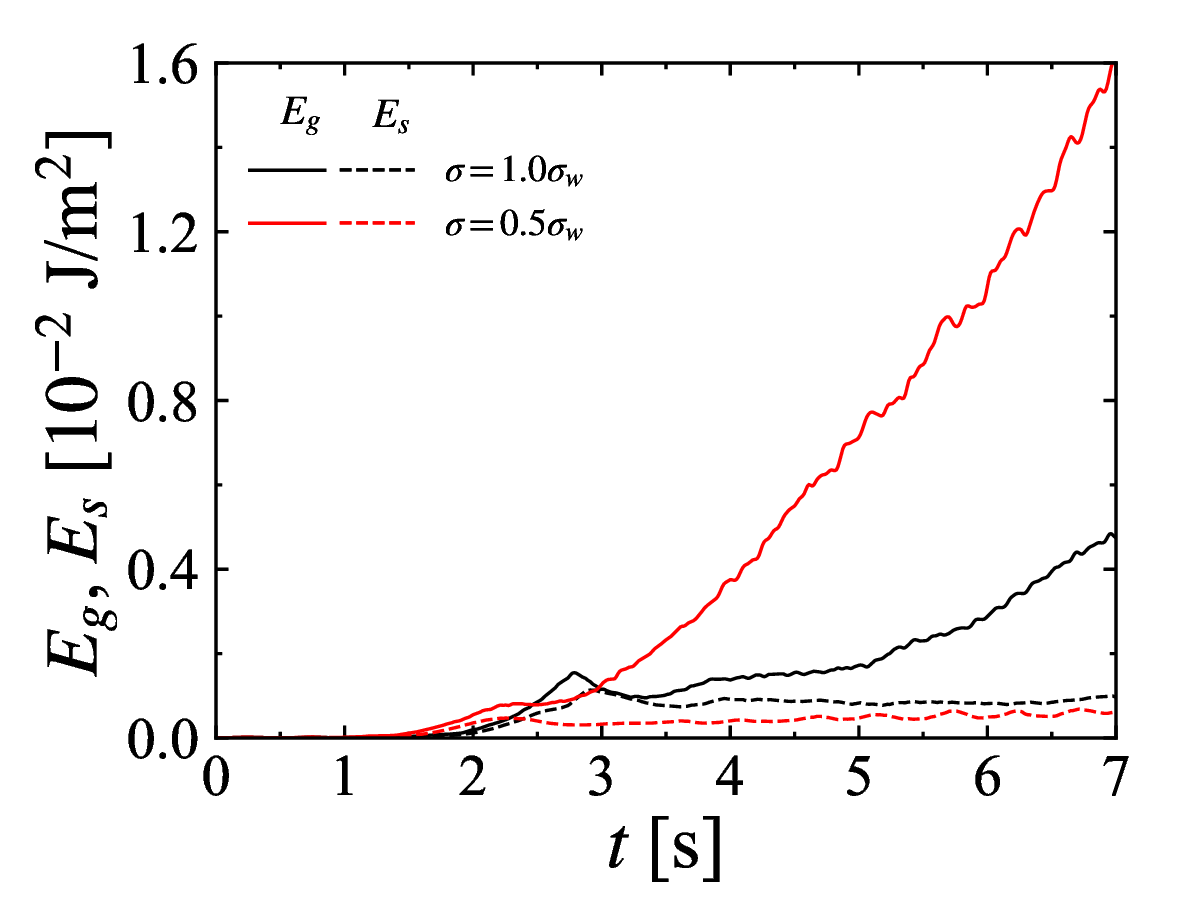} }
  \caption{Temporal variations of potential energies due to gravity and surface tension, $E_g$ and $E_s$, for the cases of $\sigma=1.0\sigma_w$ and $\sigma=0.5\sigma_w$. }
\label{fig:wave_energy}
\end{figure}

To understand the effect of surface tension reduction on the wind-wave growth mechanism, the potential 
energies due to gravity $E_g$ and surface tension $E_s$ were calculated using the following equations:
\begin{eqnarray}
  E_g &=& 
    \frac{\rho_w g}{2 L_1 L_2}
    \int_0^{L_1}\int_0^{L_2} \eta^2 {\rm d}x_1{\rm d}x_2
\label{eq:E_g} \, , \\[3pt]
  E_s &=&
    \frac{\sigma}{2 L_1 L_2}
    \int_0^{L_1}\int_0^{L_2} \left\{ 
        \left(\frac{\partial \eta}{\partial x_1}\right)^2
      + \left(\frac{\partial \eta}{\partial x_2}\right)^2
      \right\} {\rm d}x_1{\rm d}x_2
\label{eq:E_s} \, .
\end{eqnarray}
Figure \ref{fig:wave_energy} shows the temporal variations of $E_g$ and $E_s$.
The potential energies 
$E_g$ and $E_s$ are  
\MatsudaB{comparable}
in the early growth period of 1.5 s $<t<$ 2.5 s for $\sigma=1.0\sigma_w$ and 1.0 s $<t<$ 2.0 s for $\sigma=0.5\sigma_w$. 
According to the small amplitude linear wave theory, $E_g=E_s$ implies the wavelength is equal to
\MatsudaA{$L_m$}.
Figures~\ref{fig:wave_spectrum}(\textit{a}) and (\textit{b}) also show that waves with wavenumbers around $k_{\rm KH} \sim {\cal O}(k_m)$ grow predominantly in the early growth period. 
This is 
also 
consistent with the fact that the significant wavelength in the early growth period is close to $L_{\rm KH} = \sqrt{2} L_m$ as shown in figure~\ref{fig:signif_wave}(\textit{b}). 
In the early growth period, the potential energy increases exponentially, and the energy growth rate for $\sigma=0.5\sigma_w$ is larger than that for $\sigma=1.0\sigma_w$. 
These results are qualitatively consistent with the results from the linear stability analysis of \citet{Tsai&Lin(2004)}. 
\Matsuda{
\cite{Lin(2008)} reported that, based on their DNS for air--water two-phase flows, 
initial linear wave growth is followed by the exponential wave growth.
In the present DNS results, only the later exponential growth is captured.    
}
After the early growth period, $E_g$ increases, whereas $E_s$ remains almost constant. 
The difference in the temporal variation between $E_g$ and $E_s$ corresponds to the broadening of the wave height spectra on both low and high wavenumber sides as shown in figures~\ref{fig:wave_spectrum}(\textit{a}) and 4(\textit{b}). 
The surface potential energy $E_s$ shows larger values for the larger surface tension as expected from 
the definition of $E_s$. 
In contrast, the potential energy $E_g$ becomes larger for the smaller surface tension. 
The increase of $E_g$ due to surface tension reduction is more significant than the decrease of $E_s$. 
This also means that $E_g$ increases faster for the smaller surface tension.
\Matsuda{
It should be noted that \cite{Zavadsky&Shemer(2017)} reported the temporal development of the wave height measured by applying nearly impulsive wind forcing.
Their results show similar fast initial wave growth followed by the relatively slow wave growth. 
}
\Matsuda{
One would find that the growth of $H_s$ becomes even gentler for $t>5.0$ s than $t=4.0$--5.0 s, while $E_g$ for $\sigma=0.5\sigma_w$ increases almost linearly along with the time for $t>4.0$ s. 
\cite{Zavadsky&Shemer(2017)} reported that the quasi-steady equilibrium state appears when the waves are long enough to be purely gravity waves, and this stage is considered as `the principal stage’ of \cite{Phillips(1957)}. 
Therefore, the gentler growth for $\sigma=0.5\sigma_w$ at $t>5.0$ s could be relevant to this stage 
because the waves for $\sigma=0.5\sigma_w$ are closer to purely gravity waves than the waves for $\sigma=1.0\sigma_w$. 
However, clear transition to quasi-steady equilibrium state is not observed in the present DNS results.
}

To clarify the mechanism of faster increase of the potential energy $E_g$ 
after the early growth period for the case of smaller surface tension in figure~\ref{fig:wave_energy}, we focus on the energy input from air to water and the energy dissipation in the water side.
The energy fluxes due to the normal and tangential stress at the interface, $Q_{na}$ and $Q_{ta}$, respectively, are given by 
\begin{eqnarray}
  Q_{na} &=& 
   \frac{1}{L_1 L_2} 
    \int_\Gamma u_n \left( \tau_{na} - p_a - 
    \rho_a 
    g\eta \right) {\rm d}S
\label{eq:Q_na} \, , \\[3pt]
  Q_{ta} &=&
    \frac{1}{L_1 L_2} 
    \int_\Gamma \left( u_{t1}\tau_{t1} + u_{t2}\tau_{t2} \right) {\rm d}S
\label{eq:Q_ta} \, ,
\end{eqnarray}
where $u_n$ is the normal component of the interface velocity, and $u_{t1}$ and $u_{t2}$ are the tangential components of the interface velocity. $\tau_{t1}$ and $\tau_{t2}$ are the tangential components of viscous stress in the directions of $u_{t1}$ and $u_{t2}$. $\Gamma$ represents the interface area, and ${\rm d}S$ is the infinitesimal area on $\Gamma$.
$Q_{na}$ and $Q_{ta}$ represent the energy fluxes per unit horizontal area. 
$Q_{na}$ and $Q_{ta}$ can be considered as the energy fluxes attributed to the form and friction drags, respectively. 
It should be noted that, in 
(\ref{eq:Q_na}), we included the term of $u_n \tau_{na}$ but its contribution was negligibly small. 
We have defined the energy dissipation rate in the water side per unit horizontal area, $\epsilon_w$, as
\begin{equation}
  \epsilon_{w} = 
  \frac{1}{L_1 L_2}  
  \int_{V_w} \epsilon {\rm d}V
\label{eq:epsilon_w} \, ,
\end{equation}
where the local energy dissipation rate $\epsilon$ is given by
\begin{equation}
  \epsilon = 2 \rho_w \nu_w s_{ij}s_{ij}
\label{eq:epsilon} \, ,
\end{equation}
in which $s_{ij}$ is the $(i,j)$ component of the strain-rate tensor, defined as $s_{ij} \equiv \frac{1}{2}\left(\frac{\partial U_i}{\partial x_j} + \frac{\partial U_j}{\partial x_i}\right)$.
$V_w$ in (\ref{eq:epsilon_w}) represents the volume in the water side.

\begin{figure}
  \centerline{ \includegraphics[width=8cm]{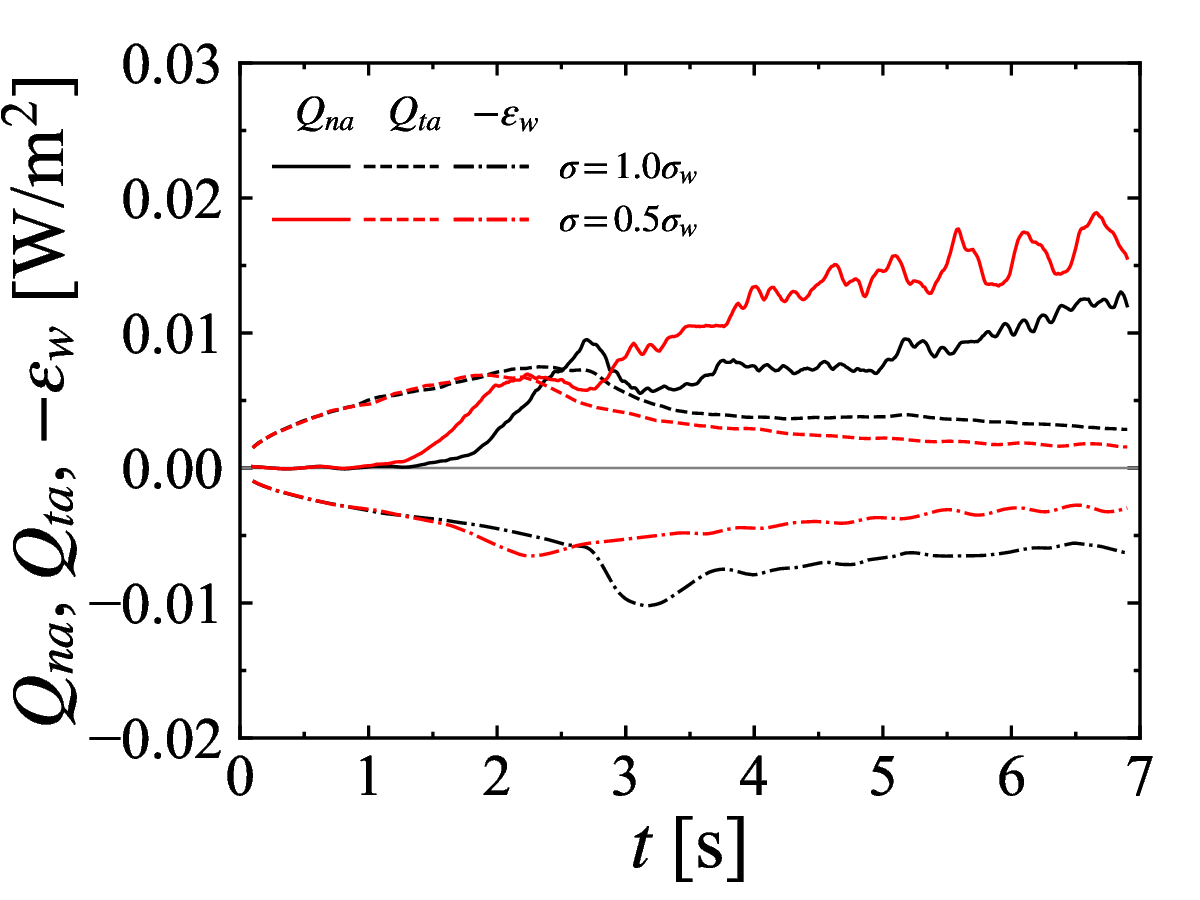} }
  \caption{Temporal variations of the energy fluxes due to the normal and tangential stresses, $Q_{na}$ and $Q_{ta}$, and the energy dissipation rate $\epsilon_w$ in the water side as negative values. 
A simple centred moving average for the period of 0.2 s is applied.}
\label{fig:energy_flux}
\end{figure}

Figure \ref{fig:energy_flux} shows the temporal variations of the energy fluxes from the air side to the water side, $Q_{na}$ and $Q_{ta}$, as well as the energy dissipation rate on the water side, $\epsilon_w$. 
The dissipation rate $\epsilon_w$ is shown as negative.
A simple centred moving average for the period of 0.2 s is applied. 
$Q_{na}$ for $\sigma=0.5\sigma_w$ is larger than that for $\sigma=1.0\sigma_w$ except the period of 
\MatsudaB{2.4 s $\lesssim t \lesssim$ 2.9 s}, and $Q_{ta}$ for $\sigma=0.5\sigma_w$ becomes smaller than that for $\sigma=1.0\sigma_w$ after 
\MatsudaB{$t \approx 1.9$ s}. 
Both $Q_{na}$ and $Q_{ta}$ supply the kinetic energy to the water side, but $Q_{ta}$ mainly accelerates the mean surface velocity and forms a shear layer under the interface. 
$Q_{na}$ contributes to the wave growth rather than $Q_{ta}$ because the energy flux to gravity-capillary waves is attributed to the form drag \citep{Melville&Fedorov(2015)}. 
Thus, the larger values of $Q_{na}$ for $\sigma=0.5\sigma_w$ means more energy input to the waves than the case of $\sigma=1.0\sigma_w$. 
It should be also noted that the energy dissipation could also have a significant effect on the wave growth because steady gravity-capillary waves can be obtained when the energy flux due to the form drag is balanced by the viscous energy dissipation \citep{Melville&Fedorov(2015)}. 
Figure~\ref{fig:energy_flux} indicates that $\epsilon_w$ is not negligibly small compared to $Q_{na}$.

\begin{figure}
  \centerline{
    (\textit{a}) \includegraphics[width=6cm]{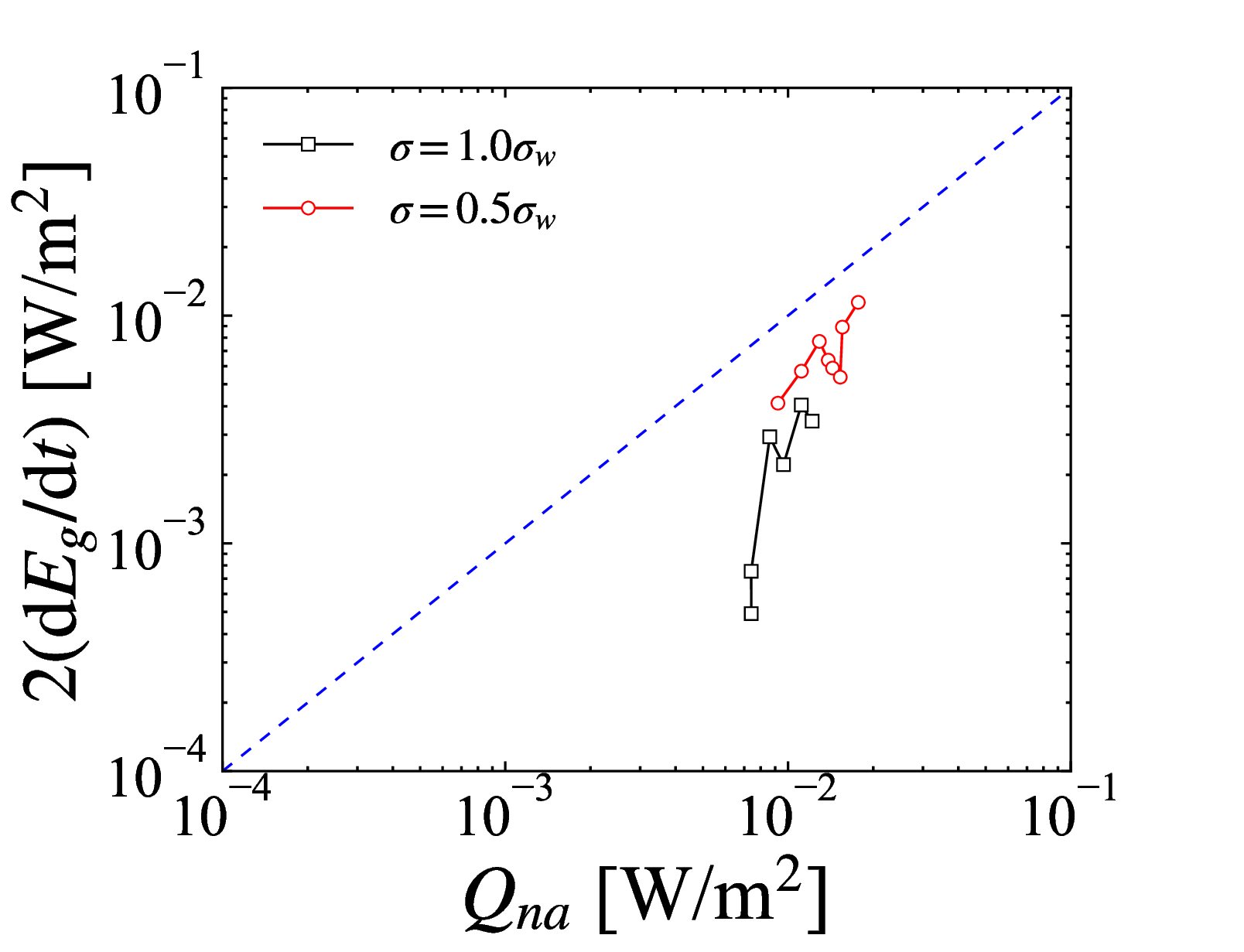}
    (\textit{b}) \includegraphics[width=6cm]{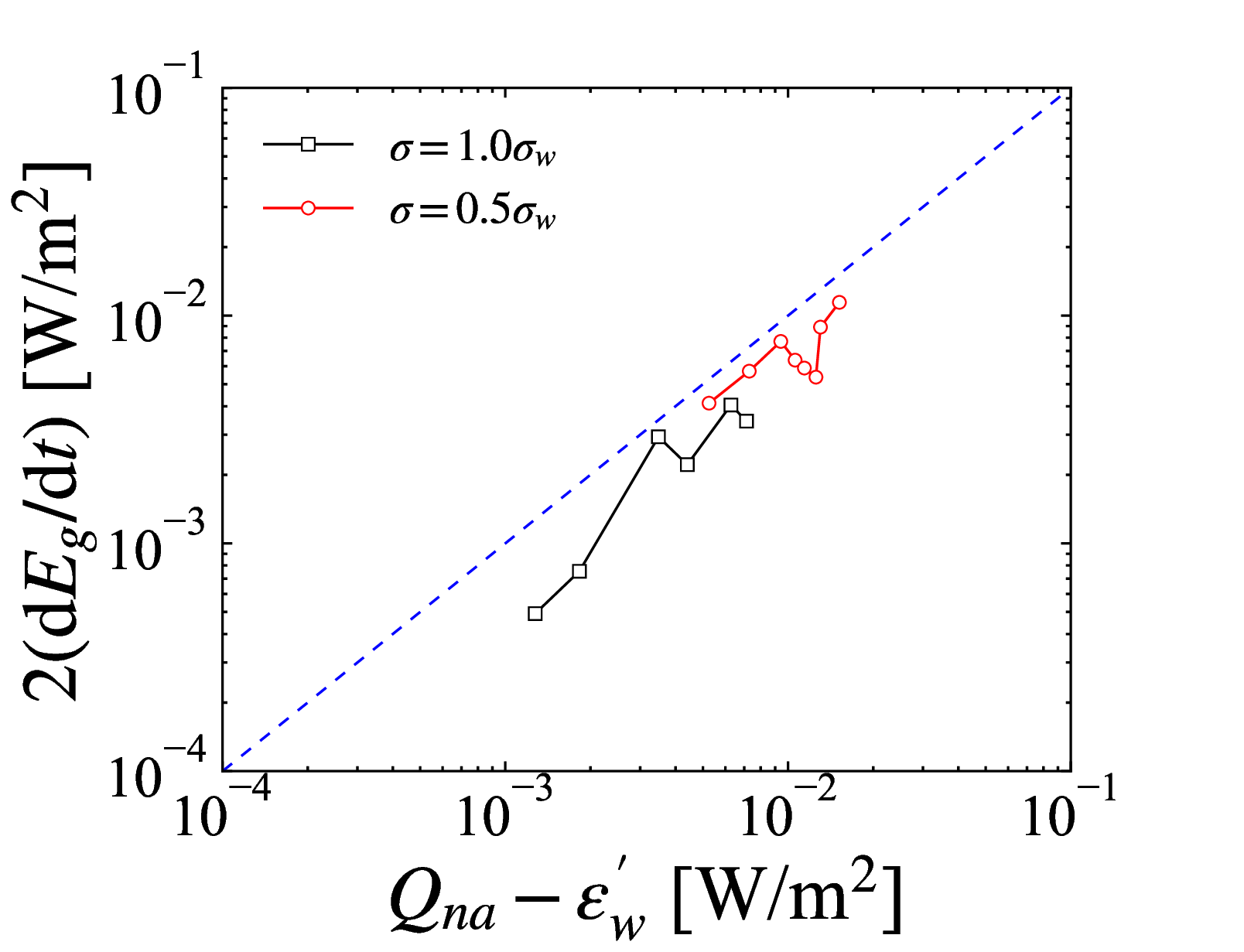}
  }
  \caption{
  Wave energy change rate  
  $2\frac{{\rm d} E_g}{{\rm d} t}$ 
  against
  (\textit{a}) $Q_{na}$ and (\textit{b}) $Q_{na}-\epsilon_w'$. $Q_{na}$ and $Q_{na}-\epsilon_w'$ are temporally averaged for every 0.5 s after $t=4.0$ s for $\sigma=1.0\sigma_w$ and $t=3.0$ s for $\sigma=0.5\sigma_w$. $\frac{{\rm d} E_g}{{\rm d} t}$ is obtained by the linear fitting to $E_g$ for every 0.5 s in the corresponding period.}
\label{fig:dEgdt}
\end{figure}

Therefore, we have examined the relationship between the energy fluxes and the rapid growth of $E_g$:
The time derivative of $E_g$ (i.e., $\frac{{\rm d} E_g}{{\rm d} t}$) has been compared with the energy input to waves and wave energy dissipation.
%
Note that the energy dissipation rate $\epsilon_w$ also contains the dissipation due to mean shear, which does not contribute to the wave energy dissipation. 
Hence, we consider the energy dissipation due to 
fluctuation of the strain rate, i.e., $\epsilon_w' \equiv \epsilon_w - \epsilon_{w,m}$, where $\epsilon_{w,m}$ is the energy dissipation rate due to the mean shear.
We evaluated $\epsilon_{w,m}$ on the boundary-fitted coordinate as
\begin{equation}
\epsilon_{w,m} \equiv 2 \rho_w \nu_w \int_{-2\delta}^0 
\Matsuda{\langle s_{ij} \rangle_{z_{g0}} \langle s_{ij} \rangle_{z_{g0}}}
{\rm d} z_{g0}
\label{eq:eps_mean},
\end{equation}
where 
\Matsuda{
$\langle s_{ij} \rangle_{z_{g0}}$ is the horizontally averaged strain rate tensor for the same $z_{g0}$, i.e., the average over the surface $\Gamma_{z_{g0}}$, which initially locates at $x_3=z_{g0}$ and moves along with the boundary-fitted coordinate. 
}
Here, we used the 
approximation  
of $\frac{\partial z_g}{\partial z_{g0}} \approx 1$ to calculate $\epsilon_{w,m}$: We assumed the volume change at each grid is negligibly small. 
We have confirmed that the error of computing $\epsilon_{w}$ due to the approximation is negligibly small (less than 4 \%). 
%
The time derivative $\frac{{\rm d} E_g}{{\rm d} t}$ has been calculated by fitting linear function to the time series of $E_g$ for every 0.5 s. 
Since we focused on the rapid growth 
of $E_g$, 
the change rate  
$\frac{{\rm d} E_g}{{\rm d} t}$ was calculated for $4.0 \le t < 7.0$ s and $3.0 \le t < 7.0$ s for $\sigma=1.0\sigma_w$ and $\sigma=0.5\sigma_w$, respectively.
When the potential energy $E_g$ increases, the kinetic energy associated with the gravity waves would also increases, and that 
the balance between energy input and dissipation 
should contribute to the increase of the total wave energy. 
However, it is difficult to extract the kinetic energy due to the wave motion from the total kinetic energy including the turbulent kinetic energy.  
Here, we assume that the kinetic energy 
of the waves 
is equivalent to the potential energy
as it is the case of monochromatic waves. 
The change rate  
of the total energy of gravity waves is then given by $2\frac{{\rm d} E_g}{{\rm d} t}$.
In figure~\ref{fig:dEgdt}, 
the change rate 
$2\frac{{\rm d} E_g}{{\rm d} t}$ for every 0.5 s is compared with the time-averaged $Q_{na}$ and $Q_{na}-\epsilon_w'$  
for every 0.5 s in the corresponding period. 
The dotted line indicates the $2\frac{{\rm d} E_g}{{\rm d} t}$ values equivalent to $Q_{na}$ or $Q_{na}-\epsilon_w'$. 
In figure~\ref{fig:dEgdt}(\textit{a}), the markers are located in lower side of the dotted line for both $\sigma=1.0\sigma_w$ and $\sigma=0.5\sigma_w$, meaning that the energy flux $Q_{na}$ is larger than $2\frac{{\rm d} E_g}{{\rm d} t}$.
In figure~\ref{fig:dEgdt}(\textit{b}), the markers are also located in lower side but closer to the dotted line. This means that the correlation of $2\frac{{\rm d} E_g}{{\rm d} t}$ with $Q_{na}-\epsilon_w'$  
is better than that with $Q_{na}$. 
Thus, the rapid growth of $E_g$ is attributed to 
the energy flux due to normal stress $Q_{na}$ minus the energy dissipation fluctuation $\epsilon_w'$. 
\Matsuda{
The wave growth is often quantified by using the wave growth rate \citep{Plant(1982),Donelan(2006),Melville&Fedorov(2015)}, which is given by the change rate of the wave energy. The comparison of the wave growth rate obtained from the DNS results with the previous measurements is described in Appendix \ref{appC}.
}

\begin{figure}
  \centerline{
    (\textit{a}) \includegraphics[width=6cm]{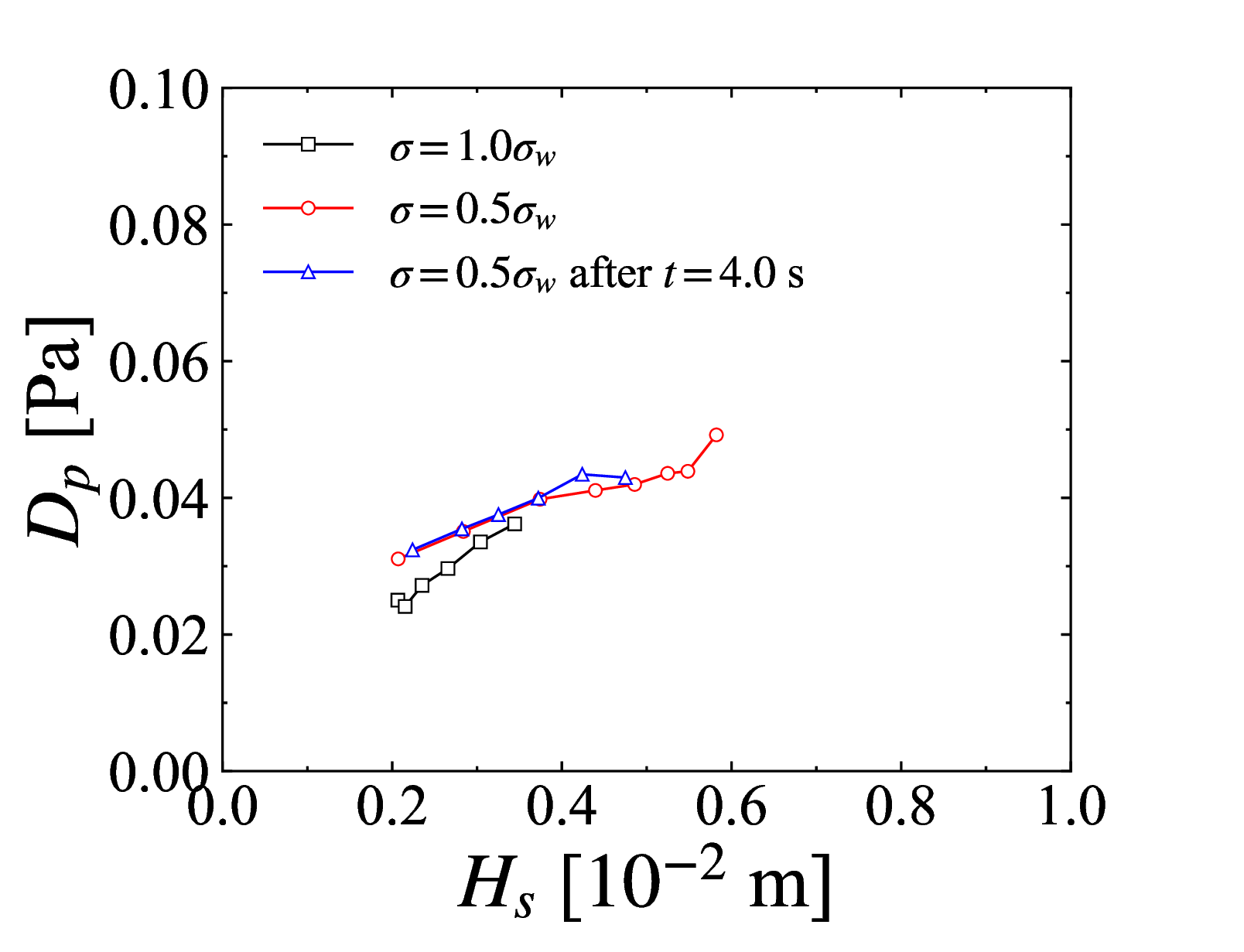}
    (\textit{b}) \includegraphics[width=6cm]{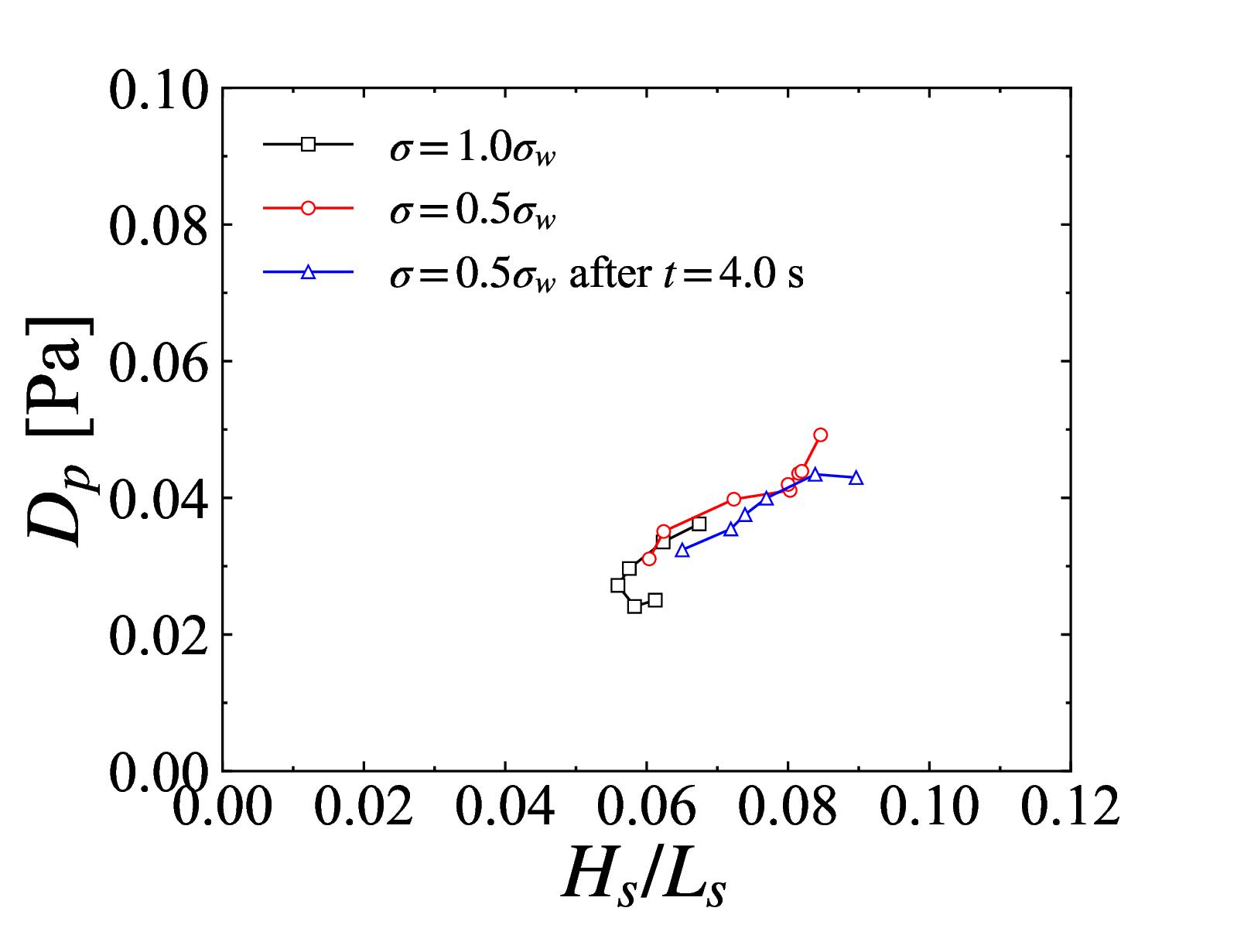}
  }
  \caption{ 
  Form drag $D_p$ 
  against
  (\textit{a}) significant wave height $H_s$ and (\textit{b}) wave slope $H_s/L_s$. $D_p$, $H_s$, and $H_s/L_s$ are temporally averaged for every 0.5 s after $t=4.0$ s for $\sigma=1.0\sigma_w$ and $t=3.0$ s for $\sigma=0.5\sigma_w$. }
\label{fig:Dp_vs_HsLs}
\end{figure}

The reason of the larger $Q_{na}$ for $\sigma=0.5\sigma_w$ is indeed not surprising because $Q_{na}$ is the energy flux attributed to the form drag, 
and it is reasonable to assume that the form drag becomes larger along with wave growth.
The form drag $D_p$ is given by 
\begin{equation}
  D_p = 
  \frac{1}{L_1 L_2}
  \int_\Gamma (p_a + \rho_a g\eta)n_1 {\rm d}S
\label{eq:Dp} \, ,
\end{equation}
where $n_1$ is the streamwise components of the normal vector to the interface.
\citet{Melville&Fedorov(2015)} evaluated the energy flux due to the normal stress by $Q_{na} \approx \mathcal{C} D_p$, where $\mathcal{C}$ is the wave velocity.
We have confirmed that $Q_{na}$ and $\mathcal{C} D_p$ are well correlated 
also 
in our results. 
Here, the obtained form drag $D_p$ is compared with the significant wave height $H_s$ and the wave slope $H_s/L_s$ in figure~\ref{fig:Dp_vs_HsLs}.
Similar to figure~\ref{fig:dEgdt}, the time-averaged values for every 0.5 s are plotted for $4.0 \le t < 7.0$ s and $3.0 \le t < 7.0$ s for $\sigma=1.0\sigma_w$ and $\sigma=0.5\sigma_w$, respectively. 
In figure~\ref{fig:Dp_vs_HsLs}(\textit{a}), the form drag $D_p$ for the same wave height $H_s$ is larger for the smaller surface tension case. 
The difference in $D_p$ between the cases of $\sigma=1.0\sigma_w$ and $\sigma=0.5\sigma_w$ is robust because $D_p$ calculated for the case of sudden change from $\sigma=1.0\sigma_w$ to $\sigma=0.5\sigma_w$ at $t = 4.0$ s also becomes larger than $D_p$ for the case where the surface tension remained $\sigma=1.0\sigma_w$.
Figure~\ref{fig:Dp_vs_HsLs}(\textit{b}) shows that the form drag $D_p$ is rather correlated with the wave slope $H_s/L_s$ because $D_p$ for the three cases of surface tension well collapse. 
This means that the significant waves become steeper due to the surface tension reduction and that causes the increases of the form drag $D_p$. 
Therefore, the increase of $Q_{na}$ due to the surface tension reduction can be explained by the increase of the wave slope $H_s/L_s$.

To investigate the relationship between the energy fluxes and the surface tension, we have evaluated the energy fluxes for the gravity and capillary waves separately. 
Here, the energy fluxes due to the normal stress at gravity and capillary wave scales ($Q_{na,g}$ and $Q_{na,c}$ respectively) are defined as 
\begin{eqnarray}
  Q_{na,g} &=& 
    \int_{0}^{k_{\rm cut}} S_{Qna}(k) \, {\rm d}k
\label{eq:Q_nag} \, , \\[3pt]
  Q_{na,c} &=&
    \int_{k_{\rm cut}}^{k_{\rm max}} S_{Qna}(k) \, {\rm d}k
\label{eq:Q_nac} \, ,
\end{eqnarray}
where $k$ is the magnitude of the two-dimensional horizontal wavenumber vector ${\bm k}=(k_1,k_2)$, $k_{\rm max}$ is the maximum wavenumber, and $S_{Qna}(k)$ is the Fourier spectrum of the energy flux due to the normal stress at the interface. 
$S_{Qna}(k)$ can be given by 
\begin{equation}
  S_{Qna}(k) = \frac{1}{\Delta k} \sum_{k-\Delta k/2 \le |\bm k| < k+\Delta k/2} \mathfrak{R}\left[ - \widehat{(p_a + \rho_a g\eta)} \widehat{\frac{\partial \eta}{\partial t}}^* \right]
  \label{eq:Pqna},
\end{equation}
where $\widehat{A}(k_1,k_2)$ indicates the Fourier coefficient of a given horizontal distribution $A(x_1,x_2)$,  
and $\mathfrak{R}[\cdot]$ indicates the real part of the complex number.
$Q_{na,g}$ and $Q_{na,c}$ can be interpreted as sharp-cut low- and high-pass filtered values of local energy flux $- (p_a + \rho_a g\eta)\frac{\partial \eta}{\partial t}$ for the cut-off wavenumber $k_{\rm cut}$.
We set $k_{\rm cut}$ to $k_m$ for $\sigma=1.0\sigma_w$. 
We have confirmed that the above energy fluxes satisfy $Q_{na} = Q_{na,g} + Q_{na,c}$.
Similarly, the energy dissipation at gravity and capillary wave scales, $\epsilon_{w,g}$ and $\epsilon_{w,c}$ respectively, are defined as 
\begin{eqnarray}
  \epsilon_{w,g} &=& 
    \int_{0}^{k_{\rm cut}} \int_{-2\delta}^0 S_{\epsilon w}(k,z_{g0}) \, {\rm d}z_{g0} \, {\rm d}k
\label{eq:epsilon_wg} \, , \\[3pt]
  \epsilon_{w,c} &=&
    \int_{k_{\rm cut}}^{k_{\rm max}} \int_{-2\delta}^0 S_{\epsilon w}(k,z_{g0}) \, {\rm d}z_{g0} \, {\rm d}k
\label{eq:epsilon_wc} \, ,
\end{eqnarray}
where $S_{\epsilon w}(k,z_{g0})$ is the Fourier spectrum of the energy dissipation rate as a function of initial vertical position $z_{g0}$. 
$S_{\epsilon w}(k,z_{g0})$ can be given by 
\begin{equation}
  S_{\epsilon w}(k,z_{g0}(m_3)) = \frac{2 \rho_w \nu_w}{\Delta k} \sum_{k-\Delta k/2 \le |\bm k| < k+\Delta k/2} \mathfrak{R}\left[ \widehat{s_{ij}}({\bm k},z_{g0}(m_3)) \widehat{s_{ij}}^*({\bm k},z_{g0}(m_3)) \right]
  \label{eq:Dspect} ,
\end{equation}
where $\widehat{s_{ij}}({\bm k},z_{g0}(m_3))$ is obtained by the Fourier transform in the $x_1$ and $x_2$ direction for $s_{ij}(x_g(m_1),y_g(m_2),z_g(m_1,m_2,m_3))$ 
for 
each $m_3$. 
We used the 
approximation  
of $\frac{\partial z_g}{\partial z_{g0}} \approx 1$ again 
to define $\epsilon_{w,g}$ and $\epsilon_{w,c}$ as 
(\ref{eq:epsilon_wg}) and (\ref{eq:epsilon_wc}), respectively. 
We have confirmed that sum of these energy dissipation rates is approximately equal to $\epsilon_{w}'$; i.e., $\epsilon_{w,g} + \epsilon_{w,c} \approx \epsilon_{w}'$. 

\begin{figure}
  \centerline{
    (\textit{a}) \includegraphics[width=8cm]{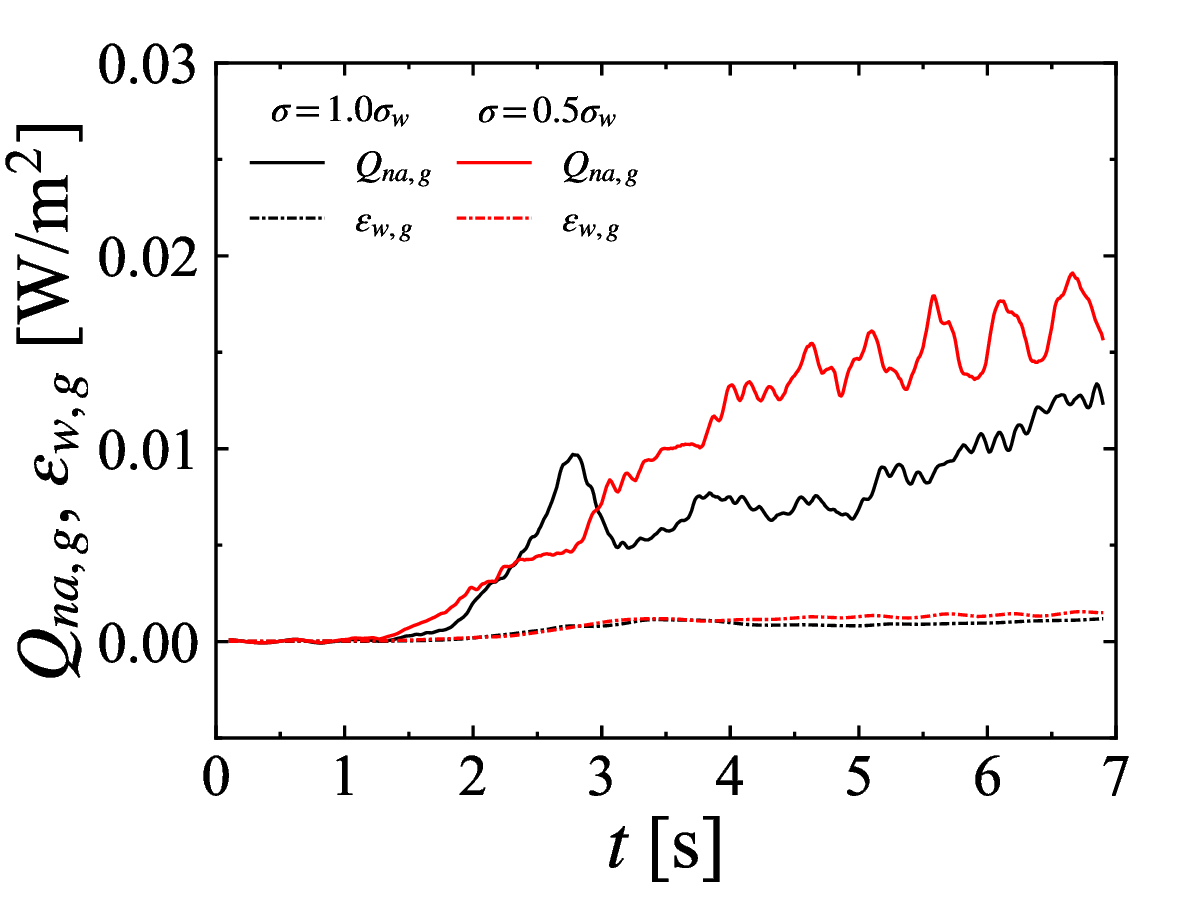}
  }
  \centerline{
    (\textit{b}) \includegraphics[width=8cm]{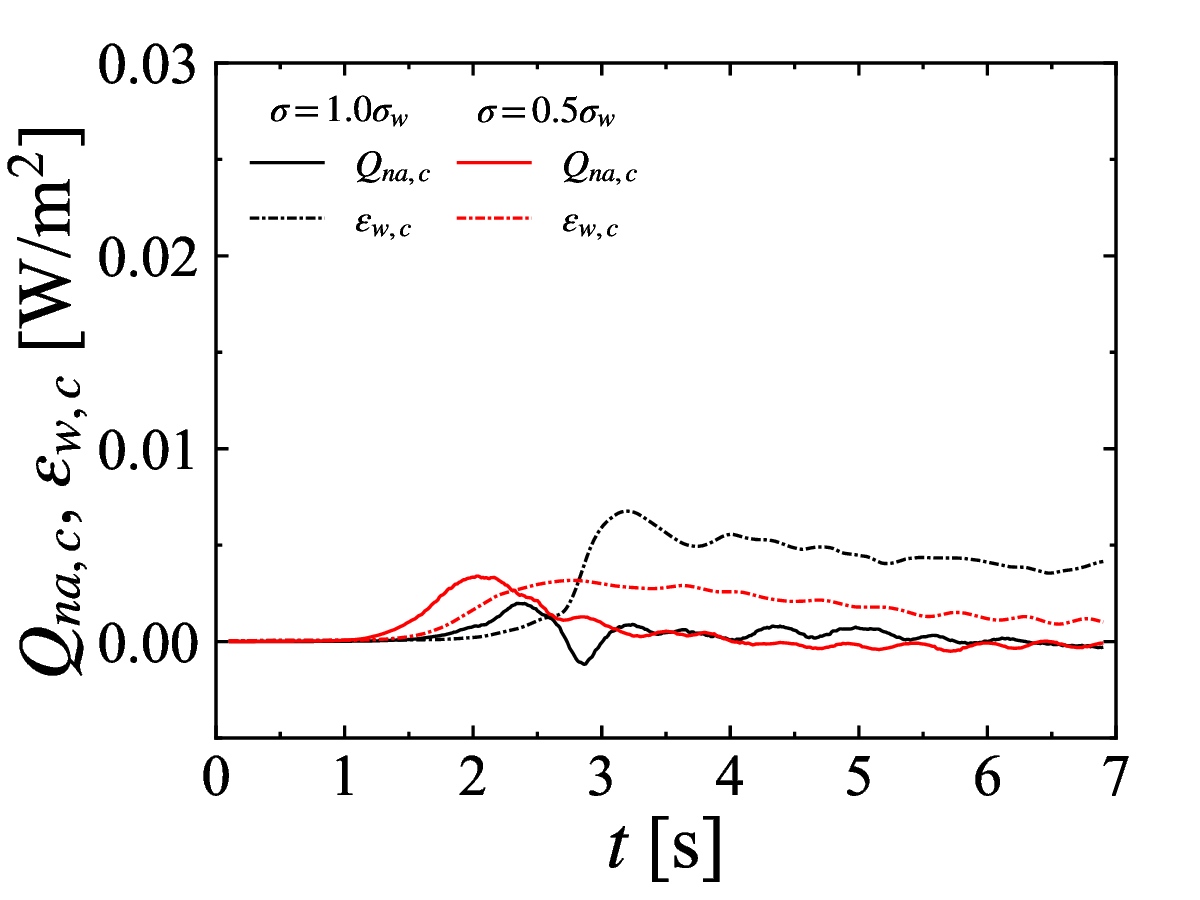}
  }
  \caption{Energy flux due to normal stress and energy dissipation rate at (\textit{a}) gravity wave scales and (\textit{b}) capillary wave scales. A simple centred moving average for the period of 0.2 s is applied.}
\label{fig:Qna_epsw}
\end{figure}

Figure~\ref{fig:Qna_epsw} shows the energy flux due to normal stress and energy dissipation rate at gravity and capillary wave scales.
After $t=2.5$ s, $Q_{na,g}$ is larger than $Q_{na,c}$, indicating that the energy flux due to the normal stress at the gravity wave scales has dominant influence on the total energy flux $Q_{na}$. This is consistent with our intuition that the contribution of the form drag is dominated by the significant gravity waves. 
Concerning the dissipation rate, $\epsilon_{w,c}$ is 
far 
larger than $\epsilon_{w,g}$. Therefore, the energy dissipation at the capillary wave scales is dominant for 
$\epsilon_{w}'$. 
The energy dissipation for capillary wave scales $\epsilon_{w,c}$ for $\sigma=1.0\sigma_w$ is significantly larger than that for $\sigma=0.5\sigma_w$. 
It is also observed in figure~\ref{fig:Qna_epsw}(\textit{b}) that $\epsilon_{w,c}$ is larger than $Q_{na,c}$. This means that the energy dissipation in the water side exceeds the energy input from the air side at the capillary wave scales, and there must be the energy transfer from the gravity wave scales to the capillary wave scales. 
The difference between $\epsilon_{w,c}$ and $Q_{na,c}$ is smaller for the smaller surface tension case.
This indicates that the energy transfer from the gravity wave scales decreases due to the surface tension reduction.
There are several mechanisms that can result in the inter-scale energy transfer; the four-wave resonant interaction, harmonic resonance of nonlinear waves, and turbulent energy transfer. 
Unfortunately, it is difficult to specify the dominant mechanism because of the difficulty of evaluating the energy transfer.  
However, comparison of $Q_{ta}$ 
(in figure~\ref{fig:energy_flux}) 
and $\epsilon_{w,c}$ suggests that change in the resonant condition due to the surface tension reduction results in the decrease of the energy transfer from the gravity wave scales because $\epsilon_{w,c}$ (i.e., the energy transfer from the gravity wave scales) is larger than $Q_{ta}$, which produces shear-induced turbulence in the water side, for $\sigma=1.0\sigma_w$.

\Matsuda{
To fully understand the wave growth mechanism, quantitative evaluation of wave--wave interaction is crucial. However, it is difficult to evaluate the energy flux between waves directly from our simulation data. 
The present results of the energy fluxes at gravity and capillary wave scales indirectly prove the energy transfer between these scales and the effect of the surface tension reduction on the energy transfer.
}
These results indicate that the significant gravity waves receive kinetic energy from the air side with contributions of form drag, i.e., $Q_{na}$, and the energy is partially transferred to capillary waves and dissipated by the viscosity of water. 
When the surface tension decreases, the energy transfer to the capillary waves decreases, and the significant waves retain more energy, resulting in the faster growth of the wave height. 
The faster wave growth can further cause the increase in the wave slope and, therefore, the increase of the energy received from the air side with contributions of form drag.


\subsection{Effect on scalar transfer}
\label{sec:scalar}

\begin{figure}
  \centerline{
    \includegraphics[width=8cm]{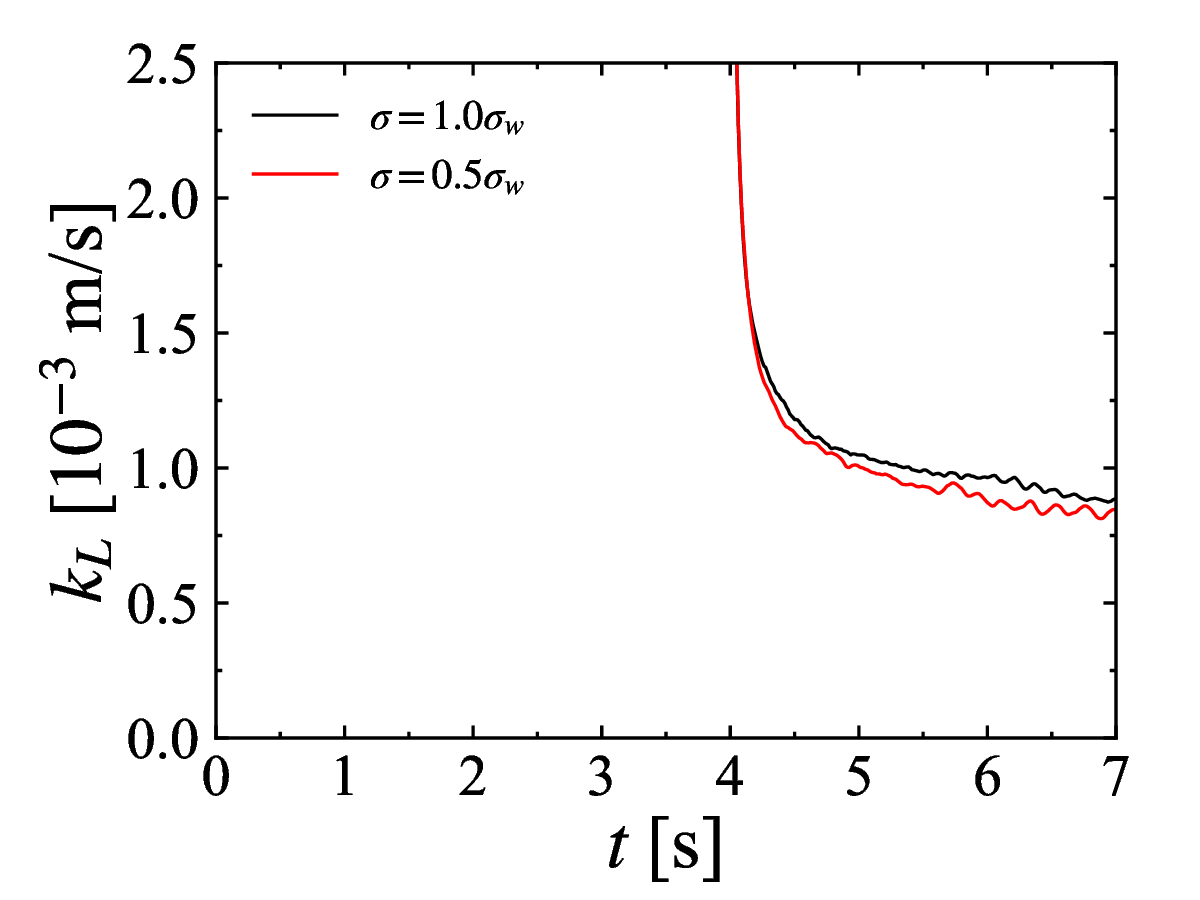}
  }
  \caption{Temporal variations of scalar transfer 
  coefficient
  on the water side $k_L$. }
\label{fig:kL}
\end{figure}

\begin{figure}
  \centerline{
    \includegraphics[width=8cm]{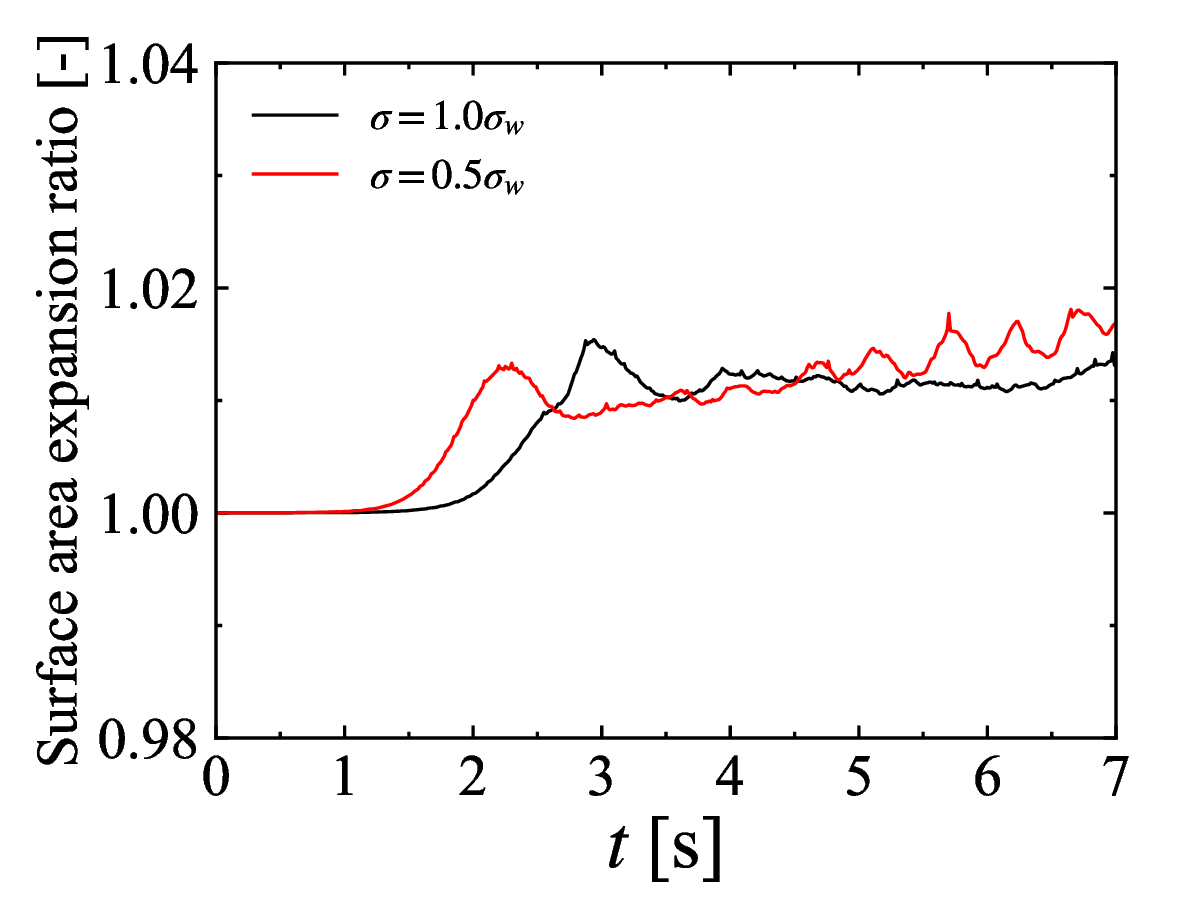}
  }
  \caption{Temporal variations of surface area expansion ratio.}
\label{fig:surf_area_ratio}
\end{figure}

The effect of the surface tension reduction on the scalar transfer across the air--water interface has been investigated by solving the scalar transport equation (\ref{eq:scalar}) for $t \ge 4.0$ s.
The scalar transfer was evaluated by the scalar transfer coefficient
in the water side $k_L$, defined as 
\begin{equation}
  F = k_L (C_i-C_b)
\label{eq:kL} \, .
\end{equation}
Here, $C_i$ is the scalar concentration on the interface ($C_i = 1$), $C_b$ is the bulk scalar concentration on the water side ($C_b = 0$), and $F$ is the 
horizontally averaged 
scalar flux defined as
\begin{equation}
  F = 
  \frac{1}{L_1 L_2}
  \int_\Gamma D \frac{\partial C}{\partial x_n} {\rm d}S
\label{eq:scalar_flux} \, ,
\end{equation}
where $x_n$ is the normal distance from the interface. 
Figure \ref{fig:kL} shows the scalar transfer coefficient 
$k_L$ for $\sigma=1.0\sigma_w$ and $0.5\sigma_w$. 
Immediately after the start of the scalar transport calculation at $t=4.0$ s, $k_L$ shows large values because  large concentration gradient near the air--water interface is formed in this transient period. 
However, for $t>4.5$ s, the scalar concentration boundary layer is sufficiently developed to achieve the local equilibrium of scalar transfer near the interface, and $k_L$ for $\sigma=0.5\sigma_w$ is smaller than that for $\sigma=1.0\sigma_w$. 
That is, the scalar transfer decreases as the surface tension decreases. 
The difference is about 
\MatsudaB{6.5 \% for 5.0 s $\le t \le$ 7.0 s}. 
One could postulate that the decrease of $k_L$ is due to the difference in the surface area of the air--water interface. 
Figure~\ref{fig:surf_area_ratio} shows the temporal variation of the surface area expansion ratio, defined as the ratio of the surface area of the interface to the horizontal area.
The change in the expansion ratio is less than or approximately equal to 2 \%, and the difference of the expansion ratio is even smaller; i.e., the difference of the expansion ratio is too small to explain the difference of $k_L$.
These results indicate that the difference in the surface area of the interface does not explain the 
\MatsudaB{6.5 \%} decrease in $k_L$ due to the surface tension reduction.

The transfer of gases such as CO$_2$ across a wind-driven wavy air--water interface is strongly affected by the turbulence under the interface \citep{Komori(1993a)}. 
\begin{figure}
  \centerline{
    (\textit{a}) \includegraphics[width=6cm]{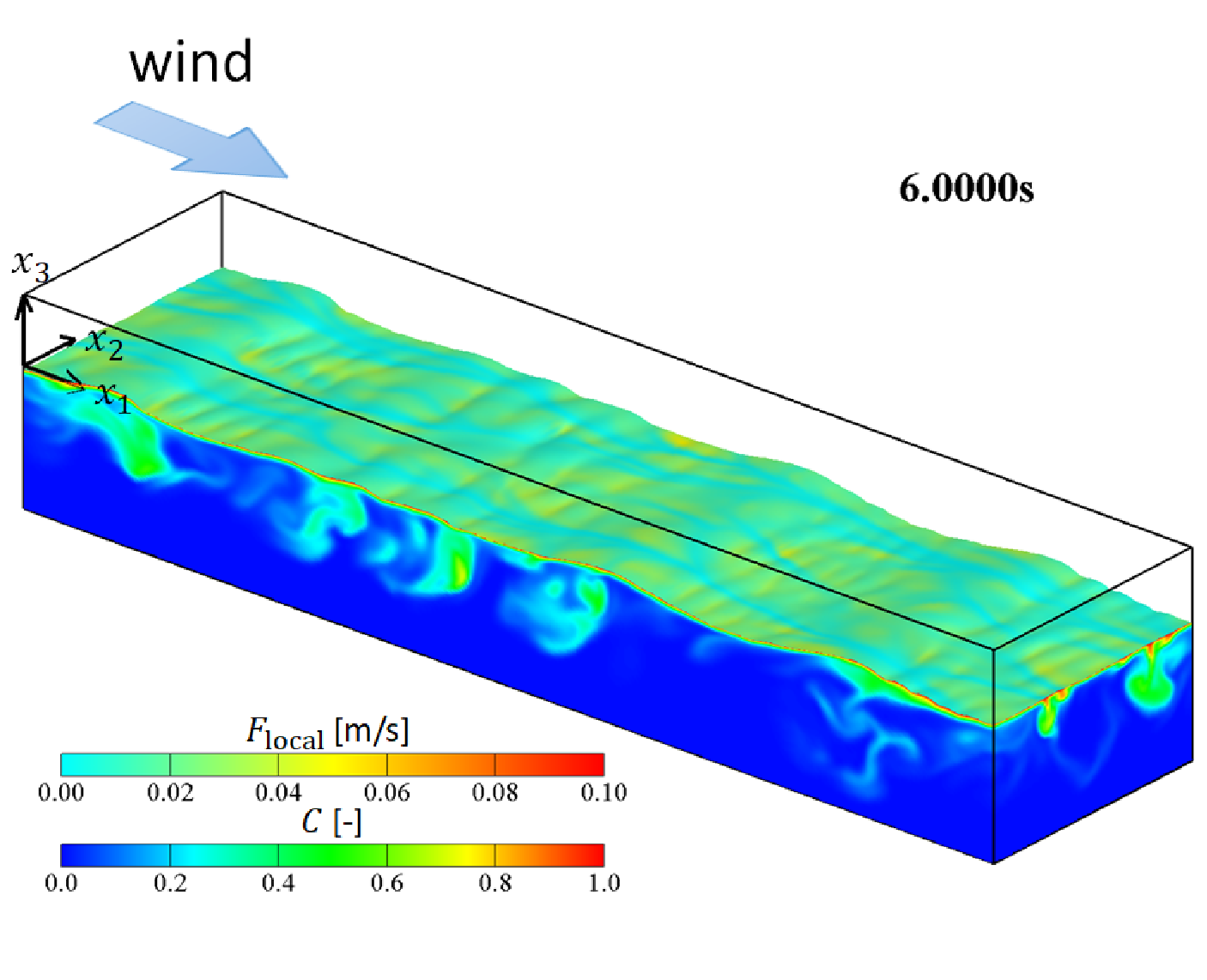}
    (\textit{b}) \includegraphics[width=6cm]{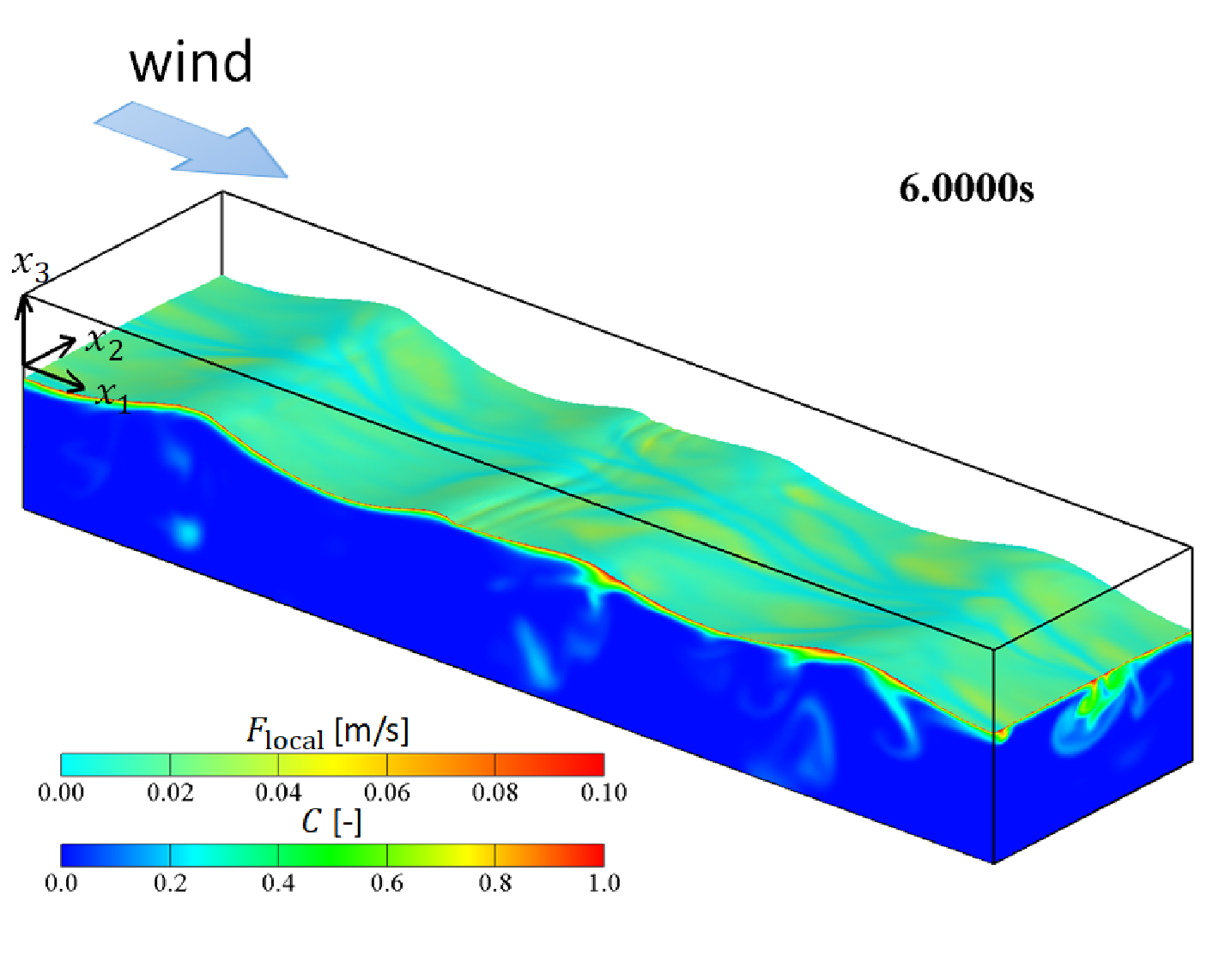}
  }
  \centerline{
    (\textit{c}) \includegraphics[width=6cm]{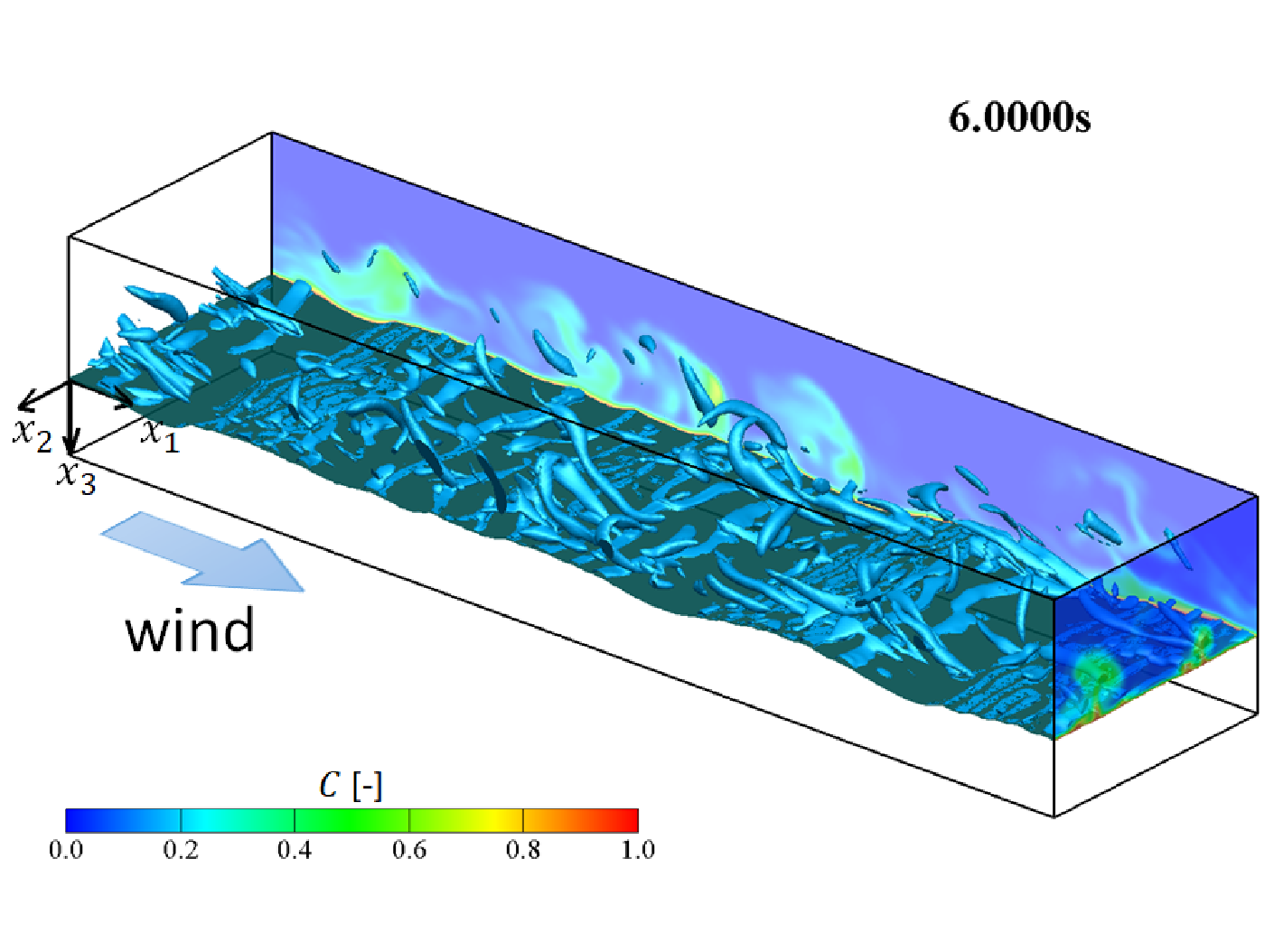}
    (\textit{d}) \includegraphics[width=6cm]{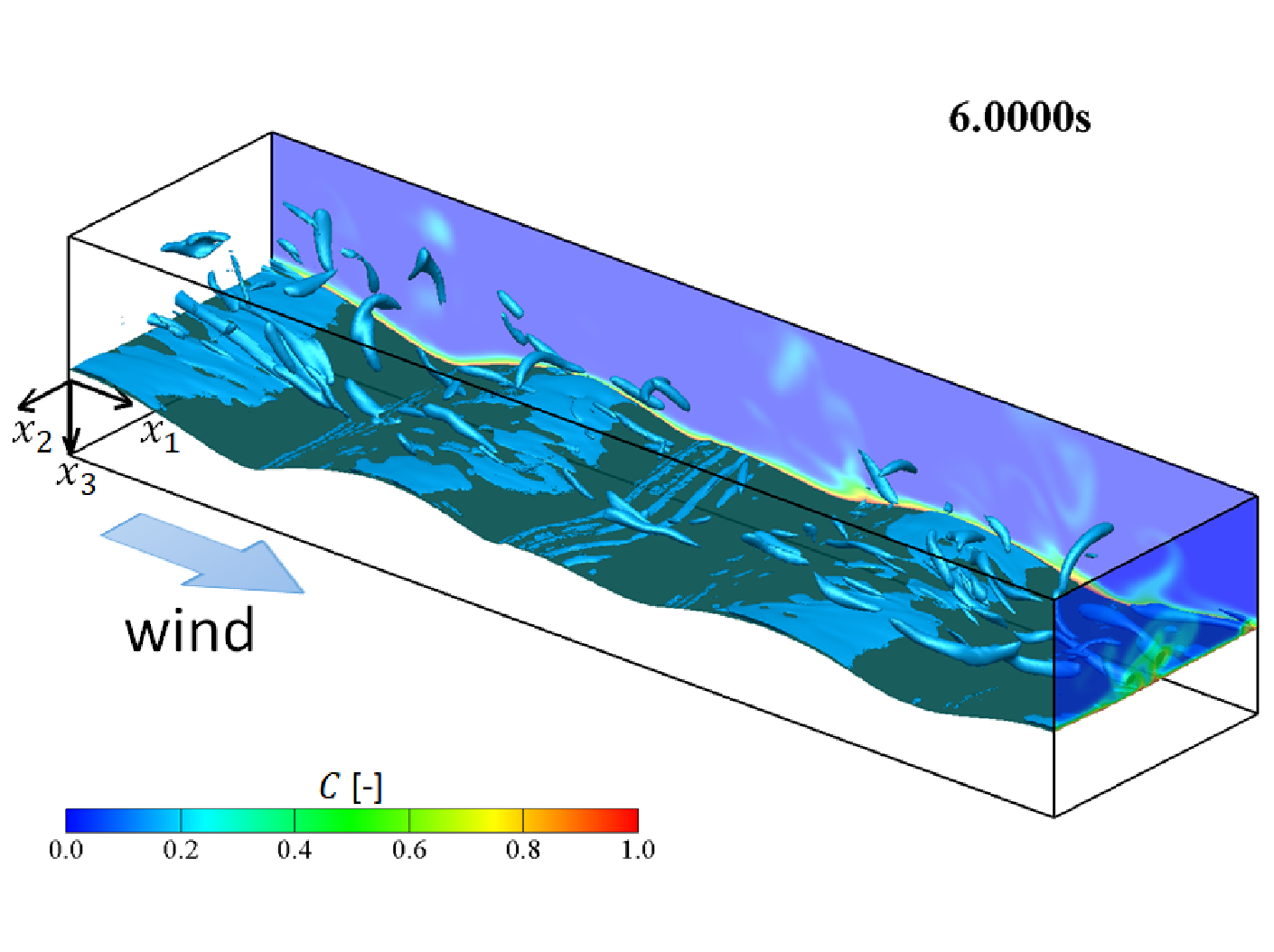}
  }
  \caption{(\textit{a}, \textit{b}) Colour contours of the local scalar flux $F_{local}$ on the interface at $t=6.0$  s and (\textit{c}, \textit{d}) the isosurfaces of the second invariant of velocity gradient tensor ($Q=0.002$ s$^{-2}$) in the water side, viewed from the water side 
  (bottom-side up), at $t=6.0$ s; (\textit{a}, \textit{c}) $\sigma=1.0\sigma_w$ and (\textit{b}, \textit{d}) $\sigma=0.5\sigma_w$. The colour contours on the side walls show the scalar concentration $C$. }
\label{fig:wave1_scalar}
\end{figure}
Figure~\ref{fig:wave1_scalar}(\textit{a}, \textit{b}) shows the local scalar flux $F_{local} \equiv D \frac{\partial C}{\partial x_n}$ on the interface. The colour contours on the side walls shows the scalar concentration in the water side. 
For the case of $\sigma=1.0\sigma_w$ in figure~\ref{fig:wave1_scalar}(\textit{a}), there are streamwise streaky structures of the local scalar flux on the interface, which are formed due to the 
\MatsudaA{turbulent flow structure}
in the water side \MatsudaA{\citep{Komori(2010),Takagaki(2015),Tejada-Martnez(2020)}}.
The similar streaky structures of the local scalar flux are also observed for the case of $\sigma=0.5\sigma_w$ in figure~\ref{fig:wave1_scalar}(\textit{b}), suggesting that the turbulence significantly affects to the scalar transfer even when the surface tension is reduced. 

\Matsuda{
To evaluate the scalar transfer below the moving interface, we calculate the downward scalar flux on the continuous surface $\Gamma_{z_{g0}}$, 
introduced in Section \ref{sec:growth}.
The surface $\Gamma_{z_{g0}}$ can be also considered as the surface for a fixed vertical index $m_3$, and the surface $\Gamma_{z_{g0}}$ for $z_{g0}=0$ corresponds to the air--water interface $\Gamma$. 
The scalar budget is evaluated for the volume below the surface $\Gamma_{z_{g0}}$. 
The scalar budget equation can be described for each level of $z_{g0}$ as
\begin{equation}
\frac{{\rm d} \chi(z_{g0})}{{\rm d} t} 
= F_{\rm diff}(z_{g0}) + F_{\rm turb}(z_{g0})
\label{eq:scalar_budget},
\end{equation}
where $\chi(z_{g0})$ is the total scalar per unit horizontal area integrated for the volume below the surface $\Gamma_{z_{g0}}$, and $F_{\rm diff}(z_{g0})$ and $F_{\rm turb}(z_{g0})$ are the downward molecular diffusion and turbulent scalar fluxes across the surface $\Gamma_{z_{g0}}$, respectively. 
$\chi(z_{g0})$, $F_{\rm diff}(z_{g0})$, and $F_{\rm turb}(z_{g0})$ are defined as
\begin{eqnarray}
\chi(z_{g0}) &=& \frac{1}{L_1L_2}\int_0^{L_1}\int_0^{L_2} \left\{ \int_{-2\delta}^{z(x_1,x_2,z_{g0})} C {\rm d}x_3 \right\} {\rm d}x_1 {\rm d}x_2
\label{eq:chi}, \\
F_{\rm diff}(z_{g0}) &=& \frac{1}{L_1L_2}\int_{\Gamma_{z_{g0}}}  D  \frac{\partial C}{\partial x_j} n_j {\rm d}S_{z_{g0}}
\label{eq:Fdiff}, \\
F_{\rm turb}(z_{g0}) &=& - \left\langle cw_3 \right\rangle_{z_{g0}}
\label{eq:Fturb},  
\end{eqnarray}
respectively, where $z(x_1,x_2,z_{g0})$ is the vertical position of the surface $\Gamma_{z_{g0}}$, 
$n_j$ is $j$th component of the unit normal vector to $\Gamma_{z_{g0}}$, 
${\rm d}S_{z_{g0}}$ is infinitesimal area on $\Gamma_{z_{g0}}$,
$c$ is the scalar concentration fluctuation ($c \equiv C - \langle C \rangle_{z_{g0}}$), 
$w_3$ is the vertical fluid velocity relative to the velocity of the boundary-fitted coordinate 
($w_3 \equiv U_3 - U_1 \frac{\partial z_g}{\partial x_1} - U_2 \frac{\partial z_g}{\partial x_2} - V_3$, 
where $V_3 = \frac{\partial z_g}{\partial t}$ is the vertical velocity of the boundary-fitted coordinate),
and the brackets $\langle \cdot \rangle_{z_{g0}}$ denote the horizontal average over the surface $\Gamma_{z_{g0}}$, i.e., $\langle A \rangle_{z_{g0}} =  \frac{1}{L_1L_2}\int_0^{L_1}\int_0^{L_2} A(x_1,x_2,z(x_1,x_2,z_{g0})) {\rm d}x_1{\rm d}x_2$ for a given function $A$.
The derivation of the scalar budget equation (\ref{eq:scalar_budget}) is described in Appendix \ref{appD}.
}

\Matsuda{
\begin{figure}
  \centerline{
    \includegraphics[width=6cm]{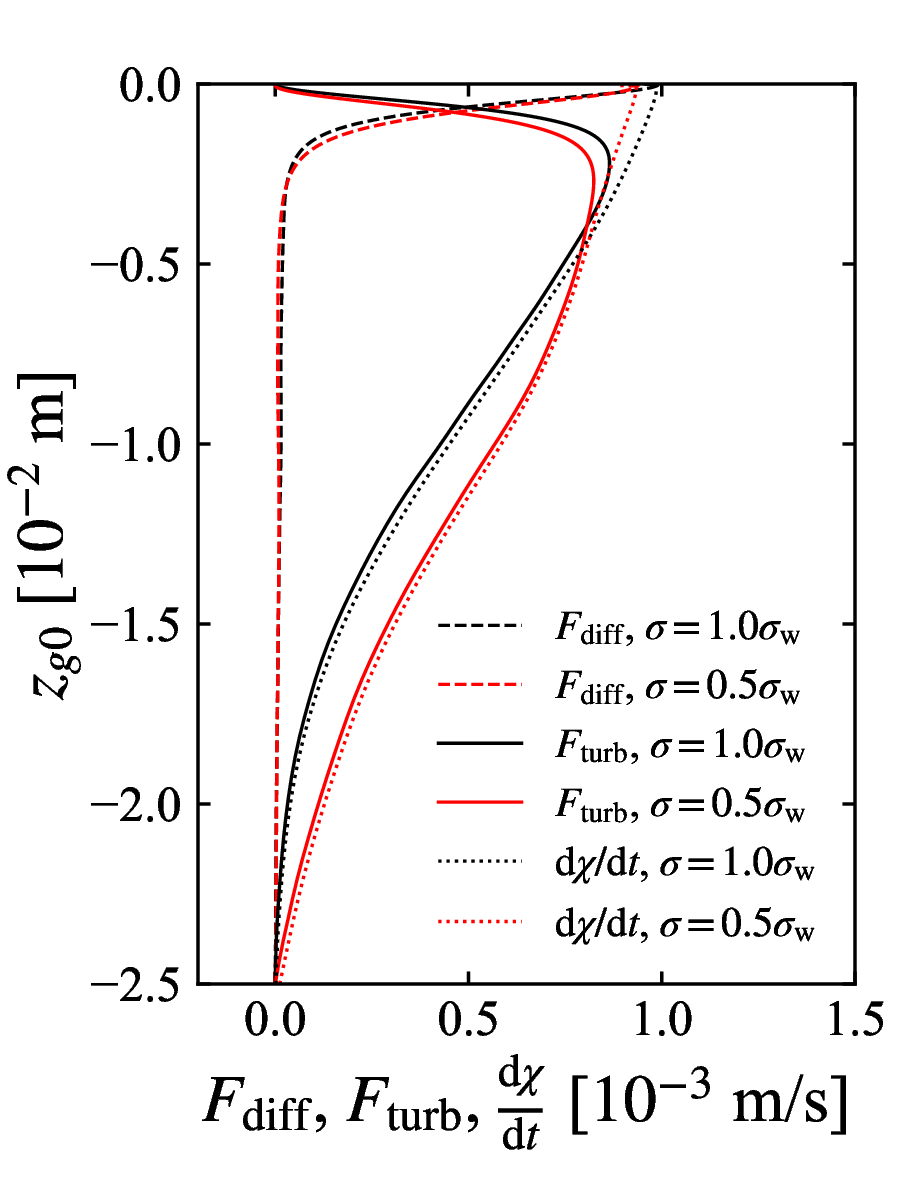}
  }
\caption{Vertical profiles of downward molecular diffusion and turbulent scalar fluxes, $F_{\rm diff}$ and $F_{\rm turb}$, respectively, and the time change of stored scalar $\chi$ below $z_{g0}$. Temporal average was taken for the period of 5.0 s $\le t \le$ 7.0 s. Horizontal average was taken on the the boundary-fitted coordinate. }
\label{fig:scalar_profile}
\end{figure}
Figure \ref{fig:scalar_profile} shows the vertical profiles of the downward molecular diffusion and turbulent scalar flux, $F_{\rm diff}$ and $F_{\rm turb}$, respectively, and the time change of $\chi$. 
The temporal average is taken for the period of 5.0 s $\le t \le$ 7.0 s, while the scalar budget equation (\ref{eq:scalar_budget}) is described for instantaneous scalar distribution.
For both cases of $\sigma=1.0\sigma_w$ and $\sigma=0.5\sigma_w$, 
$F_{\rm diff}$ is dominant for the scalar transfer only near the air--water interface.
The profile of $F_{\rm turb}$ shows a peak at $z_{g0} \approx -2.2 \times 10^{-3}$ m,
and $F_{\rm turb}$ is dominant below the $z_{g0}$. 
The peak value of $F_{\rm turb}$ for $\sigma=0.5\sigma_w$ is approximately 
\MatsudaB{4.7 \%} smaller than that for $\sigma=1.0\sigma_w$.
The magnitude and the difference of the peak values are consistent with the difference in $\frac{{\rm d}\chi}{{\rm d}t}$ \MatsudaB{near the surface} and the difference in $k_L$ in figure~\ref{fig:kL}.
These results indicate that the decrease in $k_L$ due to surface tension reduction is caused by the decrease in the turbulent scalar transfer.
}

In figure~\ref{fig:energy_flux},
it is observed that $Q_{ta}$ for $\sigma=0.5\sigma_w$ is smaller than that for $\sigma=1.0\sigma_w$. This means that the energy input for the shear-induced turbulence weakens due to the surface tension reduction. 
To confirm the turbulence structures in the water side, the turbulent vortices have been visualized by using the second invariant of the velocity gradient tensor, 
$Q \equiv \frac{1}{2} (\omega_{ij}\omega_{ij} - s_{ij}s_{ij})$, where $\omega_{ij}$ is the vorticity tensor defined as $\omega_{ij} \equiv \frac{1}{2}\left(\frac{\partial U_j}{\partial x_i} - \frac{\partial U_i}{\partial x_j}\right)$, and $s_{ij}$ is the strain-rate tensor defined above. 
Figures~\ref{fig:wave1_scalar}(\textit{c}, \textit{d}) show isosurfaces of $Q=0.002$ s$^{-2}$, viewed from the water side 
(bottom-side up). 
Strong turbulent vortices are distributed near the interface for both surface tension cases:
Spanwise vortices are observed in the vicinity of the interface, and streamwise vortices are observed below the wave troughs.
The structure similar to the spanwise vorticity was reported by \cite{Hung&Tsai(2009)}. They computed the evolution of two-dimensional gravity-capillary waves and observed that vortices shed from wave troughs distribute in the vicinity of the interface.  
The streaky structure of the scalar flux on the interface is attributed to trains of the streamwise vortices. 
It is also observed that the streamwise vortices for $\sigma=0.5\sigma_w$ are less than those for $\sigma=1.0\sigma_w$.

\begin{figure}
  \centerline{
    \includegraphics[width=6cm]{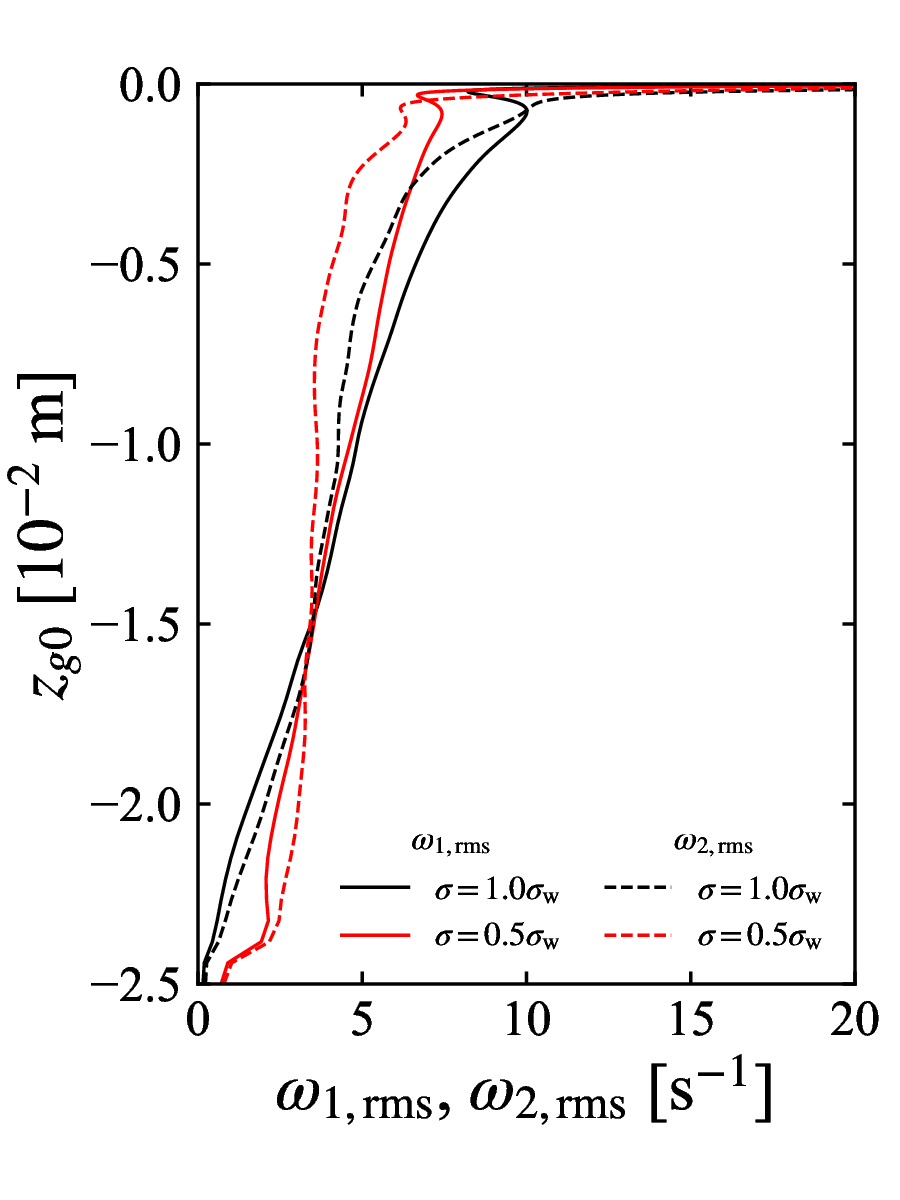}
  }
  \caption{Vertical profiles of rms values for streamwise and spanwise vorticity fluctuations, $\omega_{1,{\rm rms}}$ and $\omega_{2,{\rm rms}}$, respectively. Temporal average was taken for the period of 5.0 s $\le t \le$ 7.0 s.} 
\label{fig:u2rms}
\end{figure}

\MatsudaB{To compare the turbulent fluctuation, 
we have calculated the rms values of the streamwise and spanwise vorticities, 
because it is difficult to separate the effects of wave motion and turbulent flow on the velocity fluctuation. }
Figure~\ref{fig:u2rms} 
shows the rms values of streamwise and spanwise vorticity fluctuations, $\omega_{1,{\rm rms}}$ and $\omega_{2,{\rm rms}}$, respectively.
Large values of $\omega_{1,{\rm rms}}$ and $\omega_{2,{\rm rms}}$ are observed in the vicinity of the interface, and these large values correspond to the spanwise vortices in figures~\ref{fig:wave1_scalar}(\textit{c}, \textit{d}).
At the vertical position of $z_{g0} \approx -2.2 \times 10^{-3}$ m, where the peak of 
\Matsuda{$F_{\rm turb}$}
is located, the streamwise vorticity fluctuation $\omega_{1,{\rm rms}}$ exceeds the spanwise vorticity fluctuation $\omega_{2,{\rm rms}}$. Thus, the streamwise vortices play major roles for the turbulent scalar transfer quantified by 
\Matsuda{$F_{\rm turb}$}.
We can also find that the streamwise vorticity fluctuation is weaker for the case of $\sigma=0.5\sigma_w$. This also indicates that the decrease in the turbulent scalar transfer is due to the decrease in the turbulent fluctuation.  

\begin{figure}
  \centerline{
    \includegraphics[width=8cm]{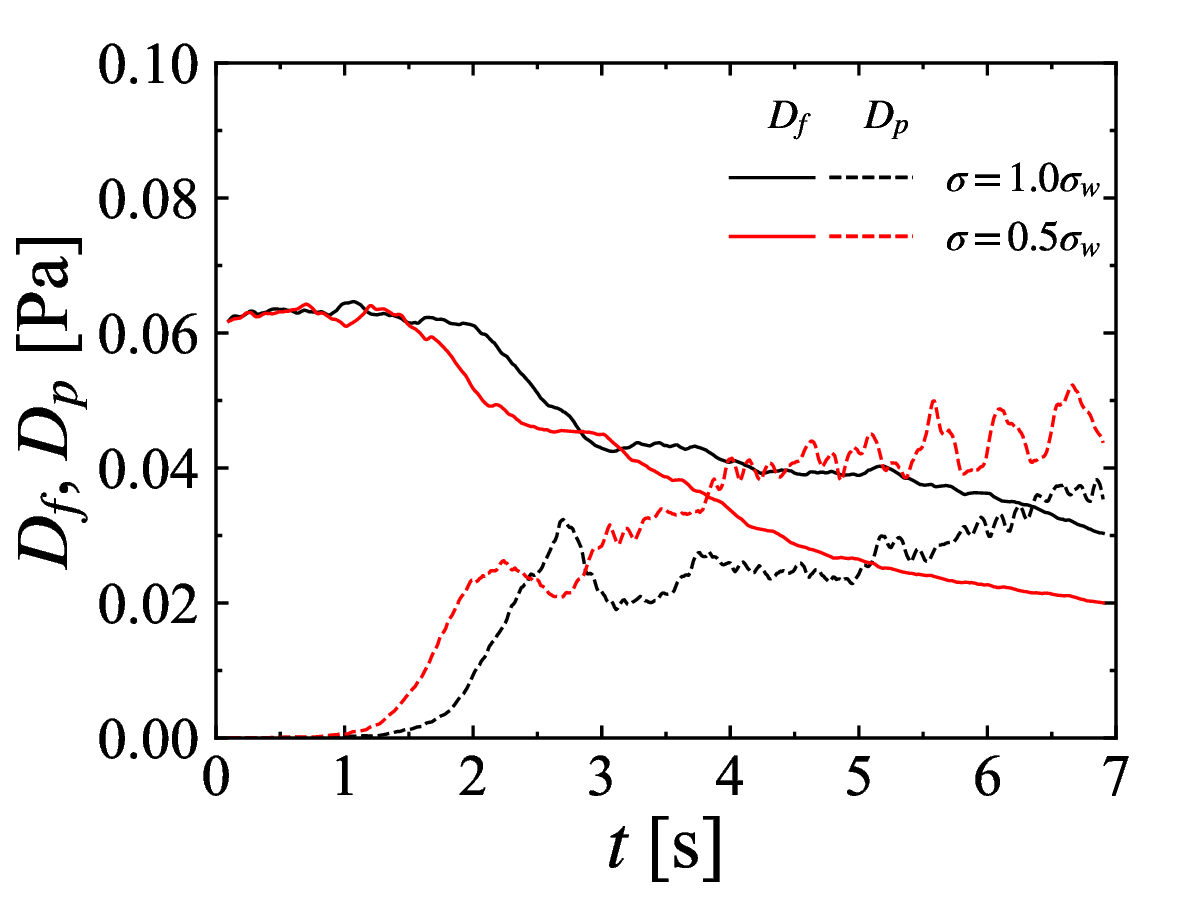}
  }
  \caption{Temporal variations of friction and form drags, $D_{\rm f}$ and $D_{\rm p}$. A simple centred moving average for the period of 0.2 s is applied.}
\label{fig:drag_force}
\end{figure}

The shear in the water side is produced by the energy flux $Q_{ta}$ as explained above. 
Thus, the smaller $Q_{ta}$ can result in the weaker shear-induced turbulence. 
The decrease of $Q_{ta}$ is attributed to the decrease of the friction drag.
Note that the friction and form drags satisfy the relationship of $D_f + D_p = \rho_a u_{*a}^2 \approx {\rm const.}$, 
where $D_f$ and $D_p$ are the friction and form drags, respectively. $D_p$ is given by 
(\ref{eq:Dp}), and $D_f$ is given by 
\begin{equation}
  D_f = 
  \frac{1}{L_1 L_2}
  \int_\Gamma \tau_{ta}t_1 {\rm d}S
\label{eq:Df} \, , 
\end{equation}
where $t_1$ is the streamwise components of the tangential vector to the interface.
Figure \ref{fig:drag_force} shows the friction and form drags, $D_f$ and $D_p$.
For $t \ge 4.0$ s, the form drag $D_p$ increases with time, and $D_p$ for the case of smaller surface tension is larger than that for the larger surface tension.
As discussed in the previous subsection, the increase in $D_p$ is attributed to the increase in the wave slope $H_s/L_s$ associated with the wave growth. 
In contrast, the friction drag $D_f$ decreases with time, and $D_f$ for the smaller surface tension case is smaller than that for the larger surface tension.
This suggests that the shear-induced water turbulence, which promotes the scalar transfer across the interface, can be suppressed by the decrease in $D_f$ due to the surface tension reduction. 

These results indicate that, when the surface tension is reduced, the wind wave becomes higher and the surface area becomes slightly larger. However, the scalar transfer across the air--water interface decreases because the turbulence under the interface is suppressed due to the decrease in the friction drag.

\section{Conclusions} 
Effects of uniform surface tension reduction on the wind-wave growth and the scalar transfer across the air--water interface have been investigated 
\MatsudaC{
for finite-amplitude and non-breaking gravity--capillary waves
}
using direct numerical simulation (DNS) of a wind-driven air--water two-phase turbulent flow. 
The incompressible Navier-Stokes equations for both air and water sides 
have been 
solved simultaneously using an arbitrary Lagrangian--Eulerian (ALE) method with boundary-fitted moving grids. 
\MatsudaC{Finite difference schemes were adopted for the computation of spatial derivatives.}
The periodic boundary condition was applied in the lateral directions, and the slip condition is applied to the top and bottom boundaries. 
Initially, the lower water side was static and the free surface was flat, 
while
a wall-bounded turbulent air flow developed for a flat and rigid surface 
was imposed in the upper air side. 
\MatsudaC{The wind speed 
was comparatively low (approximately 5 m/s) so that no intensive wave breaking occurs.}
\MatsudaC{The DNS was performed for the time integration period of 7 seconds to examine growth of gravity--capillary waves
with the wavelength of less than about 0.07 m}

Growth of wind waves was simulated for two values of the surface tension: the value for water and half of that. 
The results show that the significant waves for the smaller surface tension grow faster, and their wave height becomes higher than that for the larger surface tension.
\MatsudaC{
Although wind waves were not well captured during the first 1 second in the present DNS results 
owing to the initial impact of imposing the wall-bounded turbulent air flow
on the air--water interface,
the initial wave development is not critical for faster wave growth for the case of smaller surface tension.
}
\MatsudaC{
This is supported by the result that 
the wave growth speed is increased when the surface tension is suddenly reduced from the wind waves at $t = 4.0$ s developed with the large surface tension.
}
The effect of surface tension reduction on the wave height
has also been confirmed by the wave-height measurements in a small wind-wave tank (Appendix~\ref{appA}).

The growth of wind waves have been also analyzed by the wave height spectra.
For both surface tension cases, the spectra grow around the wavenumber for the weakly nonlinear Kelvin-Helmholtz instability \MatsudaA{for finite amplitude waves} in the early growth period. After that, the spectra broaden 
for both low wavenumbers (gravity waves) and high wavenumbers (capillary waves), having a peak on the low wavenumber side. 
The broadening of the wave height spectra is consistent with the harmonic resonance, which forms ripple-like capillary waves. 
The spectrum also shows that, when the surface tension is reduced, 
a bump at the resonant capillary wave scale moves to higher wavenumber and becomes smaller, indicating that 
the resonant capillary waves become 
shorter and weaker.

\MatsudaC{
The numerical convergence of the DNS results have been discussed in Appendix~\ref{appNC}. 
At the present grid resolution, the temporal variations of the significant wave height and wavelength and the wave height spectrum are weakly dependent on the resolution. 
However, the resolution dependence is 
smaller than the difference due to the surface tension reduction. Therefore, the DNS results under the present computational condition are 
enough to discuss the effects of surface tension reduction.
}

The temporal variations of wave energy show that the potential energies due to gravity and surface tension are comparable during the early growth period.
After this period, the potential energy due to gravity increases rapidly, whereas that due to surface tension remains almost constant.
The surface tension reduction results in the increase 
in 
the potential energy due to gravity and the decrease 
in 
that due to surface tension. 
The results also show that the growth of the potential energy due to gravity is faster for the smaller surface tension.

The energy flux from air to water has been also investigated to understand the faster growth.
The energy flux due to the normal stress, which contributes to wave growth, becomes larger due to the surface tension reduction.
The surface tension reduction also causes the decrease in the energy dissipation in the water side.  The 
relationship 
between the energy flux 
and the temporal wave-energy change shows that the rapid increase of the wave energy is attributed to 
the balance between energy flux from the air flow and energy dissipation in the water. 
The energy flux and energy dissipation filtered for gravity and capillary wave scales show that the energy flux from the air flow is received mostly at the gravity wave scales, and the energy dissipation in the water occurs mostly 
at 
the capillary wave scales. This indicates that there is an energy transfer mechanism from the gravity wave scales to the capillary wave scales.
The energy dissipation for the capillary wave scales becomes smaller due to the surface tension reduction. This means the energy transfer from the gravity wave scales to the capillary wave scales decreases due to the surface tension reduction. 
The analyzed results also show that the energy flux due to the tangential stress, i.e., the energy source of shear-induced turbulence, is smaller than the energy dissipation. 
This suggests that the wave energy due to the normal stress at the gravity wave scales is partially transferred to smaller scales by resonant wave interactions and dissipated at the capillary wave scales. 
In conclusion, the significant gravity waves receive kinetic energy from the air side with contributions of form drag, and the energy is partially transferred to the capillary waves and dissipated by the viscosity of water. 
When the surface tension decreases, the energy transfer to the capillary wave scales decreases, and the significant waves retain more energy,   resulting in the faster growth of the wave height. 

The scalar transfer across the air--water interface has been further discussed based on the DNS results. 
The scalar transfer coefficient 
(scalar transfer velocity) 
for the gas absorption decreases due to the surface tension reduction. 
The decrease in scalar transfer coefficient 
cannot be explained by the difference of the interface surface area expansion because 
influence of 
the surface tension reduction 
on  
the surface area 
is negligibly small. 
Since the turbulent scalar flux and the 
\MatsudaB{vorticity fluctuation} in the water side becomes smaller due to the surface tension reduction, the decrease in the transfer coefficient is caused by the suppression of turbulence under the interface, particularly turbulence associated with streamwise vortices. 
The suppression of turbulence can be attributed to the decrease of the friction drag, which drives the shear layer beneath the interface.

\bigskip

%
This work was supported by JSPS KAKENHI Grant Number JP19K21937.
SK would like to thank the partial 
support by JST COI Grant Number JPMJCE1316. 
The authors would like to thank A. Kimura, K. Takane, and R. Hayashi for their help in conducting the experiments.
The numerical simulations presented in this paper were carried out on the Earth Simulator supercomputer system in the Japan Agency for Marine-Earth Science and Technology. 

\bigskip

\noindent
{\bf Declaration of interests.} The authors report no conflict of interest.

\appendix
\section{Comparison with wind-wave tank experiments}\label{appA}

\begin{figure}
  \centerline{
    \includegraphics[width=12cm]{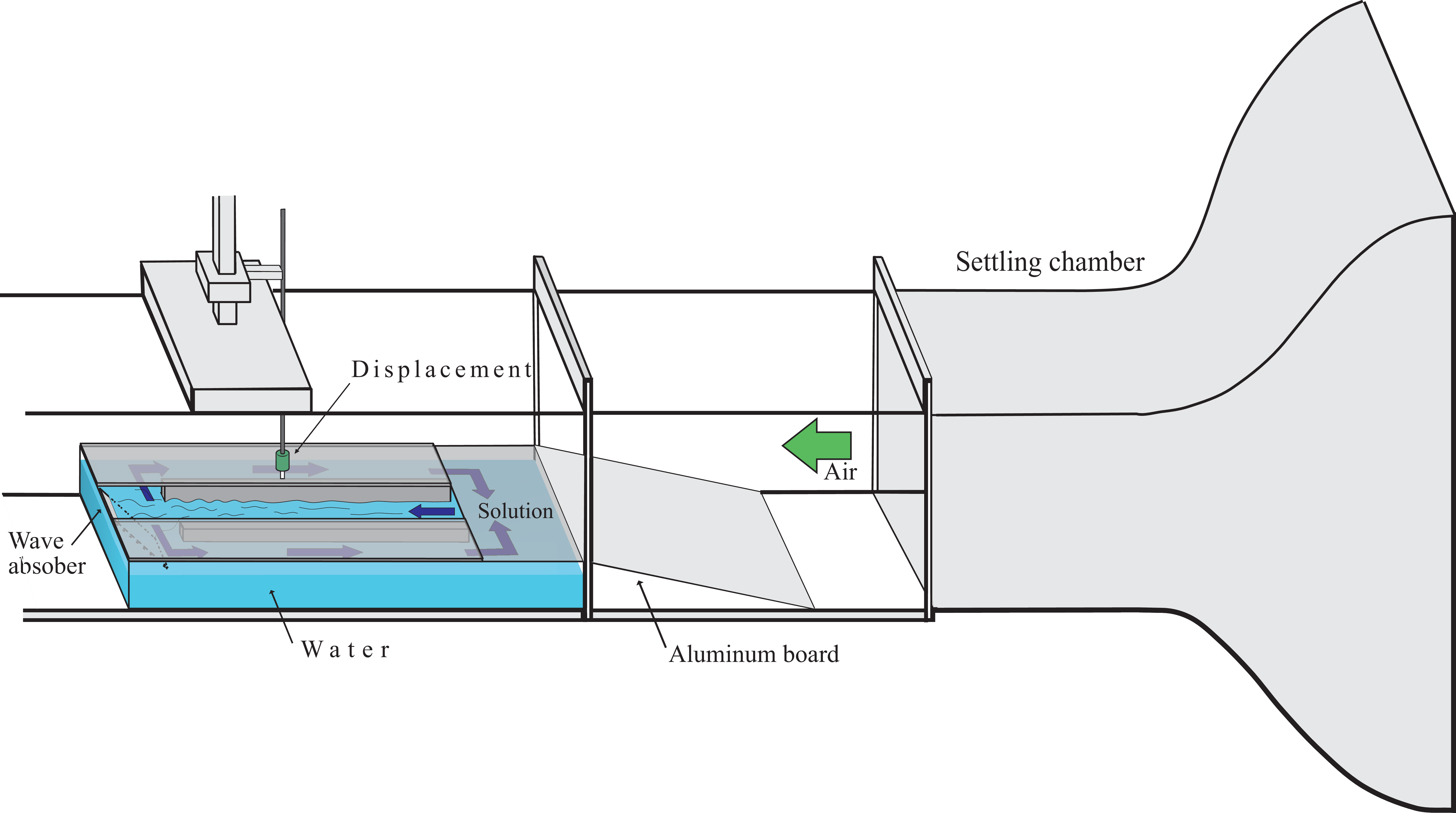}
  }
  \caption{Schematic diagram of the experimental setup: A small wind-wave tank placed in the wind tunnel and the water-level fluctuation measurement system.}
\label{fig:wind_tunnel}
\end{figure}

Laboratory experiments have been conducted to 
examine 
the DNS results, in which the wave height becomes higher due to the uniform surface tension reduction.
%
Figure \ref{fig:wind_tunnel} shows the schematic diagram of the experimental apparatus. 
The experiments were conducted using a small recirculation wind-wave tank (length 1.05 m, width 0.20 m and depth 0.05 m), located in a wind tunnel \citep{Nagata(2007)} (length 5.0 m, width 0.30 m and height 0.30 m).
A straightened air flow was introduced over the wind-wave tank. 
The water surface was open to the air flow only at the central test section (length 0.90 m and width 0.09 m) of the tank.
The surface water flow induced by the wind shear recirculated through the covered side sections so that a counter flow was avoided in the lower part of the central test section.
A wave absorber was fixed at the end of the test section to prevent the reflection of wind waves and counter flow.
The water-level fluctuation was measured using a capacitance-type displacement meter (Iwatsu Electric Co. Ltd., ST-3572) capable of non-contact measurement. The probe (Iwatsu Electric Co. Ltd., ST-0717A) was set at a fetch length of $x=0.70$ m and at an elevation of 0.0020 m or 0.0030 m above the static water level. 
The sampling time and frequency were 300 s and 1000 Hz respectively.
The measurements were conducted for five cases of the free-stream wind speed $U_\infty$ between 1 and 6 m/s.
In order to quantify the wave height, significant waves were identified using the zero-up 
crossing 
method.

Liquids used in the experiments were filtered tap water and two aqueous solutions of ethanol and glycerine. Table~\ref{tab:solutions} shows the surface tension 
$\sigma$ 
and the viscosity 
$\mu$ 
of each solution. The surface tension of the 30 wt\% ethanol solution was approximately half of the filtered tap water. However, the viscosity of the aqueous ethanol solution was about twice to the filtered tap water. 
Thus, the concentration of the aqueous glycerine solution was adjusted to approximately 30 wt\% so that the viscosity was close to the aqueous ethanol solution while the surface tension remained close to the filtered tap water.

\begin{table}
  \begin{center}
\def~{\hphantom{0}}
  \begin{tabular}{lcc}
    Solution              & $\sigma$ ($10^{-3}$ N/m) & $\mu$ ($10^{-3}$ Pa s) \\[3pt]
    Filtered tap water    & 72.0                     & 0.95 \\
    Ethanol solution      & 36.3                     & 2.12 \\
    Glycerine solution    & 69.0                     & 2.41 \\
  \end{tabular}
  \caption{Measured values of surface tension $\sigma$ and dynamic viscosity $\mu$.}
  \label{tab:solutions}
  \end{center}
\end{table}

\begin{figure}
  \centerline{
    \includegraphics[width=9cm]{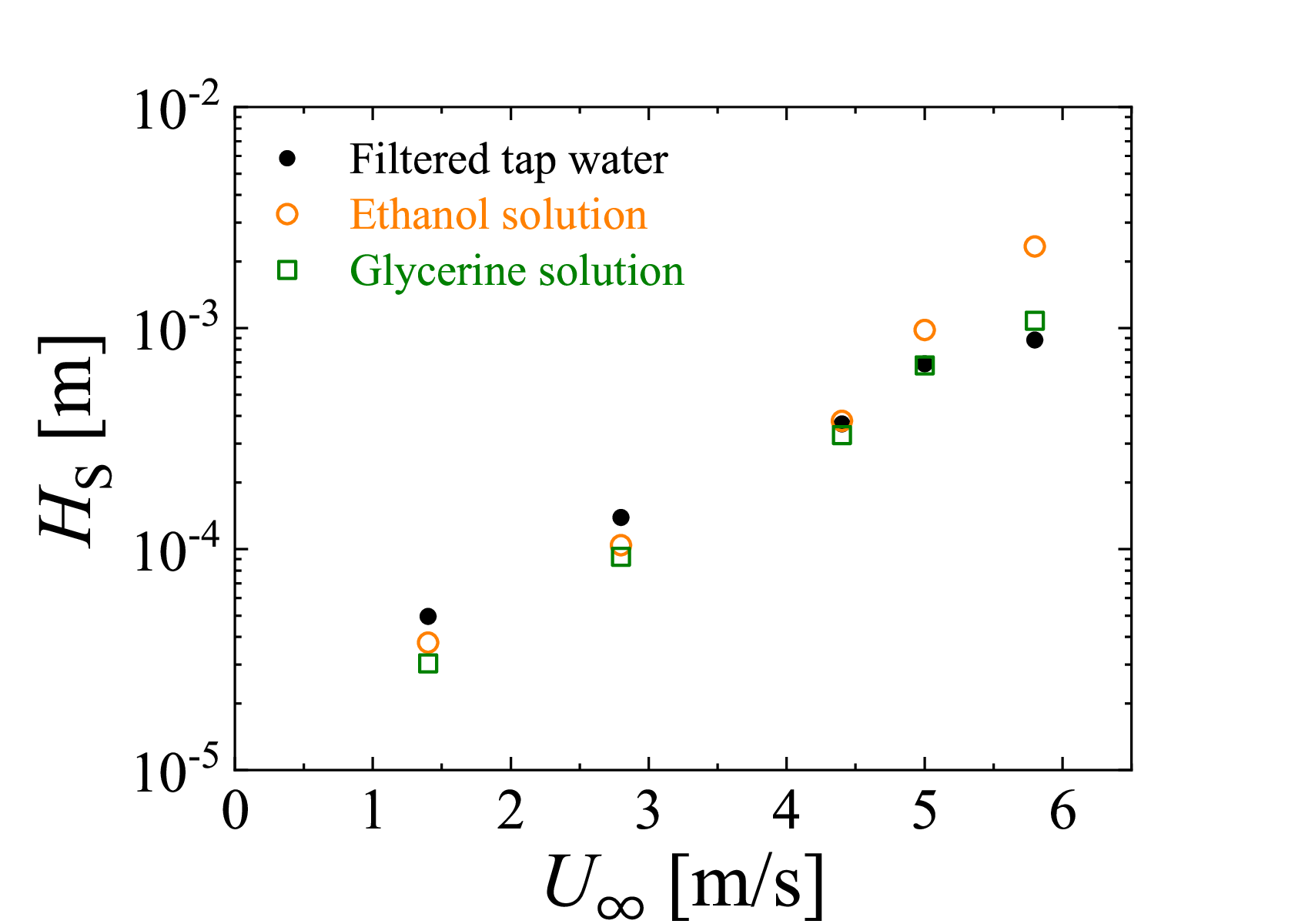}
  }
  \caption{Significant wave height $H_s$ 
  against
  free-stream wind speed $U_\infty$. }
\label{fig:Hs_exp}
\end{figure}

Figure~\ref{fig:Hs_exp} shows the measurements of the signiﬁcant wave height, $H_s$, with the free-stream wind speed, $U_{\infty}$, for the filtered tap water and two aqueous solutions of ethanol and glycerine. When the wind speed $U_{\infty}$ is higher than 5 m/s, the significant wave height of the aqueous ethanol solution is larger than that of aqueous glycerine solution. The difference in the significant wave height is due to the difference of surface tension and not to the difference in viscosity because the significant wave height for the aqueous glycerine solution is almost the same as that of the filtered tap water with the half viscosity at $U_{\infty} > 5$ m/s.
\Matsuda{
\cite{Paquier(2015),Paquier(2016)} investigated the effect of viscosity on the wind waves experimentally and reported the presence of "the wrinkle regime" under large viscosity or weak wind conditions and the transitions to the regular wave regime under small viscosity or strong wind speed conditions.
For $\nu \approx 1.0$--$2.0\times 10^{-6}$ m$^2$/s, the winkle regime should be observed for approximately $u_{*a} \lesssim 0.19$--0.21 or $U_{\infty} \lesssim 3.6$--4.0 m/s.
Thus, the difference in the effect of surface tension reduction between low and high wind speed could be due to the difference in the wave regimes. 
Note that, apparently from the visualization in figure \ref{fig:DNS_interface}, the wind waves obtained by the DNS are in the regular wave regime.
}

\begin{figure}
  \centerline{
    \includegraphics[width=8cm]{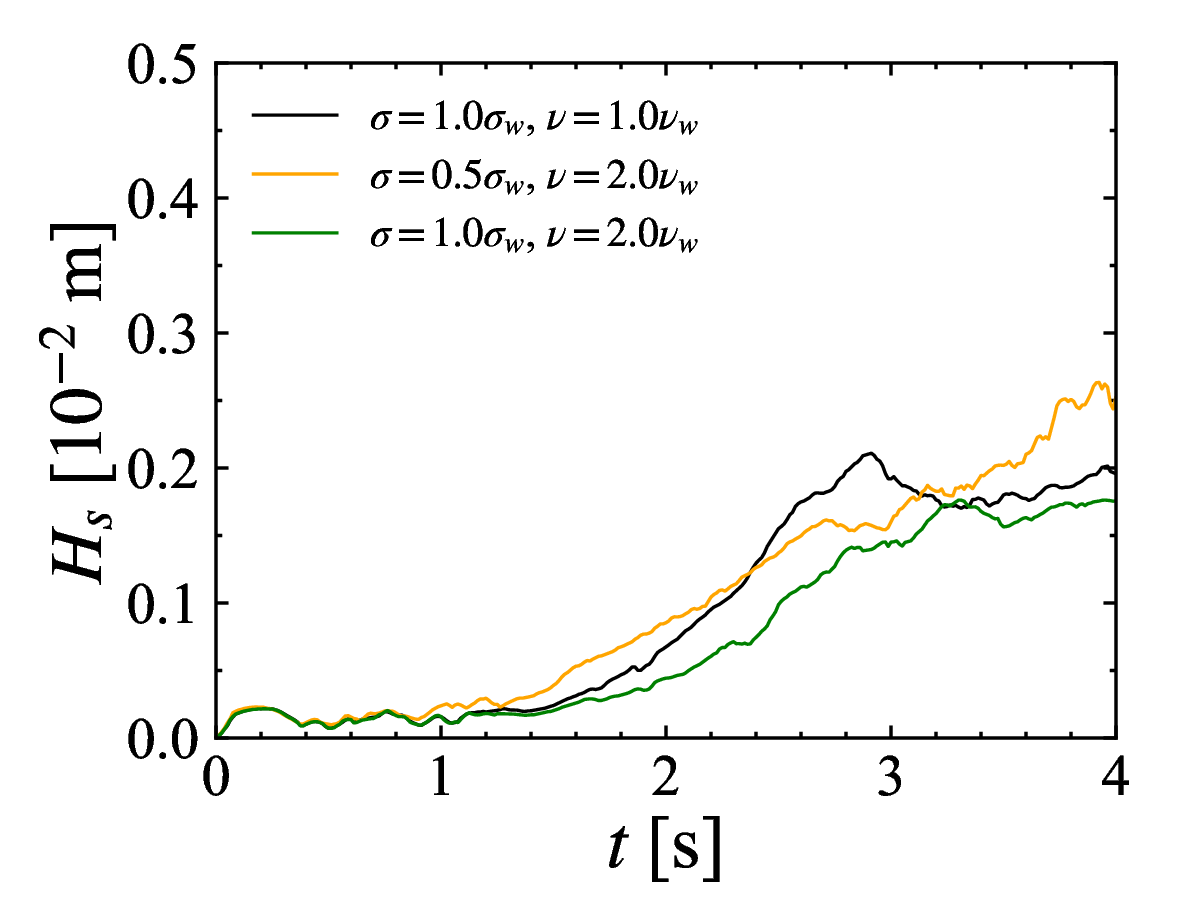}
  }
  \caption{
  Temporal variations of significant wave height $H_s$ obtained by DNS for the filtered tap water ($\sigma = 1.0\sigma_w$, $\nu = 1.0\nu_w$), the aqueous ethanol solution (
  \Matsuda{$\sigma = 0.5\sigma_w$}, $\nu = 2.0\nu_w$), and the aqueous glycerine solution ($\sigma = 1.0\sigma_w$, $\nu = 2.0\nu_w$).}
\label{fig:A3}
\end{figure}

The experimental condition 
for 
the filtered tap water for $U_{\infty} \approx 5$ m/s approximately corresponds to the DNS condition of $\sigma = 1.0\sigma_w$, where the viscosity was set to $\nu = \nu_w$ and the free-stream wind speed was $U_{\infty} \approx 5.2$ m/s. 
Here, we have performed the DNS for additional cases for the aqueous ethanol and glycerine solutions. 
For the case of the aqueous ethanol solution, the surface tension and  the kinematic viscosity were set to $\sigma = 0.5\sigma_w$ and $\nu = 2.0\nu_w$, respectively, and for the aqueous glycerine solution, $\sigma = 1.0\sigma_w$ and $\nu = 2.0\nu_w$, respectively. 
Other computational conditions were the same as those in Subsection \ref{sec:DNScond}. 
Figure~\ref{fig:A3} shows the temporal variations of the significant wave height, $H_s$, for the cases of the filtered tap water ($\sigma = 1.0\sigma_w$, $\nu = 1.0\nu_w$), the aqueous ethanol solution (
\Matsuda{$\sigma = 0.5\sigma_w$}, $\nu = 2.0\nu_w$), and the aqueous glycerine solution ($\sigma = 1.0\sigma_w$, $\nu = 2.0\nu_w$). 
To compare the DNS 
results 
with the 
laboratory  
measurements, we estimate the wave growth time $t$ in the DNS that corresponds to the fetch of $x = 0.7$ m in the wind wave tank. According to the fetch law \citep[e.g.,][]{Wilsonn(1965)}, the fetch is formulated as the function of a wave velocity $\mathcal{C}_s$, 
which 
is defined as 
$\mathcal{C}_s \equiv \sqrt{gL_s/2\pi + 2\pi\gamma/L_s}$ for deep-water gravity-capillary waves with a wavelength of $L_s$. 
This suggests that the significant wave velocity $\mathcal{C}_s$ or wavelength $L_s$ is a suitable parameter for comparison between the DNS 
results 
and 
laboratory 
measurements. 
At the fetch of $x=0.7$ m, the measured values of the wave velocity and the wavelength are $\mathcal{C}_s=0.29$ m/s and $L_s=0.033$ m, respectively. 
In figure~\ref{fig:signif_wave}(\textit{b}), the significant wavelength obtained by the DNS is close to the measured value for 3 s $\lesssim t \lesssim$ 4 s. We have also confirmed that the significant wave velocity averaged over the period of 3 s $\le t \le$ 4 s is \MatsudaB{$\mathcal{C}_{s} = 0.279$} m/s and close to the laboratory measurement.
$\mathcal{C}_s$ for the half surface tension case can be smaller than the filtered tap water case but its effect on $\mathcal{C}_s$ is smaller than 6 \% for the measured wavelength. 
In the growth stage of 
3 s $\lesssim t \lesssim$ 4 s, 
the DNS in figure~\ref{fig:A3} well supports the experimental results that the homogeneous surface tension reduction enhances the wave height 
even when the viscosity is doubled, 
and the viscosity effect 
on the wave growth 
is 
smaller than the surface tension effect. 
It is also observed that the wave heights in the DNS results are higher than the measured values. This could be due to the difference in the friction velocity, which was not measured in the experiments. Thus, we do not compare the DNS results with the laboratory measurements quantitatively. 
It 
would be 
also interesting to note that the viscosity effects of surface tension reduction on wind-wave growth. However, we omit the detailed discussion about the sensitivity to the viscosity because it is beyond the scope of this paper.

\Matsuda{
\section{Interface curvature}\label{appB}
The interface curvature $\kappa$ is defined as the divergence of the 
normal vector on the interface, $\kappa = - \nabla_\Gamma \cdot {\bm n}$, where $\bm n$ is the unit normal vector to the interface, and $\nabla_\Gamma$ denotes the divergence operator in the interface. To compute the curvature $\kappa$, we consider the generalized coordinate $(\xi_1,\xi_2,\xi_3)$, instead of the Cartesian coordinate $(x_1,x_2,x_3)$.
In the generalized coordinate, the interface $\Gamma$ is defined as the surface where $\xi_3$ equals to a constant value $\xi_{\Gamma}$, and the curvature $\kappa$ is given by the derivative of $\bm n$ in $\xi_1$ and $\xi_2$ directions:
\begin{equation}
    \kappa = -\left[ \frac{{\bm a}_2\times{\bm n}}{|{\bm a}_1\times{\bm a}_2|} \frac{\partial {\bm n}}{\partial \xi_1} 
                   + \frac{{\bm n}\times{\bm a}_1}{|{\bm a}_1\times{\bm a}_2|} \frac{\partial {\bm n}}{\partial \xi_2}\right]
\end{equation}
where ${\bm a}_i \equiv \frac{\partial x_j}{\partial \xi_i} {\bm e}_j$, and ${\bm e}_j$ is the basis vector of the Cartesian coordinate. The normal vector is given by ${\bm n}=\frac{{\bm a}_1\times{\bm a}_2}{|{\bm a}_1\times{\bm a}_2|}$.
The relationship of ${\bm n}\cdot{\bm a}_1={\bm n}\cdot{\bm a}_2=0$ yields 
\begin{equation}
    \kappa = \frac{h_{11}g_{22} + h_{22}g_{11} - 2h_{12}g_{12}}{g_{11}g_{22}-g_{12}g_{12}}
\end{equation}
where $h_{ij}\equiv{\bm n}\cdot\frac{\partial {\bm a}_j}{\partial \xi_i}$ and $g_{ij}\equiv{\bm a}_i\cdot{\bm a}_j$.
}

\Matsuda{
\section{Numerical convergence}\label{appNC}
The numerical convergence of the DNS results have been examined by changing the streamwise resolution. 
The DNS results obtained for $M_1=400$, where $M_1$ is the number of grid points, were compared with the additional DNS results for $M_1=200$, 300, and 532. The grid spacing was scaled for the fixed domain size. 
Figure \ref{fig:wall_HRx} shows the vertical profiles of the mean streamwise air velocity $U_1$ and the root-mean-square of the streamwise air velocity fluctuations $u'$ of the initial wall-bounded turbulent flows obtained for different resolutions.
The vertical axis is normalized by $u_{*a}$, i.e., $U_1^+ = U_1/u_{*a}$ and $u'^+ = u'/u_{*a}$.
The horizontal axis is the vertical position in the viscous unit $z^+ = z u_{*a}/\nu_a$.
As shown in figure \ref{fig:wall_HRx}, for both $U_1^+$ and $u'^+$, the profiles for different resolutions are well converged. 
\begin{figure}
  \centerline{
    (\textit{a}) \includegraphics[width=6cm]{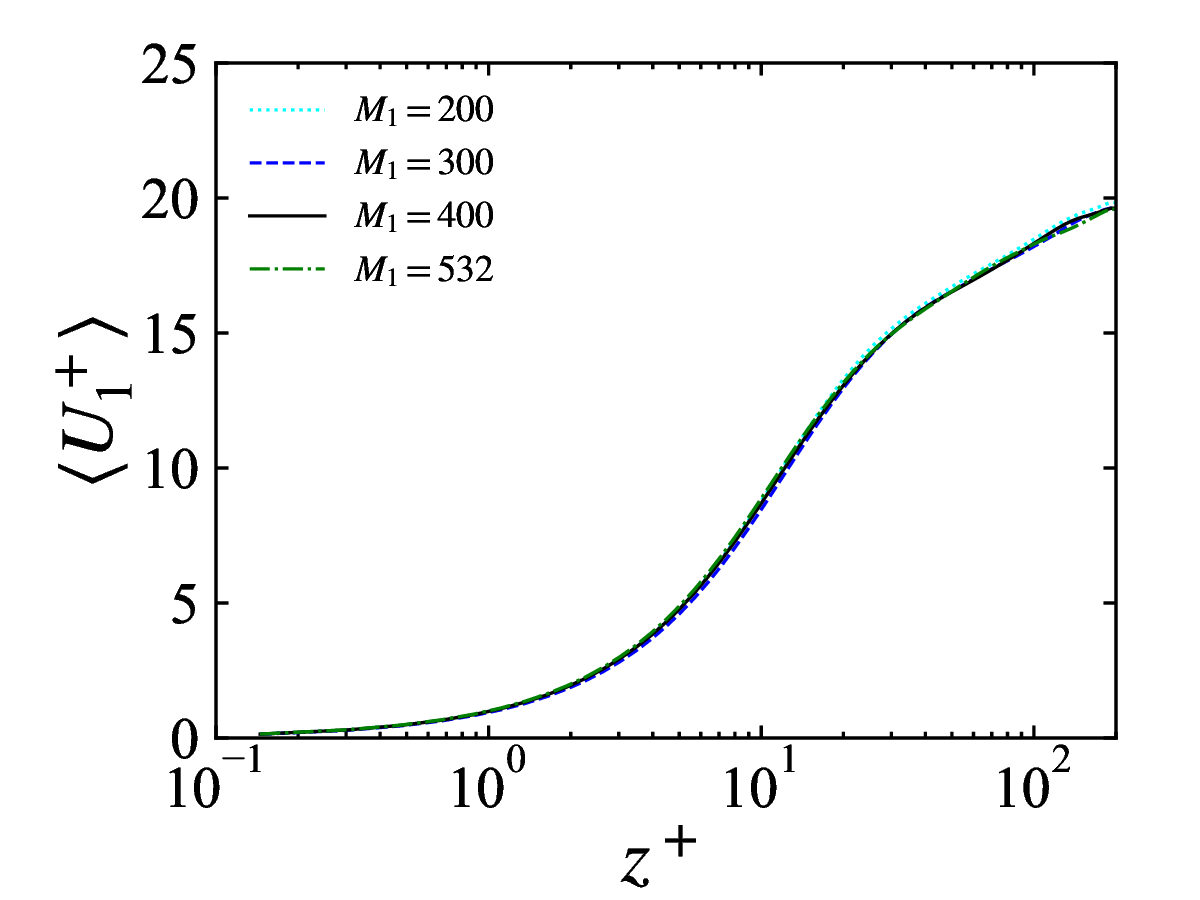}
    (\textit{b}) \includegraphics[width=6cm]{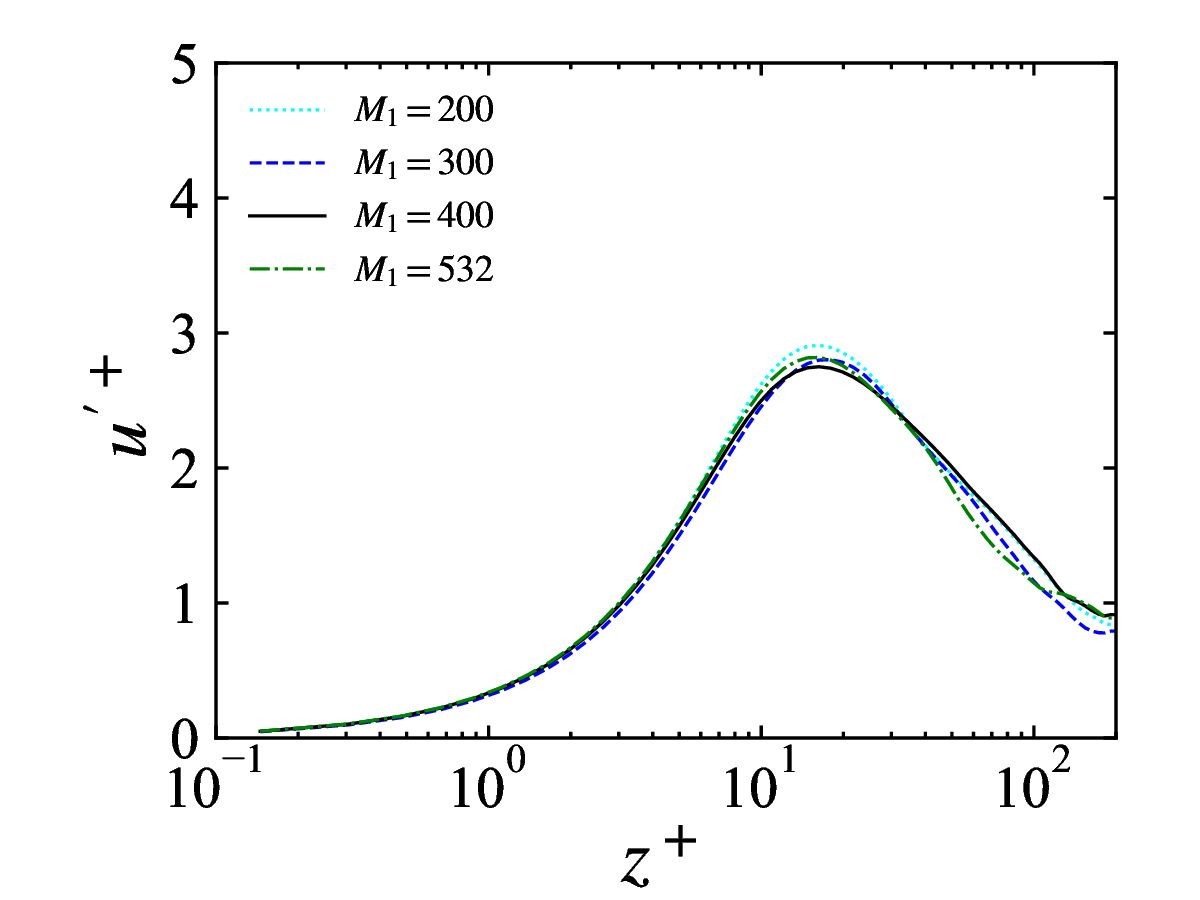}  }
  \caption{Vertical profile of (\textit{a}) mean streamwise air velocity, $U_1^+$, and (\textit{b}) streamwise air velocity fluctuation, $u'^+$, of the initial wall-bounded turbulent flows obtained with different resolutions.}
\label{fig:wall_HRx}
\end{figure}
Figure \ref{fig:signif_wave_HRx} shows the temporal variations of significant wave height and wavelength, $H_s$ and $L_s$, respectively, for different resolutions.
The growth of $H_s$ for $M_1=200$ is slower than the growth for higher resolution particularly for $t<3$ s, but the wave growth for the other resolutions are well converged. 
For $M_1\ge300$, the difference in $H_s$ between the different resolutions increases gradually along with time. 
The small difference between the different resolution cases for $M_1\ge300$ is smaller than the difference due to the surface tension reduction. 
The convergence is similarly confirmed for temporal variation of $L_s$.
\begin{figure}
  \centerline{
    (\textit{a}) \includegraphics[width=6cm]{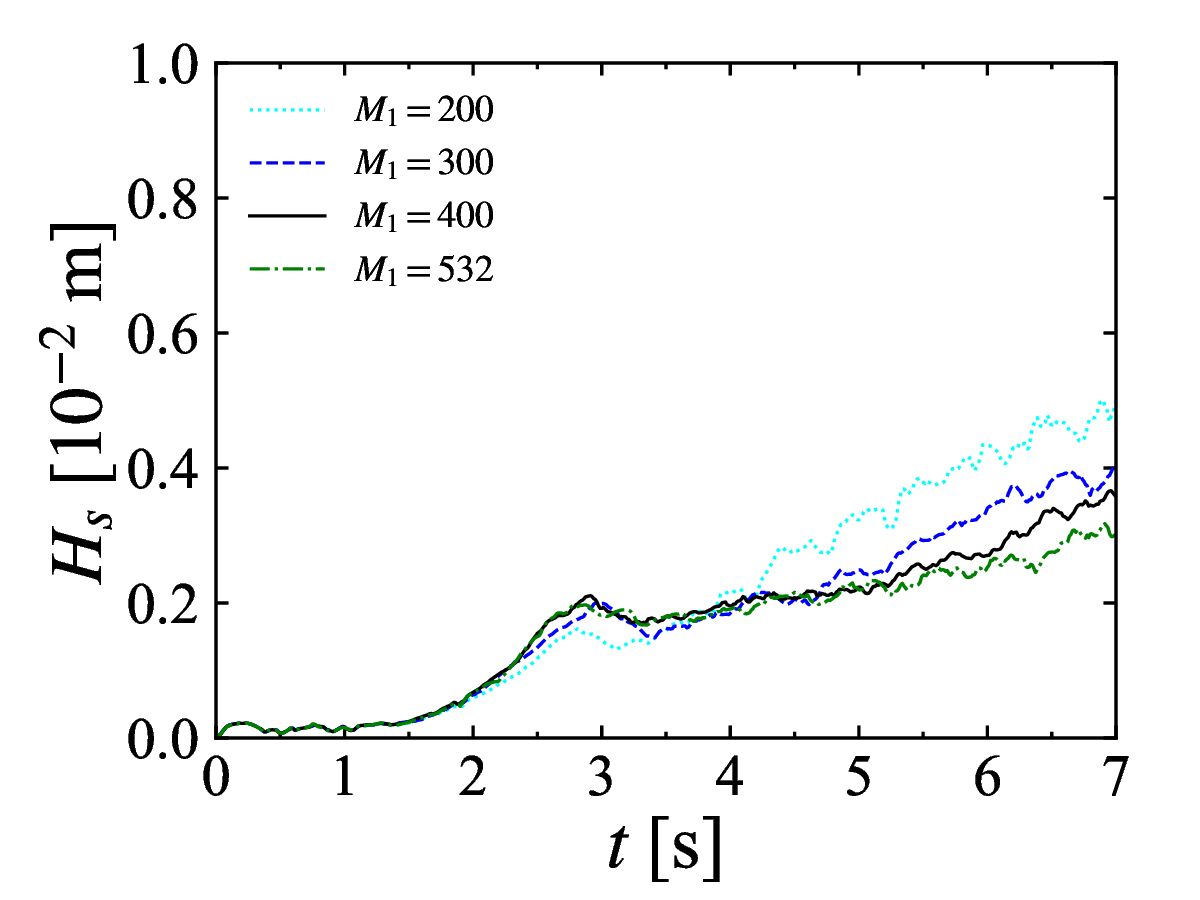}
    (\textit{b}) \includegraphics[width=6cm]{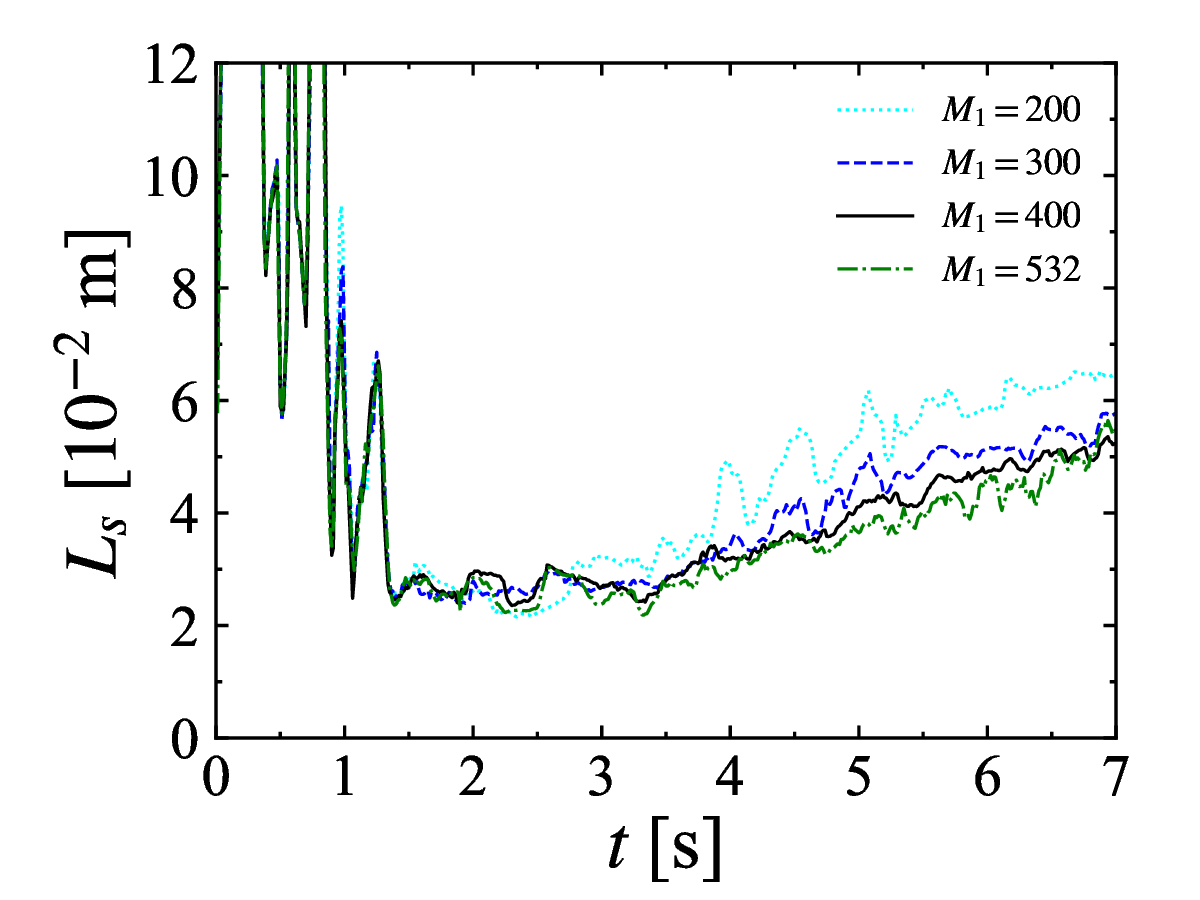}
  }
  \caption{Temporal variations of (\textit{a}) significant wave height, $H_s$, and (\textit{b}) significant wavelength, $L_s$, obtained for $\sigma=1.0\sigma_w$ with different resolutions.}
\label{fig:signif_wave_HRx}
\end{figure}
}
\MatsudaC{
Figure~\ref{fig:wave_spectrum_HRx} shows the time-averaged wave height spectra for the cases of $\sigma=1.0\sigma_w$ and $\sigma=0.5\sigma_w$ obtained for different resolutions. As expected from the temporal variations of the significant wave height and wavelength, the wave spectra for low wavenumbers ($k_1<k_m$) are well converged. For high wavenumbers ($k_1>k_m$), the wave spectra are almost converged. 
We can still see weak dependence on the resolution but the resolution dependence is sufficiently smaller than the difference due to the surface tension reduction. 
Thus, 
the present DNS is sufficiently converged for all over the scales to discuss the surface tension effect.
}

\begin{figure}
  \centerline{ 
    \includegraphics[width=8cm]{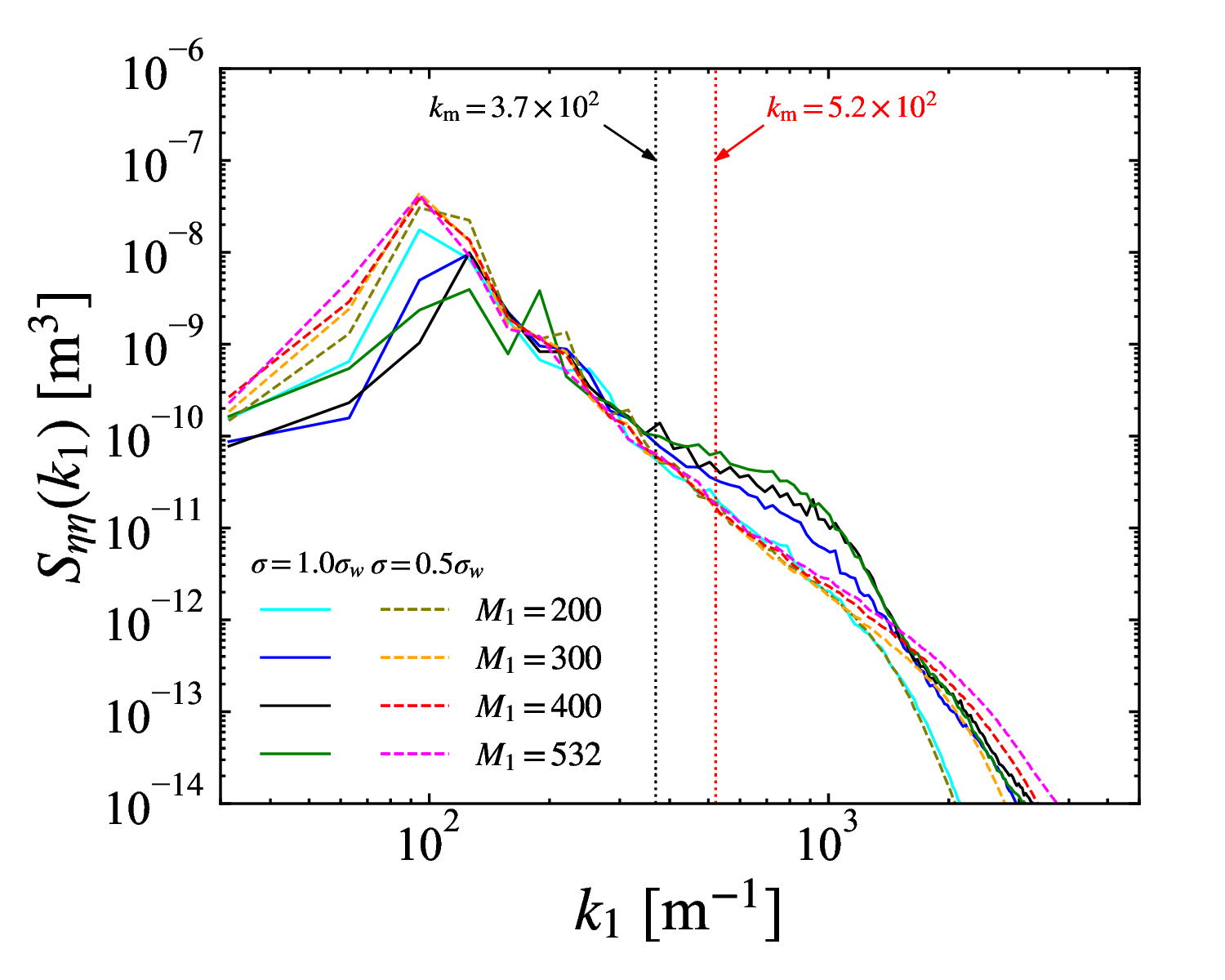} 
  }
  \caption{
  \MatsudaC{
  Wave height spectra averaged for the period of $4.0<t\le7.0$ s, obtained for $\sigma=1.0\sigma_w$ and $\sigma=0.5\sigma_w$ with different resolutions. 
  Vertical dotted lines in black and red are the wavenumbers $k_m$ for $\sigma=1.0\sigma_w$ and $\sigma=0.5\sigma_w$, respectively.
  }
  }
\label{fig:wave_spectrum_HRx}
\end{figure}

\Matsuda{
\section{Wave growth rate}\label{appC}
The wave growth was often evaluated by using the wave growth rate \citep{Donelan(2006)}.
Here, the wave growth rate in our DNS is compared with the measurements in the literature \citep{Plant(1982)}. 
We use the wave growth rate $\beta$ defined as 
\begin{equation}
\beta \equiv \frac{1}{E}\frac{{\rm d}E}{{\rm d}t}, 
\end{equation}
where $E$ is the total wave energy. As described in Section 3.1, it is difficult to extract the kinetic energy of the wave motion from the total kinetic energy including the turbulent kinetic energy. Here, we assume again that the the kinetic energy of the wave motion is equivalent to the potential energy, and the wave growth rate is obtained by $\beta=\frac{{\rm d}\log(E_g+E_s)}{{\rm d}t}$. 
Figure \ref{fig:beta_Plant} shows $\beta$ calculated by fitting linear function to $\log(E_g+E_s)$ for every 0.5 s.  
The gray symbols are 
the measurements summarized in figure 2 of \cite{Plant(1982)}. The solid and dashed lines are the relation obtained by \cite{Plant(1982)}, and given by: 
\begin{equation}
    \beta = \frac{(0.04\pm0.02)u_{*a}^2\omega \cos \theta}{{\cal C}^2}
    \label{eq:beta_Plant},
\end{equation}
where $u_{*a}$ is the friction velocity on the air side, $\omega=2\pi f$ is the radian wave frequency, $\theta$ is the angle between wind and wave directions, and $\cal C$ is the phase speed.
As shown in figure \ref{fig:beta_Plant}(\textit{a}), the relationship between $\beta/f$ and $u_{*a}/{\cal C}$ agrees with the 
measurements.
Note that $\rho_a u_{*a}^2$ is the sum of form and friction drags, $D_p$ and $D_f$, respectively, while $u_{*a}$ in (\ref{eq:beta_Plant}) is relevant to the contribution of the form drag $D_p$ \citep{Melville&Fedorov(2015)}.
Since the energy flux due to the normal stress approximately satisfies $Q_{na}\approx {\cal C}D_p$, the wave growth rate $\beta/f$ should be compared with $Q_{na}/\rho_a{\cal C}^3$. 
In figure \ref{fig:beta_Plant}(\textit{b}), the wave growth rate obtained from our DNS results is plotted against $Q_{na}/\rho_a{\cal C}^3$, 
and the correlation with $Q_{na}/\rho_a{\cal C}^3$ is better than that with $(u_{*a}/{\cal C})^2$. 
\begin{figure}
  \centerline{
    (\textit{a}) \includegraphics[width=6cm]{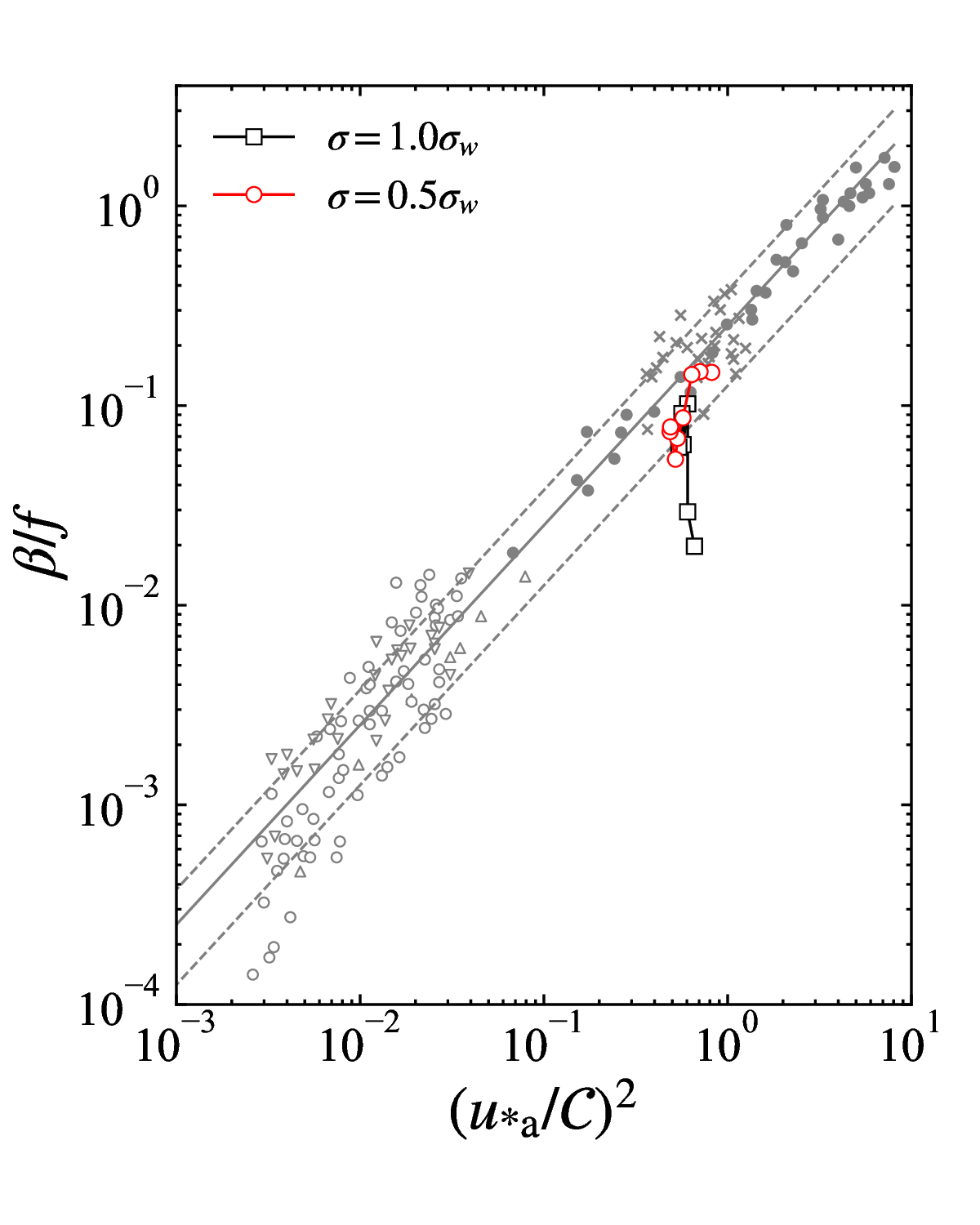}
    (\textit{b}) \includegraphics[width=6cm]{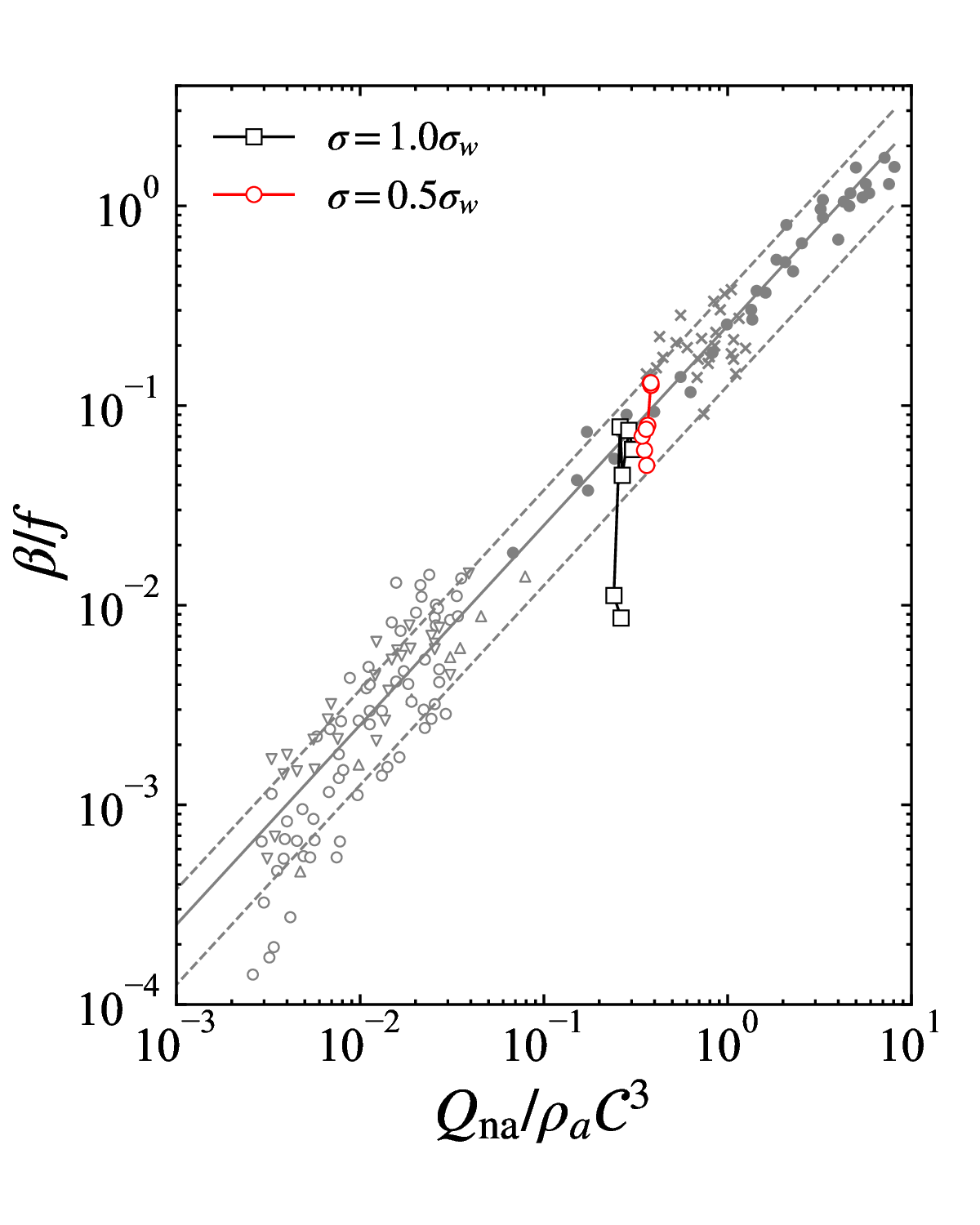} \\
  }
  \caption{Wave growth rate $\beta$ normalized by the wave frequency $f$ against (\textit{a}) $(u_{*{\rm a}}/{\cal C})^2$ and (\textit{b}) $Q_{\rm na}/\rho_a{\cal C}^3$. 
  $u_{*a}^2$ and $Q_{na}$ are temporally averaged for every 0.5 s after $t=4.0$ s for $\sigma=1.0\sigma_w$ (black open squares) and $t=3.0$ s for $\sigma=0.5\sigma_w$ (red open circles). $\beta$ is obtained by the linear fitting to $\log(E_g+E_s)$ for every 0.5 s in the corresponding period.
  The other symbols are the measurements summarized in figure 2 of \cite{Plant(1982)}. The solid and dashed lines represent (\ref{eq:beta_Plant}).}
\label{fig:beta_Plant}
\end{figure}
}

\Matsuda{
\section{Scalar budget equation}\label{appD}
To evaluate the scalar transfer below the moving interface, we consider the continuous surface $\Gamma_{z_{g0}}$.
Therefore, we consider the coordinate transform from Cartesian coordinate $(x_1,x_2,x_3)$ to the boundary fitted coordinate $(\Xi_1,\Xi_2,\Xi_3)$. 
The all computational grids are fixed on $(\Xi_1,\Xi_2,\Xi_3)$, and $(\Xi_1,\Xi_2,\Xi_3)$ is equivalent to $(x_1,x_2,x_3)$ at $t=0$.
In the boundary fitted coordinate, the surface $\Gamma_{z_{g0}}$ is represented by $\Xi_3=z_{g0}$.
The scalar transport equation in the boundary fitted coordinate can be described as 
\begin{equation}
\frac{\partial C|{\bm J}|}{\partial t} 
+ \frac{\partial C(U_j-V_j)K_{jk}}{\partial \Xi_k}
= \frac{\partial}{\partial \Xi_k} \left( D \frac{\partial C}{\partial x_j} K_{jk} \right)
,
\end{equation}
where ${\bm J}$ is the Jacobian matrix (${\bm J}=\frac{\partial (x_1,x_2,x_3)}{\partial (\Xi_1,\Xi_2,\Xi_3)}$),
$|{\bm J}|$ is the determinant of ${\bm J}$, $K_{ij}$ is $(i,j)$ component of the ajugate matrix of ${\bm J}$, 
and $V_i$ is $i$th component of the velocity of grid points on the boundary-fitted coordinate.
Since the boundary-fitted coordinate deforms only in $x_3$ direction, the components of Jacobian matrix are $J_{11}=J_{22}=1$, $J_{13}=\frac{\partial z_g}{\partial x_1}$, $J_{23}=\frac{\partial z_g}{\partial x_2}$, $J_{33}=\frac{\partial z_g}{\partial z_{g0}}$, and $J_{12}=J_{21}=J_{31}=J_{32}=0$. 
Thus, the determinant becomes $|{\bm J}|=|J_{33}|$, and $K_{ij}$ is given by $K_{11}=K_{22}=J_{33}$, $K_{13}=-J_{13}$, $K_{23}=-J_{23}$, $K_{33}=1$, and $K_{12}=K_{21}=K_{31}=K_{32}=0$.
The components of the velocity of the boundary-fitted coordinate are $V_1=V_2=0$ and $V_3=\frac{\partial z_g}{\partial t}$.
Applying the integral over the volume for $0 \le \Xi_1 \le L_1$, $0 \le \Xi_2 \le L_2$, and $-2\delta \le \Xi_3 \le z_{g0}$ in the boundary fitted coordinate, 
and using the Gauss' divergence theorem for the advection and diffusion terms, we obtain
\begin{equation}
\frac{{\rm d} \chi(z_{g0})}{{\rm d} t} 
= \left\langle D \frac{\partial C}{\partial x_j} K_{j3}  \right\rangle_{z_{g0}}
- \left\langle C(U_j-V_j)K_{j3} \right\rangle_{z_{g0}}
,
\end{equation}
where $\chi(z_{g0})$ is given by (\ref{eq:chi}).
In the boundary-fitted coordinate, $\chi(z_{g0})$ and the bracket $\langle\cdot\rangle_{z_{g0}}$ are described as $\chi(z_{g0}) = \frac{1}{L_1L_2}\int_0^{L_1}\int_0^{L_2}\int_{-2\delta}^{z_{g0}} C |{\bm J}| {\rm d}\Xi_1 {\rm d}\Xi_2 {\rm d}\Xi_3$ and $\langle A \rangle_{z_{g0}} =  \frac{1}{L_1L_2}\int_0^{L_1}\int_0^{L_2} A(\Xi_3=z_{g0}) {\rm d}\Xi_1{\rm d}\Xi_2$ for a given function $A$, respectively.
By using the scalar concentration fluctuation $c = C-\langle C \rangle_{z_{g0}}$ 
and the vertical fluid velocity relative to the velocity of boundary-fitted coordinate $w_3$, the full budget equation is yielded as 
\begin{equation}
\frac{{\rm d} \chi(z_{g0})}{{\rm d} t} 
= F_{\rm diff} 
- \left\langle cw_3 \right\rangle_{z_{g0}} - \langle C \rangle_{z_{g0}} \langle w_3 \rangle_{z_{g0}}
,
\end{equation}
where $F_{\rm diff}$ is given by (\ref{eq:Fdiff}).
Note that the third term vanishes because $\langle w_3 \rangle_{z_{g0}}$ vanishes. 
}

\bibliographystyle{jfm}
\bibliography{jfm-reference}

\end{document}